\documentclass[smallextended]{svjour3}       
\smartqed  
\usepackage{graphicx}
\usepackage{amsmath,amssymb}
\usepackage{color}
\usepackage{layouts}
\usepackage[table]{xcolor}

\usepackage[markup=underlined]{changes}

\definechangesauthor[name=Chris Crabtree,color=red]{cc}

\newcommand{\subsubsubsection}[1]{\bigskip\noindent\textbf{#1}}

\journalname{arxiv.org}
\begin{document}

\title{Behavior of Compressed Plasmas in Magnetic Fields
}


\author{Gurudas Ganguli         \and
        Chris Crabtree$^{*}$           \and
        Alex Fletcher           \and 
        Bill Amatucci 
}


\institute{G. Ganguli \and C. Crabtree$^{*}$ \and A. Fletcher \and B. Amatucci \at
              Plasma Physics Division \\
              Naval Research Laboratory\\
              Washington, DC 20375 \\
              Tel.: +01-202-767-2401\\
              Fax: +01-202-767-3553\\
              $^{*}$\email{chris.crabtree@nrl.navy.mil}
}

\date{Received: date / Accepted: date}

\maketitle

\begin{abstract}
Plasma in the earth's magnetosphere is subjected to compression during geomagnetically active periods and relaxation in subsequent quiet times.  Repeated compression and relaxation is the origin of much of the plasma dynamics and intermittency in the near-earth environment.  An observable manifestation of compression is the thinning of the plasma sheet resulting in magnetic reconnection when the solar wind mass, energy, and momentum floods into the magnetosphere culminating in the spectacular auroral display.  This phenomenon is rich in physics at all scale sizes, which are causally interconnected.   This poses a formidable challenge in accurately modeling the physics.  The large-scale processes are fluid-like and are reasonably well captured in the global magnetohydrodynamic (MHD) models, but those in the smaller scales responsible for dissipation and relaxation that feed back to the larger scale dynamics are often in the kinetic regime.  The self-consistent generation of the small-scale processes and their feedback to the global plasma dynamics remains to be fully explored.   Plasma compression can lead to the generation of electromagnetic fields that distort the particle orbits and introduce new features beyond the purview of the MHD framework, such as ambipolar electric fields, unequal plasma drifts and currents among species, strong spatial and velocity gradients in gyroscale layers separating plasmas of different characteristics, \textit{etc.}  These boundary layers are regions of intense activity characterized by emissions that are measurable.  We study the behavior of such compressed plasmas and discuss the relaxation mechanisms to understand their measurable signatures as well as their feedback to influence the global scale plasma evolution.

\end{abstract}

\section{Introduction}
\label{intro}

The holy grail of much of modern science is the comprehensive knowledge of the coupling between the micro, meso, and macro scale processes that characterize physical phenomena.  This is particularly important in magnetized plasmas which typically have a very large degree of freedom at all scale sizes.  The statistically likely state involves a complex interdependence among all of the scales.  In the unbounded space plasma undergoing global compression during geomagnetically active periods the multiplicity of spatio-temporal scale sizes is astoundingly large.  The statistically likely state has mostly been addressed by global magnetohydrodynamic (MHD) or fluid models, which ignore the contributions from the small-scale processes that can be locally dominant.  This was understandable in the past when the early space probes could hardly resolve smaller scale features.  Also, single point measurements from a moving platform made in evolving plasma are not ideal for resolving the small-scale details of a fast time scale process.  Statistical ensembles generated through measurements from repeated satellite visits in a dynamic plasma washes out many small-scale features that evolve rapidly.  Therefore, the need for understanding the contributions from the small-scale processes was not urgent.  

However, there are pitfalls in relying on global fluid models alone for an accurate assessment of satellite measurements that essentially represent the local physics.  These models ignore the kinetic physics, which often operate at faster time scales at the local level and are necessary for dissipation, which is important for relaxation and feedback to form a steady state that satellites measure.  For example, the large-scale MHD models cannot account for the ambipolar effects and hence they are inadequate for the physics at ion and electron gyroscales, which are now being resolved by multi-point measurements from modern space probe clusters, such as NASA’s Magnetospheric Multi-Scale Satellite (MMS) \cite{Burch:2016fu} , the Time History of Events and Macroscale Interactions during Substorms (THEMIS) mission \cite{Angelopoulos2008}, and the European Space Agency’s Cluster mission \cite{Escoubet:1997}.  

Global scale kinetic simulations that can resolve gyroscales are still not practical. These simulations suffer perennial issues such as insufficient mass ratios, insufficient particles per cell, or use implicit algorithms that ignore the small scale features. Thus, these simulations are incapable of accurately resolving the gyroscales for capturing ambipolar effects, which as we show in Sec. \ref{sec:Equilibrium}, can be critical to the comprehensive understanding of the physics necessary for interpreting satellite observations.  With technological breakthroughs in the future, resolution of gyroscales in global models will become possible.  It is, therefore, necessary to assess the origin of small-scale processes responsible for relaxation and their feedback mechanisms for a deeper understanding and also to motivate future space missions with improved instrumentation to search for them in nature.  The objective of this article is to highlight the fundamental role of plasma compression in the inter-connectedness of physical processes at local and global levels in general, and in particular in the earth’s immediate plasma environment through specific examples. 

Although the large-scale models are not yet suitable for addressing the smaller scale physics, they are necessary for understanding the global morphology and global transport of mass, energy, and momentum that creates the compressed plasma layers when plasmas of different characteristics interface.  In the near term, before first principles kinetic global models become practical, the large-scale fluid models should be extended to include small scale (sub-grid) kinetic physics that is discussed in this article so that the effects of natural saturation and dissipation of compression can be accounted for on a larger scale. Clearly, therefore, the knowledge of large and small scale processes are like the proverbial two sides of a coin, both equally necessary for a comprehensive understanding of the salient physics.   Since the role of smaller scale processes was not central to most previous studies we focus our analysis here to their self-consistent origin and their contributions to the overall plasma dynamics. Arguably, small-scale structures will be increasingly resolved by future technologically-advanced space probes, so there is now a need to accurately understand their cause and effect.

\section{Equilibirium Structure of Compressed Plasma Layers}
\label{sec:Equilibrium}

To understand the physics of compressed layers it is best to consider specific examples of such layers that arise naturally.  Weak compressions, which are characterized by scale sizes much larger than an ion gyrodiameter and affect both the ions and electrons similarly, are not of interest here.  Large-scale models can address them.  The focus of this article is on stronger compressions, characterized by scale sizes comparable to an ion gyrodiameter or less, which affect ions and electrons differently and lead to ambipolar effects that are beyond the scope of electron-MHD (eMHD) frameworks \cite{GORDEEV1994215}.  To address such conditions, we construct the equilibrium plasma distribution function within the compressed layers and analyze the field and flow structures they support in the metastable equilibrium with self-consistent electric and magnetic fields as well as their inherent spatial and velocity gradients.  This specifies the background plasma condition, which can then be used as the basis to study their stability, evolution, and feedback to establish steady state structures.  Such small-scale structures, with scale sizes comparable to ion and electron gyroscales, are being resolved with modern space probes \cite{Fu:2012fv}.  

We use relevant constants of motion to construct the appropriate distribution function subject to Vlasov-Poisson or Vlasov-Maxwell constraints as necessary.  Given the background parameters the solutions provide the self-consistent electrostatic and vector potentials, which then fully specify the equilibrium distribution function, $f_{0}(\mathbf{v},\Phi_{0}(x),\mathbf{A}(x))$ where $\Phi_{0}(x)$ and $\mathbf{A}(x)$ are electrostatic and vector potentials.  In effect, the potentials are Bernstein-Green-Kruskal (BGK) \cite{Bernstein:1957hx} or Grad-Shafranov \cite{Grad:1958,Shafranov1966} like solutions.  With the distribution function fully specified, its moments readily provide the static background plasma features and their spatial profiles.  As input parameters, \textit{i.e.}, boundary conditions, we can use the output from global models if they can accurately produce them.  But since these layers are on the order of ion gyroradii and smaller, which the current generation global models cannot accurately resolve, we rely on high-resolution \textit{in situ} observations to obtain the input parameters.  Given the boundary conditions we allow the density and the potential to freely develop subject to no constraints except quasi-neutrality.  This provides the self-consistent distribution function, as was demonstrated for plasma sheaths by Sestero \cite{Sestero:1964}.

\begin{figure}[t]
	\includegraphics[width=\textwidth,height=.3\textwidth]{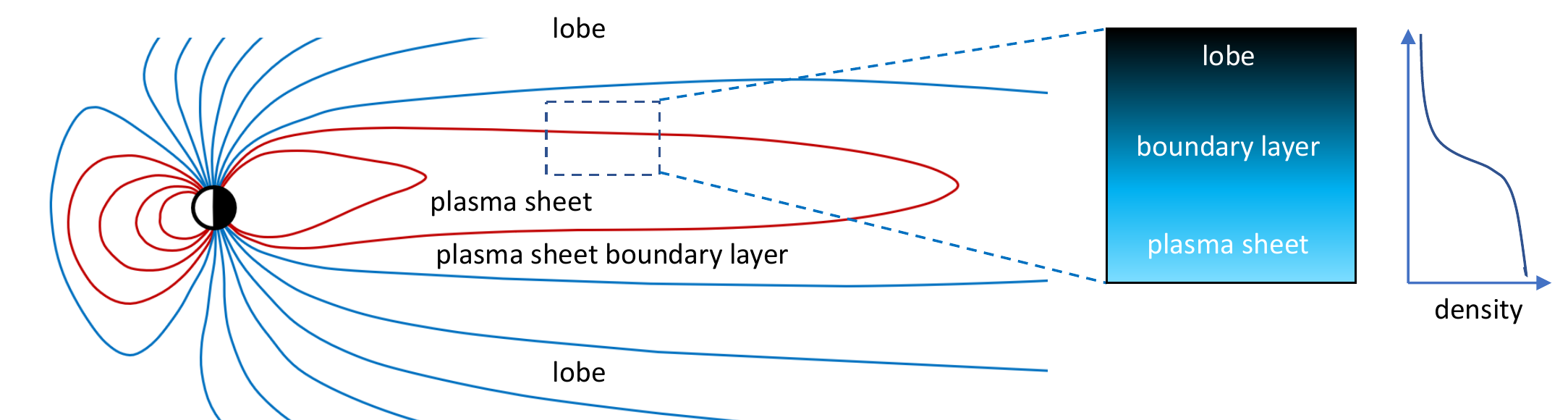}
	\includegraphics[width=0.5\textwidth]{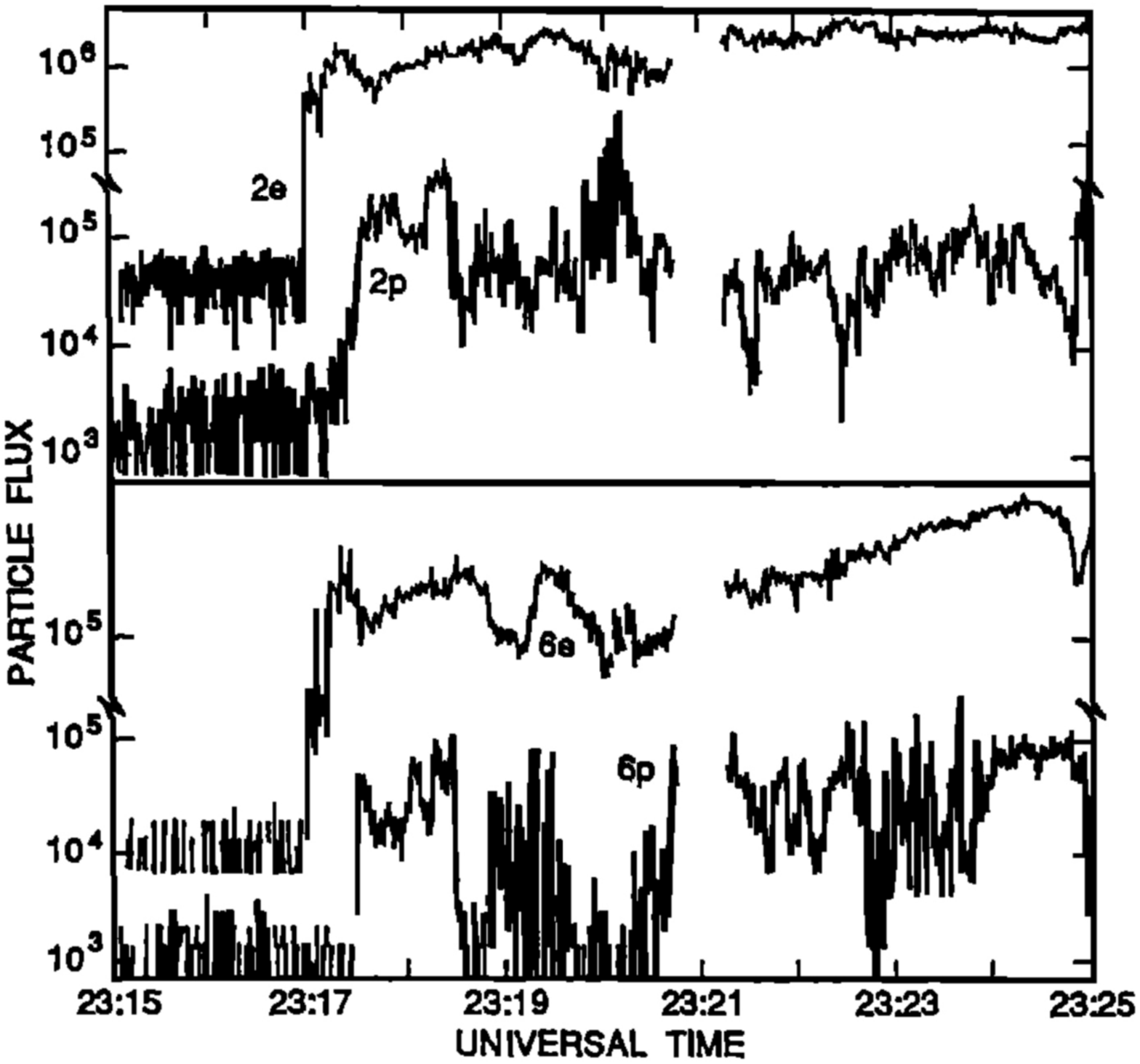}
	\caption{\label{fig:1} (a) Model profile of density at plasma sheet-lob interface. (b) Particle flux data from ISEE 1 (March 31, 1979) versus UT for two energy channels (2 keV and 6 keV).  Figure (b) reproduced from Figure 1 of Romero \textit{et~al.} \cite{Romero:1990fs} }
\end{figure}

\subsection{Vlasov-Poisson System:  Plasma Sheet-Lobe Interface}
\label{subsec:vlasovpoisson}

Consider the compressed plasma layer that is observed at the interface of the plasma sheet and the lobe in the earth’s magnetotail region \cite{Romero:1990fs} as sketched in Fig. \ref{fig:1}a.  The plasma sheet boundary layer is one of the primary regions of transport in the magnetosphere \cite{Eastman:1984}.  This layer separates the hot (thermal energies $>$ 1KeV) and dense (density $\sim$ 1cm$^{-3}$) plasma of the plasma sheet, which is embedded in closed magnetic field lines of the earth, from the cold (thermal energy $\sim$ 10’s of eV) and tenuous (density $\sim$ 0.01 cm$^{-3}$) plasma in open field lines in the lobe.  During geomagnetically active periods, known as substorms, when the coupling of the solar wind energy and momentum to the magnetosphere is strong for southward interplanetary magnetic field, the quantity of magnetic flux and the field strength in the tail lobes increases \cite{Stern:1991,Lui:1991}.  As the tail lobes grow, increasing stress is transmitted to the near earth plasma sheet and the boundary layer becomes narrow approaching gyroscales.  The narrow boundary layer is characterized by intense broadband emissions \cite{Grabbe:1984}.  Fig. \ref{fig:1}b is an example as observed by the ISEE satellite in which the layer was around half of an ion gyroradius within which the density drops by two orders of magnitude \cite{Romero:1990fs}.   

\subsubsection{Derivation of the equilibrium distribution function}
\label{subsec:esderivoff}

To obtain the equilibrium distribution function of such boundary layers we consider the region (see inset in Fig. 1a) where the magnetic field lines are nearly straight so that the curvature that exists close to the equatorial plane can be neglected.  The neglect of the curvature may be justified because its scale size, $L_{\|}$, is much larger than the gradient scale size, $L_{\perp}$, across the magnetic field, \textit{i.e.}, $L_{\perp}\sim \rho_i \ll L_{||}$, and $L_{\|}=(\partial \log(B)/\partial s)^{-1}$ where $s$ is the position along the magnetic field line and $\rho_i$ is the ion gyroradius.  This simplifies the problem by reducing it to essentially one dimension across the magnetic field in the x-direction in which the spatial variation is much stronger than it is along the magnetic field.   

To represent the pressure gradient in the x-direction we construct a distribution function using the relevant constants of motion, which are the guiding center position, $X_g=x+v_y/\Omega_\alpha$, and the Hamiltonian, $H_\alpha (x)=m_\alpha v^2/2+q_\alpha \Phi_{0} (x)$, $\Omega_{\alpha}=q_{\alpha} B/(m_{\alpha} c)$ is the cyclotron frequency where the subscript $\alpha$ represents the species, $m_\alpha$ is the mass, $q_\alpha$ is the charge and $\Phi_{0}(x)$ is the electrostatic potential, so that it is approximately a Maxwellian far away from the boundary layer on either side 
\begin{equation}
\label{eq:vp_f0}
	f_{0\alpha}(X_{g\alpha},H_{\alpha}(x)) 
	=
	\frac{N_{0\alpha}}{(\pi v_{t\alpha}^2)^{3/2}} Q(X_{g\alpha})
	\exp\left(-\frac{H_{\alpha}(x)}{T_{\alpha}}\right).
\end{equation}
The magnetic field is assumed in the $z$ direction and the pressure gradient is normal to the magnetic field in the $x$ direction. (Note that this is not the GSM coordinate system.)  The electron and ion thermal velocity is given by $v_{t\alpha}$,  $T_{\alpha}=m_{\alpha} v_{t\alpha}^2/2$ is the temperature away from the layer, and $Q_\alpha$ is the distribution of guiding centers, the shape of which is motivated by the observed density structures across the layer and is given by,
\begin{equation}
\label{eq:qdef}
	Q_{\alpha}(X_{g\alpha}) = \left\{ \begin{array}{lc}
R_{\alpha} & X_{g\alpha} < X_{g1\alpha}
 	\\
 R_{\alpha} + (S_{\alpha}-R_{\alpha})\left(\frac{X_{g\alpha}-X_{g1\alpha}}{X_{g2\alpha}-X_{g1\alpha}}\right) & 
 X_{g1\alpha}<X_{g\alpha}<X_{g2\alpha}
 \\
 S_{\alpha} & X_{g\alpha} > X_{g2\alpha}.
 \end{array}
	\right.
\end{equation}	
$N_{0\alpha} R_\alpha$ and $N_{0\alpha} S_\alpha$ are the densities in the asymptotic high (plasma sheet) and low-pressure (lobe) regions respectively, but in the transition layer the density and its spatial profile is determined self-consistently. The quantity $|S_{\alpha}-R_{\alpha} |$ is proportional to the pressure difference between the asymptotic regions and $|X_{g2\alpha}-X_{g1\alpha} |$ represents the distance over which the pressure changes. These quantities determine the magnitude and the scale-size of the electrostatic potential, which in turn determines the characteristics of the emissions that are excited at the boundary, as elaborated in Section 3.  Different values of the parameters $X_{g1\alpha} $ and $X_{g2\alpha}$ may be chosen to reproduce the observed density profile. Hence, the values of the parameters $R_{\alpha}$, $S_\alpha$, $X_{g1\alpha}$, and $X_{g2\alpha}$ are model inputs determined from observations.  These parameters reflect the global plasma condition, \textit{i.e.}, the compression.  Hence, they causally connect the small scale processes to the larger scale dynamics.  

The density structure within the boundary layer is obtained in terms of the electrostatic potential as the zeroth moment of the distribution function, Eq. (\ref{eq:vp_f0}),
\begin{equation}
	n_{0\alpha}(x)\equiv \int f_{0\alpha}(\mathbf{v},\Phi_{0}(x)) d^3\mathbf{v} 
	= N_{0\alpha} \frac{(R_{\alpha}+S_{\alpha})}{2}\exp\left(-\frac{e\Phi_{0}(x)}{T_{\alpha}}\right)
	I_{\alpha}(x)
\end{equation}
where
\begin{multline}
\label{eq:Ialphadef}
	I_{\alpha}(x) = 1\pm \left(\frac{R_{\alpha}-S_{\alpha}}{R_{\alpha}+S_{\alpha}}\right)
	\left(\frac{1}{\xi_{1\alpha}-\xi_{2\alpha}}\right)
	\times
	\\
	\left[\xi_{2\alpha}\textrm{erf}(\xi_{2\alpha})
	-\xi_{1\alpha}\textrm{erf}(\xi_{1\alpha})\right]
	+\frac{1}{\sqrt{\pi}}
	\left[\exp(-\xi_{2\alpha}^2)
	- \exp(-\xi_{1\alpha}^2)\right]
\end{multline}
$\textrm{erf}$ is the error function, $\xi_{1,2\alpha}=\Omega_{\alpha} (x-X_{g1,2\alpha})/v_{t\alpha}$, and $\pm$ refers to the species charge.  The quasi-neutrality, $\sum_{\alpha} q_{\alpha} n_{0\alpha}(x,\Phi_{0}(x))=0$ , then determines $\Phi_{0} (x)$, which in the limit that the Debye length is smaller than the plasma scale length (which is well satisfied here) is equivalent to solving Poisson’s equation. The existence of the transverse electric field reflects the strong spatial variability and nonlocal interactions that exist across the magnetic field due to the difference in the electron and ion distributions with their characteristic spatial variations. With $\Phi_{0}$ determined the distribution function is fully specified and higher moments can be obtained. This distribution function satisfies the Vlasov-Poisson system and is similar to the BGK class of solutions.

As in the previous studies \cite{Romero:1990fs,Ganguli:94b}, the temperature variation across the layer is ignored in the above.  However, there is a temperature gradient between the plasma sheet and the lobe that can affect the static background properties.  The effects of the temperature gradient can be accounted for by considering two different types of plasma population characterized by their respective temperature and density in the asymptotic regions of the plasma sheet and the lobe, assuming isothermal condition exists in both the regions away from the boundary layer.    While the plasma sheet population, denoted by subscript ‘ps’, goes to zero in the lobe, achieved by setting $R_{\alpha,ps}=1$ and $S_{\alpha,ps}=0$, the lobe population, denoted by subscript $l$, does just the opposite in the same interval $|X_{g2\alpha}-X_{g1\alpha} |$ by setting $R_{\alpha,l}=0$ and $S_{\alpha,l}=1$. 

To obtain $\Phi_{0}$ quasi-neutrality must be maintained between all populations, i.e., 
\begin{equation}
	\sum_{\alpha} q_{\alpha} (n_{0\alpha,ps}(x,\Phi_{0}(x)) + n_{0\alpha,l}(x,\Phi_{0}(x))=0 
\end{equation}
The assumption that the transition in both density and temperature takes place in the same interval is for simplicity and can be relaxed.  If the intervals differ somewhat, then the details of the spatial variation in the potential profile can be affected.  However, these are higher level details and may not be observable due to averaging by the waves that are spontaneously generated by the highly non-Maxwellian distribution functions that develop as we elaborate in Sec. \ref{sec:lineartheory}.  

In addition to the transverse electric field, the interface between the plasmasheet and the lobe is also characterized by ion and electron bi-directional beams along the magnetic field \cite{Takahashi:1988}.  In Sec. \ref{sec:vpequi_curv} we argue that the origin of these beams could be related to the curvature in the magnetic field around the equatorial region, which we ignored here, and not necessarily due to the reconnection process as it is usually assumed.

\begin{figure}[t]
	\includegraphics[width=0.6\textwidth]{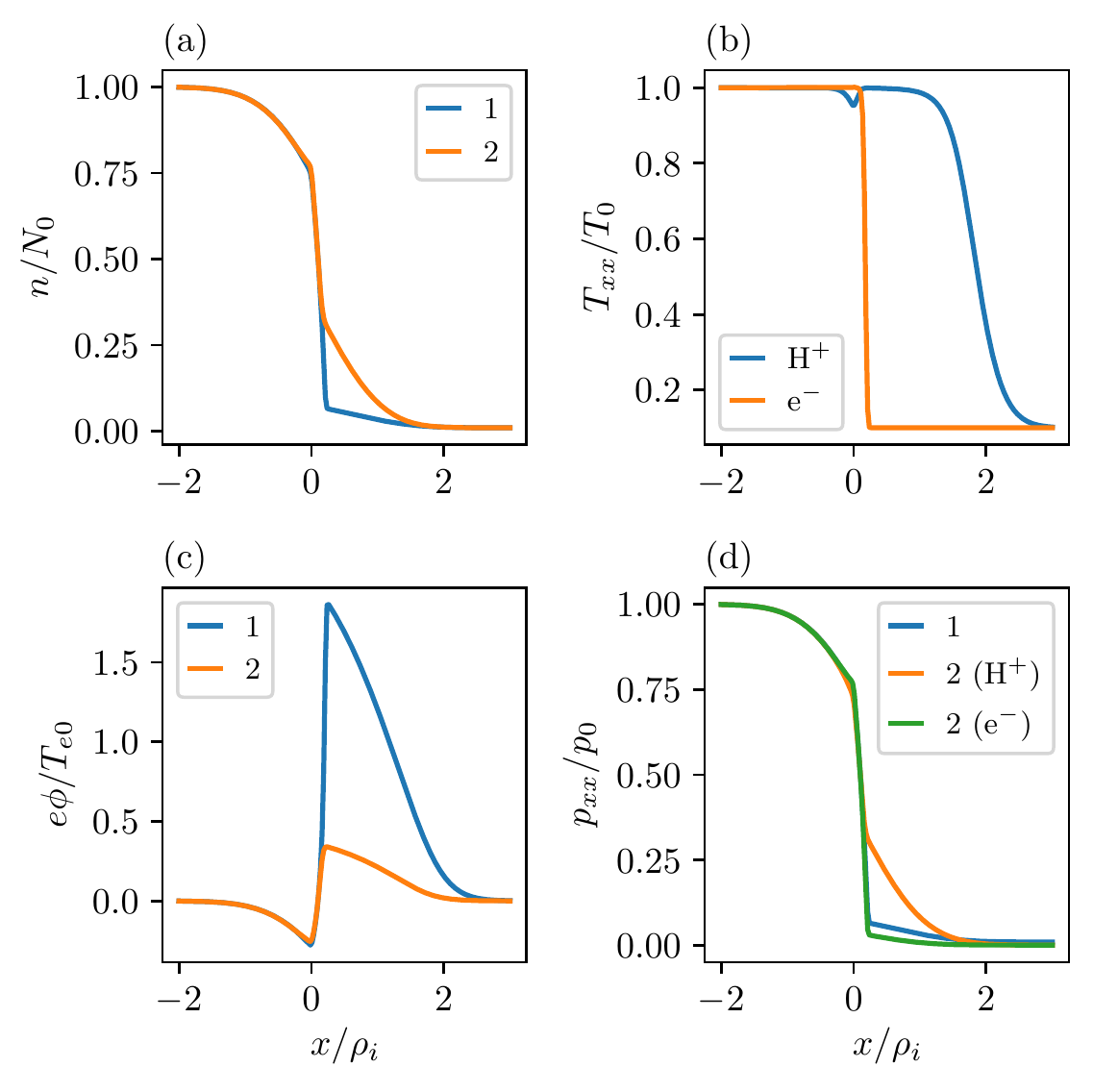}
	\caption{\label{fig:Fig2} Comparison between an equilibrium with a single uniform temperature, labeled 1 in the figure, and an equilibrium with a uniform temperature to the left of the layer and a different uniform temperature to the right of the layer, labeled 2 in the figure. (a) Density of two models.  (b) Temperatures across the layer for model 2.  (c) Electrostatic potential for both models.  (d) Pressures across the layer for both models.  For both models the parameters are as follows $x_{g1i,e}=0,0$, $x_{g2i,e}=0.2,0.2$, $R_{i,e}=1.0,1.0$, $S_{i,e}=0.01,0.01$, $T_e/T_i=1.0$, and $m_i/m_e=1836.0$ and for the two temperature model the temperature to the right of the layer is $T_{e,i1}/T_{e,i}=0.1,0.1$.  }
\end{figure}

\subsubsection{Equilibrium features}
\label{subsec:equi_features}

To understand the effects of a temperature gradient in the boundary layer we first consider a case where the density and the temperature gradients are in the same direction and then in the opposite direction.  Fig. \ref{fig:Fig2} is a comparison of the attributes for an equilibrium with only one temperature as was analyzed in Romero \textit{et al.} \cite{Romero:1990fs} and the two temperature model as described in Sec. \ref{subsec:esderivoff}, \textit{i.e.}, different populations in the lobe and the plasma sheet each characterized by their respective temperature and density. The temperature gradient of both populations is in the same direction as the density gradient.  This example underscores the kinetic origin of the equilibrium electric field.  In the two temperature model, the temperature reduces by a factor of 50 going from the high density side to the low density side, thus the total pressure drop from plasma sheet to lobe is larger.  From a fluid perspective (eMHD) one would expect that the larger pressure gradient must induce a larger electric field to maintain the pressure gradient, however, as one sees in panel (c) this is not the case.  The electrostatic potential and the magnitude of the electric field is reduced.  This is because the ambipolar effect, which scales as $(\rho_{i}-\rho_{e})$ averaged over the distribution, has been reduced by the decrease in the temperature, as ambipolar effects vanish with temperature.  The $x$-axis of both plots is normalized to the constant thermal ion gyroradius calculated to the left of the layer.  However, in the two temperature model the actual thermal ion gyroradius decreases by a factor of $\sqrt{T_{l}/T_{ps}}\simeq0.2$, where $T_{l}$ is the temperature of the lobe plasma and $T_{ps}$ is the temperature of the plasma sheet.  This means that the ratio of the ion to electron gyroradius has decreased and thus the kinetic source of the electrostatic potential has reduced. 
\begin{figure}
	\includegraphics[width=0.6\textwidth]{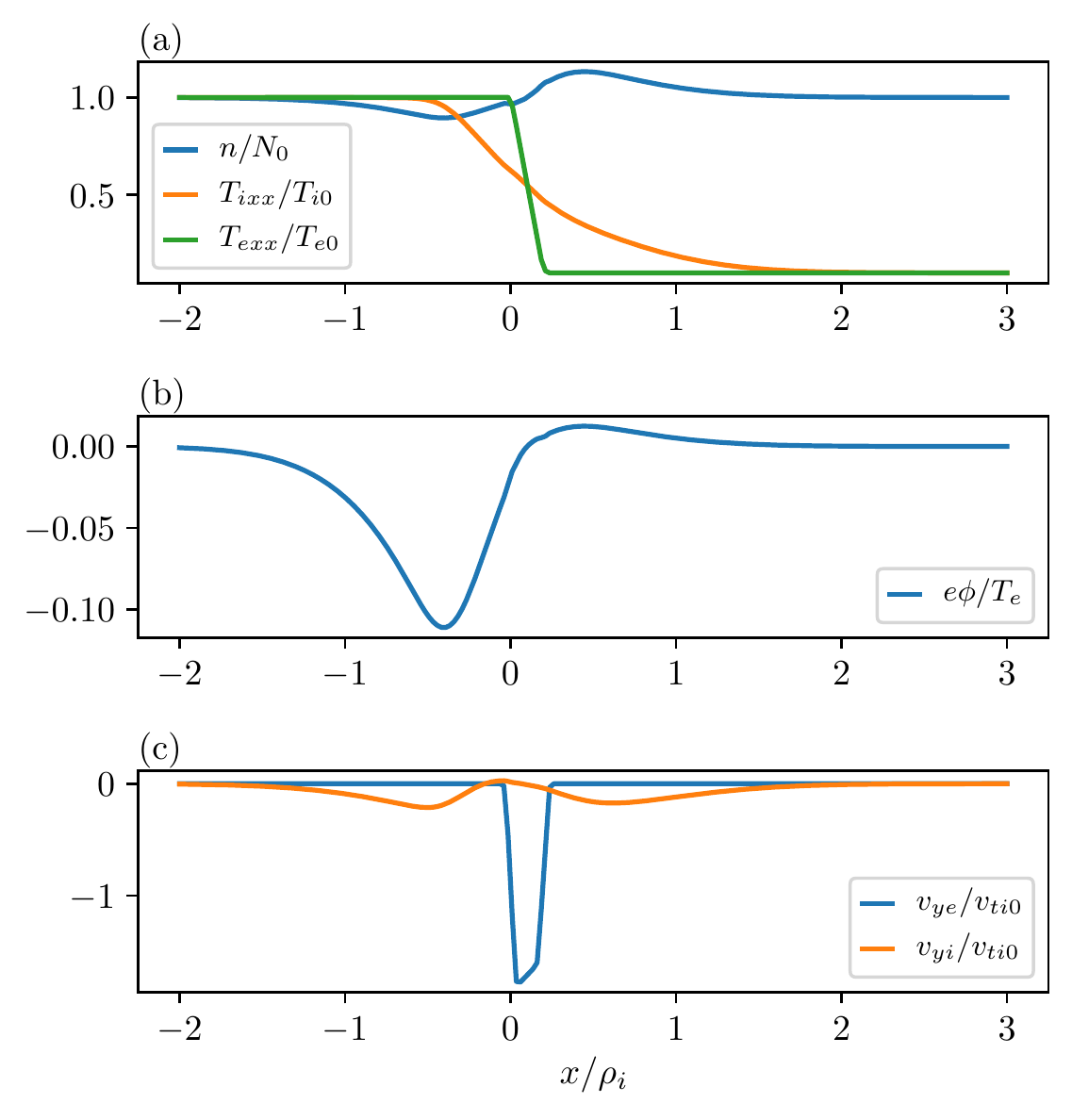}
	\caption{\label{fig:Fig3}Generation of an electric field by a temperature gradient and no imposed density gradient.  (a) Density and Temperatures across the layer. (b) Electrostatic potential.  (c) Flow velocities  normalized to the ion thermal velocity defined to the left of the layer.  The parameters are as follows $a_{1i,e}=0,0$, $a_{2i,e}=0.2,0.2$, $R_{i,e}=1,1$, $S_{i,e}=1,1$, $T_e/T_i=1.0$, and $m_i/m_e=1836.0$ and the temperatures to the right of the layer is $T_{e,i1}/T_{e,i}=0.1,0.1$.}
\end{figure}

As further illustration of the ambipolar effect we show an extreme case in Fig. \ref{fig:Fig3} where we have chosen the asymptotic density to be the same on either side of the layer by choosing the distribution of the guiding centers, $Q(X_{g})$, to be a constant but have allowed the Temperature to fall from $T_{e0}$ to $0.05T_{e0}$ across the layer.  We can see that the temperature gradient creates a change in the difference between the ion and electron gyroradius which generates the ambipolar electric field and the density in the layer adjusts to accommodate the ambipolar potential even though the guiding center distribution is constant.  We note that in this case there is a clear electron flow channel within the layer mostly due to $E\times B$ drift and sheared flow in both the ions and electrons that can be the source of instabilities as discussed in Section \ref{sec:lineartheory}.  This also implies that for the temperature gradient driven modes \cite{Rudakov:1961,Pogutse:1967,Coppi:1967} the effect of the self-consistent electric field must be examined.

\subsubsection{Bulk plasma flows in narrow layers}
\label{subsec:diff_fluid_v_kinetic}

It is important to understand the origin and nature of the flows and currents in the compressed plasma layers because they are the sources of free energy for waves that determine the nonlinear evolution of the layers.  The bulk flow characteristics change as the layer widths become less than an ion gyrodiameter.  The flows are associated with the density and temperature gradients and the ambipolar electric field that develop in the layer as a consequence of the compression.  The resulting $E\times B$ drift may not be identical for the electrons and the ions as we elaborate in the following.  


From the Vlasov equation we can calculate the equilibrium momentum balance and using the geometry of our equilibrium we can solve for the fluid (or bulk) flow in the $y$ direction as
\begin{equation}
	\label{eq:force_balance}
	V_{\alpha} = \frac{-cE_x}{B} + \frac{c}{B}\frac{1}{q_{\alpha} n_{\alpha}}\frac{d P_{\alpha xx}}{dx}
\end{equation}
where the first term is the $E\times B$ drift, $V_{E}$, and the second term is the diamagnetic drift, $V_{\nabla p}$.  While this relationship is completely general for this geometry and applies to fluid and kinetic plasmas, the relative strength of each drift may vary between fluid and kinetic approaches.  This is because individual particle orbits are important in the kinetic approach but not in the fluid approach.  It is especially important in narrow layers when the particle orbits become species dependent (Sec. \ref{sec:lineartheory}) and the ambipolar effects dominate the physics.  This leads to unique static background conditions, which influences the dynamics and hence the observable signatures, as we shall see in Sections 3 and 4.     

\begin{figure}
	\includegraphics[width=0.6\textwidth]{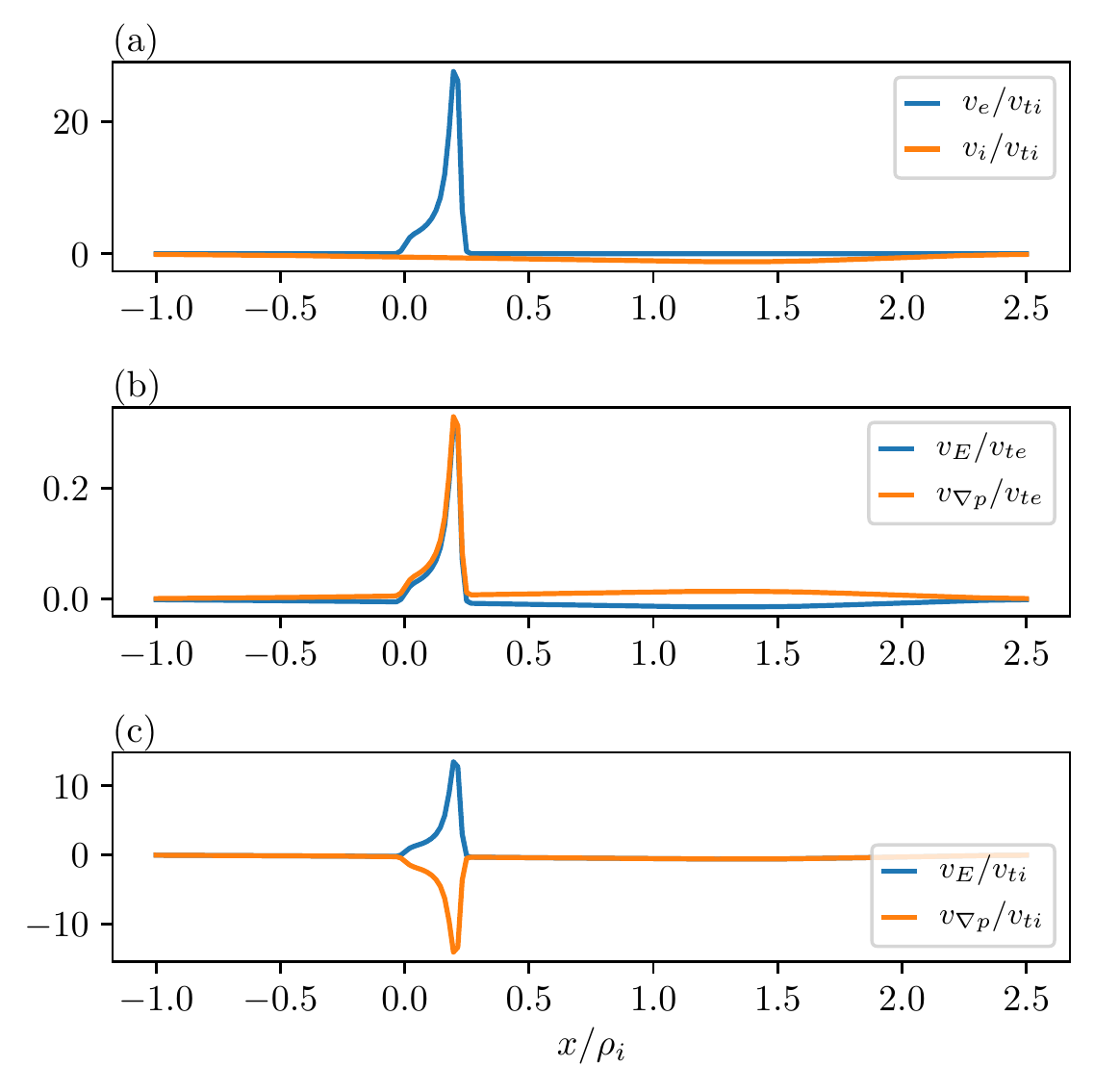}
	\caption{\label{fig:Fig4}Comparison of fluid flows and drift velocities. (a) electron and ion flows, (b) electron drifts,  (c) ion drifts.  The parameters are as follows $x_{g1i,e}=0,0$, $x_{g2i,e}=0.2,0.2$, $R_{i,e}=1.0,1.0$, $S_{i,e}=0.01,0.01$, $T_e/T_i=1.0$, and $m_i/m_e=1836.0$.}
\end{figure}

In Fig. \ref{fig:Fig4} we show the fluid flows in panel (a), the electron drift components in panel (b), and the ion drift components in panel (c) for the case presented in Fig. \ref{fig:Fig2} with no temperature gradient.  The layer width is larger than the electron gyroradius but smaller than the ion gyroradius.  Note that the fluid velocity of the electrons is far larger than the ions.  In addition, the electron $E\times B$ drift and the diamagnetic drift are in the same direction within the layer whereas for the ions these drifts are in the opposite direction.  When the ion drifts combine these components within the layer mostly cancel and the net ion fluid flow becomes negligible compared to the electrons.  Thus the Hall current is mostly generated by the electron flows and localized over electron scales.  This can be understood in the following way.  The ions have a large gyroradius compared to the scale size of the electric field and the density gradient.  So the orbit-averaged $E\times B$ drift experienced by the ions is a fraction of what is expected from the zero gyroradius limit.  This shows up in a fluid representation as in Eq. \ref{eq:force_balance}, by the development of a fluid diamagnetic drift component in the opposite direction to reduce the net ion flow.  Note that for layer widths larger than an ion gyrodiameter the ambipolar electric field will be negligible and the net current will be due to electron and ion diamagnetic drifts in the opposite directions.

\begin{figure}
	\includegraphics[width=0.6\textwidth]{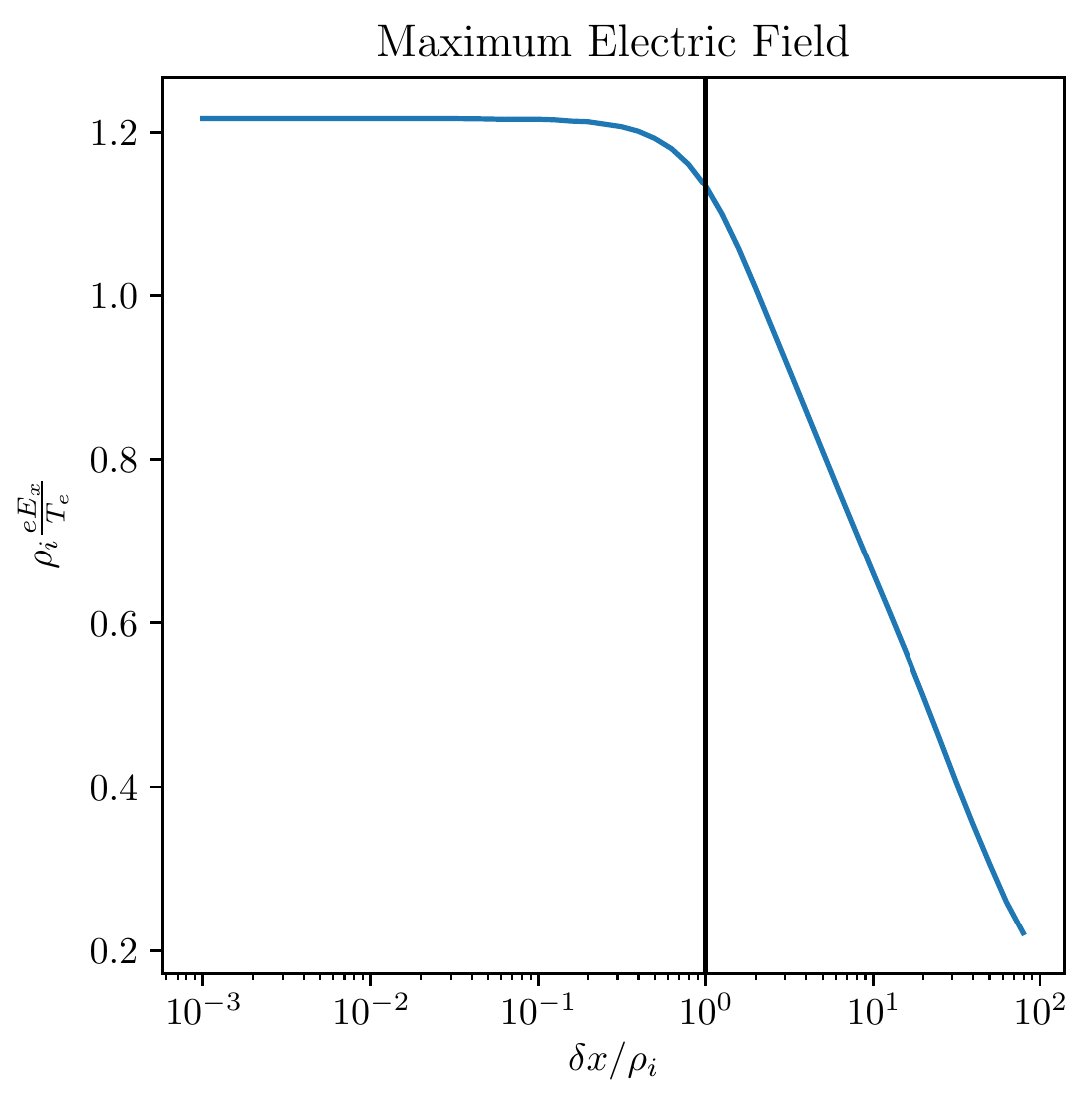}
	\caption{\label{fig:Fig5}Maximum electric field as a function of the layer width normalized to the ion gyroradius.  The ion and electron layer locations are the same.  The parameters are as follows $R_{i,e}=1.0,1.0$, $S_{i,e}=0.01,0.01$, $T_e/T_i=1$, and $m_i/m_e=1836.0$.}
\end{figure}

In narrow layers of widths comparable to the ion gyroradius but larger than an electron gyroradius the kinetic origin of the electric field from compression of a plasma is shown in Figure \ref{fig:Fig5}.  In this figure we keep all parameters of the equilibrium the same but vary the width of the layer $\delta x=X_{g1}-X_{g2}$, over which the density changes by a factor of 100.  As we decrease the layer width the maximum electric field seen in the layer increases (as one would expect from fluid theory) until the layer width gets below the ion gyroradius and then saturates asymptotically.  The  ambipolar electric field becomes strong when the density gradient scale size, $L_n$, becomes less than an ion gyrodiameter.  Consequently, on average there are insufficient electrons, with much smaller gyroradii, to charge neutralize the ions over their large gyro-orbit.  As a result, a charge imbalance is generated proportional to ($\rho_i-\rho_e$) averaged over the distribution, which leads to the electric field.  As $\delta x$ reduces, this imbalance increases because there are fewer electrons that can overlap the larger extent of the ion orbit.  When $\delta x$ falls below an ion gyroradius then there are hardly any electrons that can do the job and, as a result, the value of the averaged ($\rho_i-\rho_e$) reaches saturation asymptotically.  Hence the electric field saturates and its scale size becomes independent from $L_{n}$.  In contrast, in a fluid model (e.g. eMHD) $L/L_n=1$ remains valid throughout the layer because the electric field is directly proportional to the density gradient for constant temperature. The proportionality of the electric field with the pressure gradient breaks down as the ambipolar electric field saturates for gradient scales smaller than an ion gyroradius.   

\begin{figure}
	\includegraphics[width=0.6\textwidth]{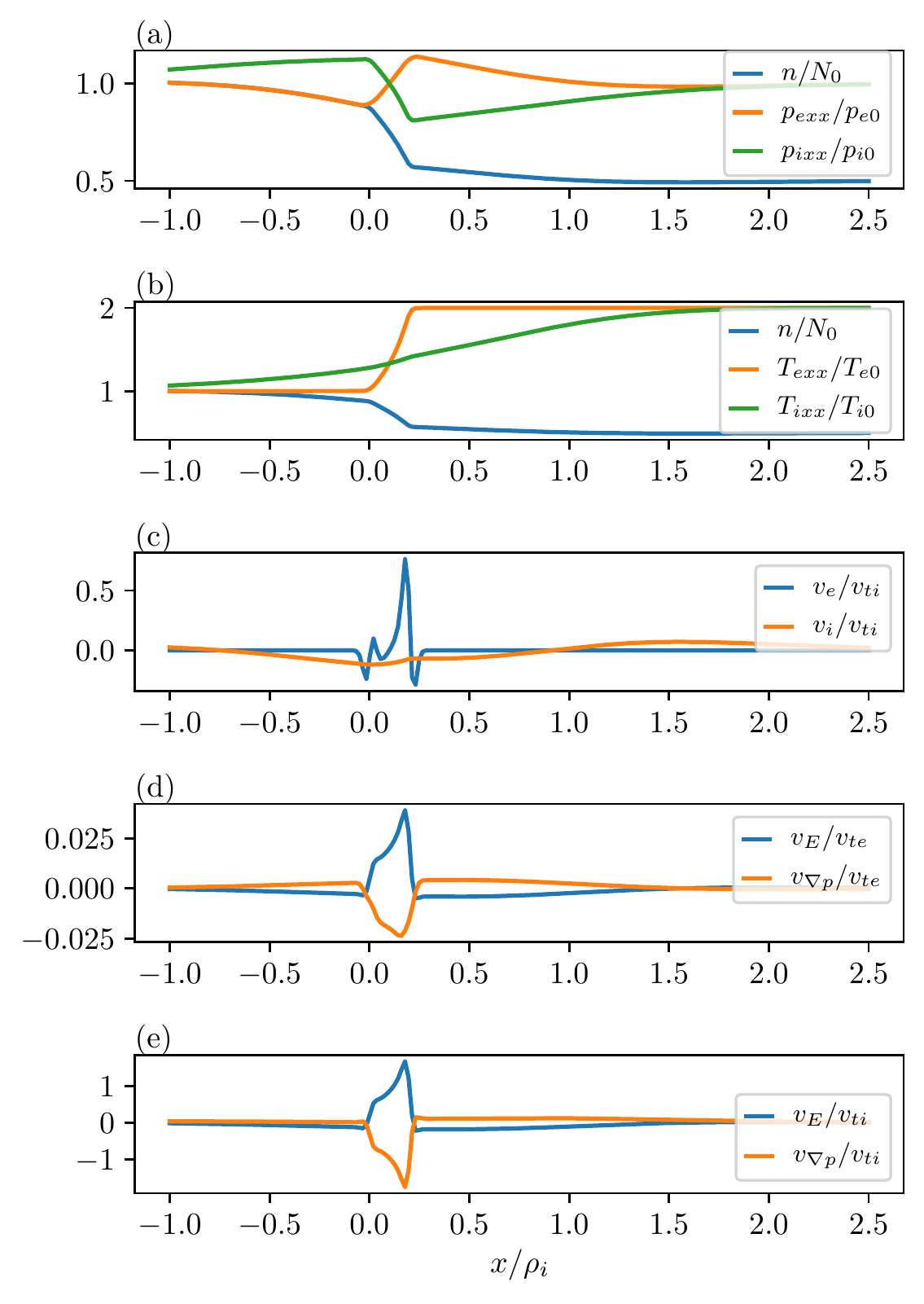}
	\caption{\label{fig:Fig6}Equilibrium where the guiding center density falls by a factor of two from the left to the right and the temperature in the right asymptotic region is twice as high as the temperature in the left asymptotic region.  (a) Density and Pressures.  (b) Density and Temperatures.  (c) Electron and ion fluid velocities.  (d) Electron drift velocities  normalized to the electron thermal velocity defined to the left of the layer.  (e) Ion drift velocities normalized to the ion thermal velocity defined to the left of the layer.  The parameters are as follows $x_{g1i,e}=0,0$, $x_{g2i,e}=0.2,0.2$, $R_{i,e}=1,1$, $S_{i,e}=0.5,0.5$, $T_e/T_i=1.0$, and $m_i/m_e=1836.0$ and the temperatures to the right of the layer is $T_{e,i1}/T_{e,i}=2.0,2.0$
}
\end{figure}

In Fig. \ref{fig:Fig6} we consider the case in which the temperature and the density gradients in the transition layer are in the opposite directions. We model this by two plasma populations in either side of the layer with characteristic density and temperatures.  While the guiding center density (\textit{i.e.}, $Q(X_g)$) of the low temperature population in the left of the transition region drops by a factor of two across a layer that has a width of $\delta x=0.2\rho_{i}$, the guiding center density of the high temperature population in the right of the layer rises by a factor of 2 in the same interval. The pressures are the same in the asymptotic regions to the left and the right of the layer. Quasi-neutrality determines the details of the spatial variation of the density and temperature of each species in the layer.  Panel (a) shows the electron and ion pressures and the densities. One can see that the ion pressure falls across the layer, while the electron pressure rises. This can be understood in the following way.  Since the layer width is much larger than the electron gyroradius the population on the left and right effectively mix only within the layer.  While the electron temperature increases across the layer, the density falls.  However, the density reduction does not fall as much as the guiding center density because it is partly compensated by the ambipolar electric field.  Consequently, the electron pressure inside the layer rises.  Since the ion gyroradius is much larger than the layer width the ions effectively mix on a scale larger than the layer width.  So the ion temperature change is much smaller than the electrons across the layer.  However, quasi-neutrality forces the ion density to be identical to the electrons, which decreases across the layer from left to right.  The combination of these two effects lowers the ion pressure in the layer.  Panel (b) shows that the net electron fluid flow dominates the net ion flow. The individual drift components are plotted in panels (c) and (d).  Both the ion and electron $E\times B$ and diamagnetic drifts are in opposite directions.  In contrast, Fig. 4 showed that in the absence of a temperature gradient the electron $E\times B$ and diamagnetic drifts were in the same direction.  This was because both the ions and electrons experienced the identical pressure gradient within the layer.  In this case, from panel (a) in Fig. 6, we see that the electron and ion pressure gradients are in the opposite directions within the layer even though asymptotically the pressure is constant on either side of the layer.

\begin{figure}
	\includegraphics[width=0.6\textwidth]{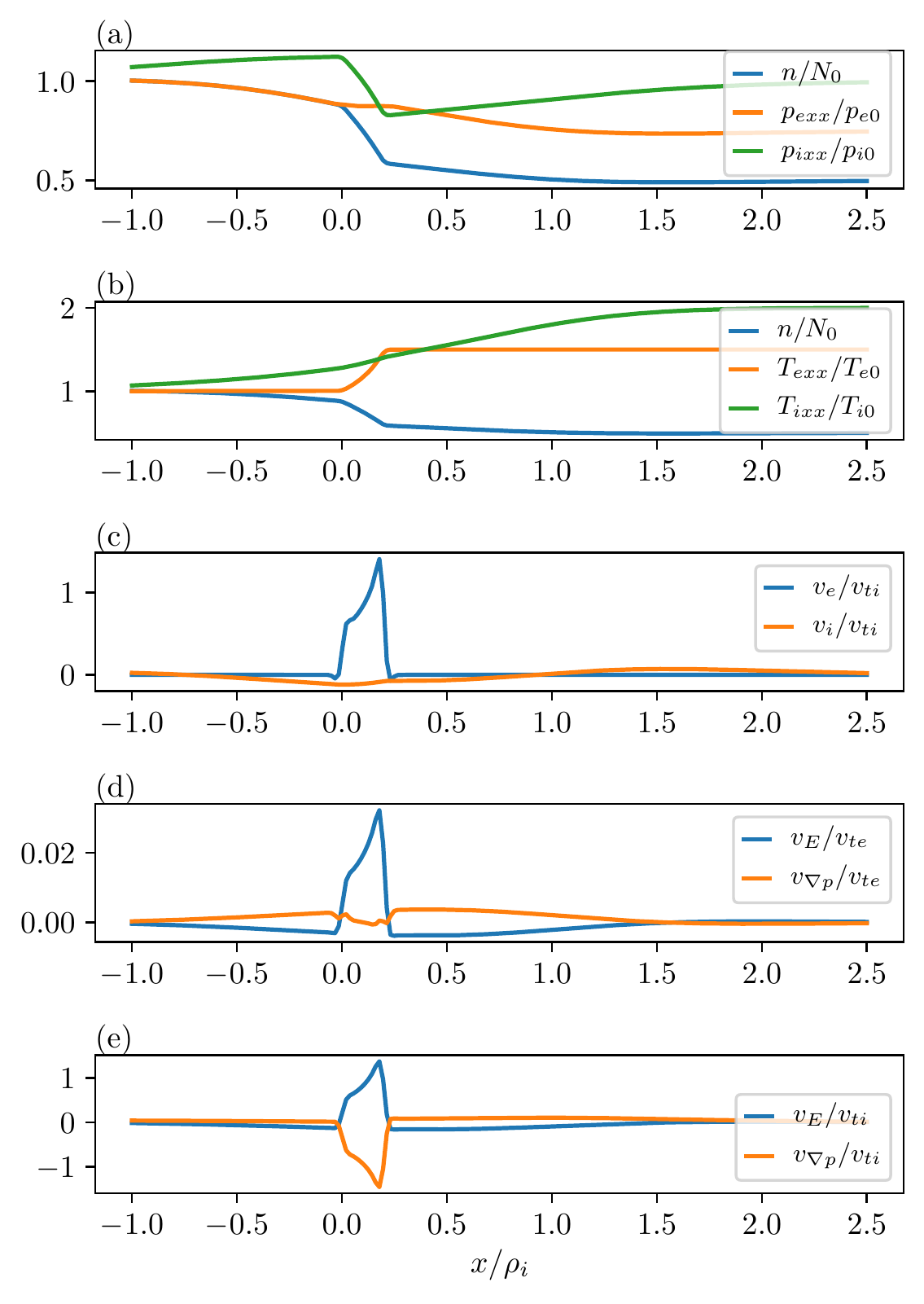}
	\caption{\label{fig:fig7}Equilibrium where the guiding center density falls by a factor of two and the ion temperature in the right asymptotic is twice as high as the temperature in the left asymptotic region but the electron temperature is only 1.5 times less.  (a) Density and Pressures.  (b) Density and temperatures.  (c) Electron and ion fluid velocities.  (d) Electron drift velocities  normalized to the electron thermal velocity defined to the left of the layer.  (e) Ion drift velocities normalized to the ion thermal velocity defined to the left of the layer.  The parameters are as follows $x_{g1i,e}=0,0$, $x_{g2i,e}=0.2,0.2$, $R_{i,e}=1,1$, $S_{i,e}=0.5,0.5$, $T_e/T_i=1.0$, and $m_i/m_e=1836.0$ and the temperatures to the right of the layer is $T_{e,i1}/T_{e,i}=2.0,1.5$
}
\end{figure}

While setting the asymptotic pressure to be equal on either side of the layer was not a sufficient condition to avoid the production of a pressure gradient in the layer, by reducing the asymptotic electron temperature (\textit{i.e.}, pressure) on one side it is possible to create a region where the electron pressure is almost constant across the layer. We illustrate this in Fig. \ref{fig:fig7}. In this case the electron pressure is almost constant across the layer and consequently the electrons have only a small diamagnetic drift as can be seen in panel (c) even though, asymptoticly, there is a pressure difference. From panel (d) we see that the ion $E\times B$  and the diamagnetic drift cancel each other leading to negligible net ion flow as seen in panel (b). Thus, the net flow within the layer is primarily due to electron EXB drift. This shows that depending on the boundary condition, as in this case with different pressures in the asymptotic regions, it is possible to generate a layer with no diamagnetic current but an electron Hall current.  This is typically, the situation in the dipolarization fronts as we shall discuss in Section 2.2 (See also Fu \textit{et~al.} \cite{Fu:2012fv}).  Also, as we will see in Section 3, this condition can lead to waves around the lower hybrid frequency driven by the gradient in the electron $E\times B$ flow that can be misinterpreted to be the lower hybrid drift instability, which results in a different nonlinear state that is measurable.  Interestingly, the eMHD description of such layers with a negligible pressure gradient would predict a stable condition.  This underscores the importance of the kinetic details of compressed plasma layers for accurately analyzing satellite data and assessing the salient physics. Satellites measure the local physics that operates in the layers where the fluid concept does not hold.

\subsection{Vlasov-Maxwell System:  Dipolarization Fronts}
\label{subsec:df_eq}

\begin{figure}
\includegraphics[width=0.6\textwidth]{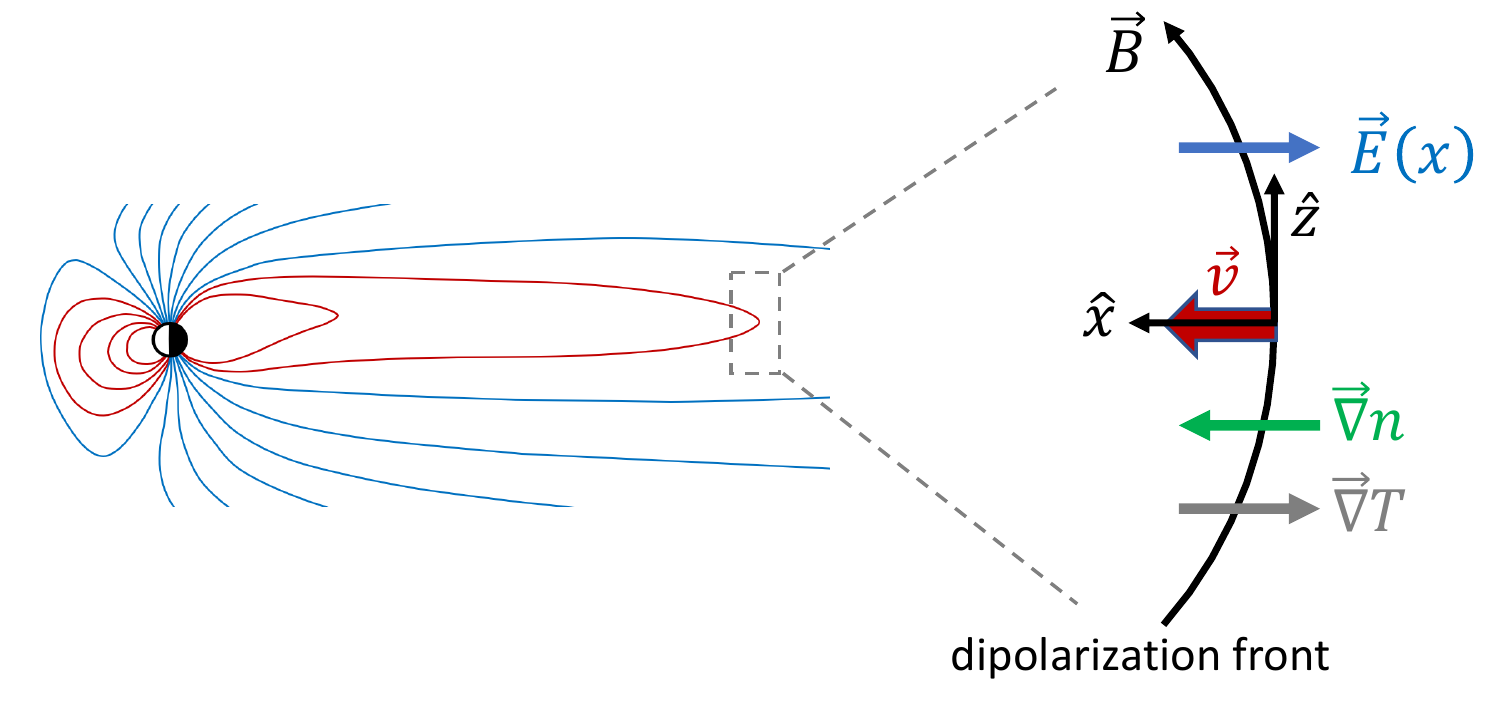}
\caption{\label{fig:fig8} Equitorial dipolarization front geometry.}
\end{figure}

\begin{figure}
\includegraphics[width=0.6\textwidth]{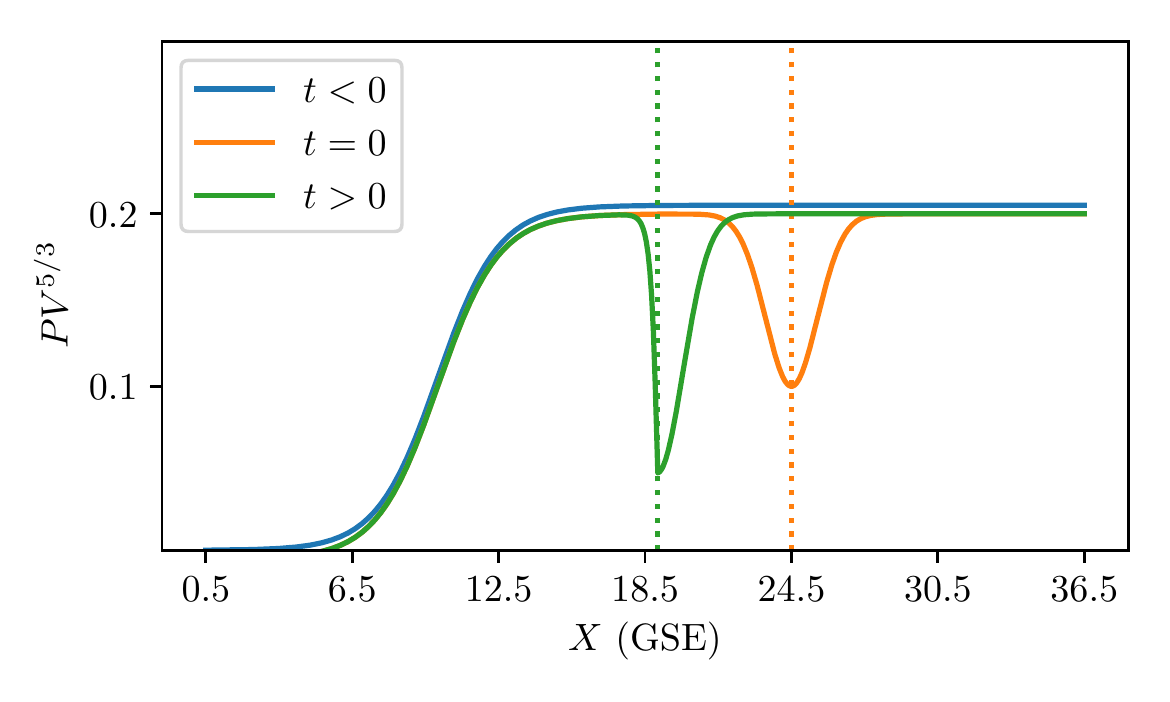}
\caption{\label{fig:fig9}.  Profile of $PV^{5/3}$ in typical magnetotail.  Some event depletes flux tubes with some maximum depletion.  The earthward speed of the DF is proportional to $\Delta PV^{5/3}$ which causes the front to steepen as it propagates.}
\end{figure}

In Sec. \ref{subsec:vlasovpoisson} we considered compressed plasmas in which electromagnetic corrections could be ignored.  This may not be possible for all compressed plasma systems, especially when the ratio of the plasma kinetic pressure to the magnetic pressure, $\beta$, is large such as a dipolarization front (DF) \cite{Nakamura:2002,Nakamura:2009,Runov:2009hl}.  The typical geometry of a DF is sketched in Fig. \ref{fig:fig8}.   DFs are observationally characterized by a rapid rise in the northward component of the magnetic field, a large earthward flow velocity, a sharp drop in the plasma density, and the onset of broadband wave activity \cite{Deng:2010kg}. These changes in plasma parameters are due to a flux tube rapidly propagating past the observing spacecraft. DFs are often observed during bursty bulk flow (BBF) events \cite{Angelopoulos:1992,Runov:2009hl}, during which large-scale magnetic flux tubes that have been depleted of plasma by some event (likely transient reconnection) propagate rapidly towards the Earth to equalize the quantity $pV^{5/3}$ \cite{ChenWolf:1993}, where $p$ is the plasma thermal pressure and $V$ is the flux tube volume.  Flux tubes that have been depleted more than neighboring flux tubes will have a larger earthward velocity, leading to a compression of the plasma at the edge as the faster moving flux tube overtakes the slower moving flux tube (See Figure \ref{fig:fig9}). This compression maintains the plasma gradients in a narrow layer with widths comparable to an ion gyroradius or smaller as the flux tube propagates Earthward.  A kinetic equilibrium solution to the Vlasov-Maxwell system is necessary since the change in the magnetic field by compression in DFs can be sufficiently large especially in high $\beta$ plasmas \cite{Fletcher:2019kq}.  

To address such conditions the model discussed in Sec. \ref{subsec:vlasovpoisson} can be generalized to include the electromagnetic effects by considering the Vlasov-Maxwell set of equations instead of the Vlasov-Poisson system of Sec. \ref{subsec:vlasovpoisson} as shown below:
\begin{equation}
	\begin{split}
		\mathbf{v}\cdot\mathbf{\nabla}_{r} f_{\alpha}(\mathbf{r},\mathbf{v})
		+ \frac{q_{\alpha}}{m_{\alpha}}\left(
		\mathbf{E} + \frac{\mathbf{v}\times\mathbf{B}}{c}\right)
		\cdot\mathbf{\nabla}_{v} f_{\alpha}(\mathbf{r},\mathbf{v}) = 0,
		\\
		\mathbf{\nabla}\cdot\mathbf{E} =
		\sum_{\alpha} 4\pi q_{\alpha} \int d^3 \mathbf{v} f_{\alpha}(\mathbf{r},\mathbf{v}),
		\\
		\mathbf{\nabla}\times\mathbf{B} = \frac{4\pi}{c}
			\sum_{\alpha} q_{\alpha} \int d^3 \mathbf{v} \,\mathbf{v} f_{\alpha}(\mathbf{r},\mathbf{v}),
	\end{split}
\end{equation}
In the frame of the DF propagating towards the earth the variation in the normal direction (with scale size of an ion gyroradius) is orders of magnitude stronger than in the orthogonal directions. Hence for small scale physics it becomes essentially a one-dimensional model, similar to the plasmasheet lobe interface discussed in Sec. \ref{subsec:vlasovpoisson}. The local magnetic field is in the $z$ direction and varies in the $x$ direction, \textit{i.e.} $\mathbf{B}=B(x)\mathbf{e}_z$, while a nonuniform electric field also varies in the $x$ direction, \textit{i.e.} $E_{x}(x)$ as sketched in Fig. \ref{fig:fig7}.  We introduce a vector potential, $\mathbf{A}$, where  $\mathbf{B}=\mathbf{\nabla}\times\mathbf{A}$ and $\mathbf{A}=A(x)\mathbf{e}_y$. The Hamiltonian is
\begin{equation}
	H_{\alpha}(x) = \frac{p_x^2}{2m_{\alpha}}
	+\frac{1}{2m_{\alpha}}\left[p_y - \frac{q_{\alpha}}{c}A(x)\right]^2
	+ \frac{p_z^2}{2m_{\alpha}}
	+ q_{\alpha}\Phi(x)
\end{equation}
where $p_x$, $p_y$, and $p_z$ are the canonical momenta.  The Hamiltonian only depends on $x$ and is independent of $t$, $y$, and $z$ so $H$, $p_y$, and $p_z$ are constants of motion, where $p_y=m_{\alpha}v_y+m_{\alpha}\Omega_{\alpha} a(x)$.  Since the system has only one degree of freedom, the dynamics is completely integrable.  With $a(x)=A(x)/B_0$ and $B_0$ is the upstream background magnetic field it follows that the guiding center position,  
\begin{equation}
	X_{g\alpha}=\frac{p_y}{m_{\alpha}\Omega_{\alpha}} = a(x)+\frac{v_y}{\Omega_{\alpha}}
\end{equation}
is a constant of motion as well.

\subsubsection{Derivation of the equilibrium distribution function}

The construction of the distribution function is similar to that described in Sec. \ref{subsec:esderivoff}, except that we now obtain the moments as a function of $a(x)$ and then solve $a(x)$ as a function of $x$ to obtain the spatial profiles of the parameters of interest \cite{Fletcher:2019kq}.  The moments of the distribution provide the physical attributes of the equilibrium configuration, in particular their spatial variations. The zeroth moment (density) is
\begin{equation}
	\label{eq:nalpha_vlasov_maxwell}
	n_{\alpha}(a) \equiv \left\langle f_{0\alpha}\right\rangle
	=
	\int d^{3}\mathbf{v}\,f_{0\alpha}(\mathbf{v},\Phi_0(a)) 
	=
	N_{0\alpha}\frac{(R_{\alpha}+S_{\alpha})}{2}
	\exp\left(-\frac{q_{\alpha}\Phi_{0}(a)}{T_{\alpha}}\right)
	I_{\alpha}(a)
\end{equation}
Note the dependence of various quantities on $a(x)$ in Eqs. \ref{eq:nalpha_vlasov_maxwell}, instead of just $x$ as in Sec. \ref{subsec:vlasovpoisson}; $a(x)$ will be determined from the first moment (\textit{i.e.} the current density). The electrostatic potential is found via quasineutrality, $n_e\simeq n_i$, as before:
\begin{equation}
	\Phi_{0}(a) = \frac{T_e T_i}{q_e T_i - q_i T_e}
	\log\left[\frac{N_{0e}(R_e+S_e)I_e}{N_0(R_i+S_i)I_i}\right]
\end{equation}
Because $\nabla n\neq 0$ and $\nabla B\neq 0$, and the electric field, $\mathbf{E}=-\nabla\Phi_0(a)$, are in the $x$ direction, the only nonzero component of the flow is in the $y$ direction.  The flow is
\begin{multline}
\label{eq:uyalpha}
	u_{y\alpha}(a)\equiv \left\langle v_y f_{0\alpha} \right\rangle/n_{\alpha}
	=
	\frac{1}{n_{\alpha}} \int d^{3}\mathbf{v}\, v_y f_{0\alpha}(\mathbf{v},\Phi_0(a)) 
	\\
	=\pm \frac{
	\exp\left(\frac{-2q_{\alpha}\phi(a)}{m_{\alpha} v_{t\alpha}^2}\right)
	N_{0\alpha} (R_{\alpha}-S_{\alpha})
	v_{t\alpha}
	\left[\textrm{erf}(\xi_{2\alpha})-\textrm{erf}(\xi_{1\alpha})\right]
	}{4n_{\alpha} (\xi_{1\alpha}-\xi_{2\alpha})}
\end{multline} 
and includes the diamagnetic drift, $\nabla B$ drift and $\mathbf{E}\times\mathbf{B}$ drift.  

The magnetic field produced by the current density inherent in the equilibrium distribution function is found by the Ampere law,
\begin{equation}
\label{eq:amperes_law1}
\frac{dB_z}{dx}=-\frac{4\pi}{c} j_y,
\end{equation}
where $j_y=\sum_{\alpha}q_{\alpha} n_{\alpha} u_{y\alpha}$ is the current density. With $B_z$, the vector potential is found via
\begin{equation}
\label{eq:adef_1}
\frac{da}{dx}=\frac{B_z}{B_0}
\end{equation}
with appropriate initial conditions. Eqs. \ref{eq:amperes_law1} and \ref{eq:adef_1} effectively forms the Grad-Shafranov equation and  may not have a readily apparent closed-form solution but can be integrated numerically. The current density in Ampere’s law can be written explicitly as a function of the vector potential $a(x)$. Thus we can numerically solve Eqs. \ref{eq:amperes_law1} and \ref{eq:adef_1} for the function $a(x)$ which then provides a mapping to $x$. All plasma parameters that have been determined as a function of $a(x)$ can now be found as a function of $x$. An electrostatic approximation is equivalent to specifying $a(x)$ explicitly (\textit{e.g.} for a uniform magnetic field, $a(x)=x$).

We can continue and consider higher order moments. For the pressure tensor all off diagonal terms vanish and $p_{\alpha xx}= p_{\alpha zz}=n_{\alpha} T_{\alpha}$. The remaining component, $p_{\alpha yy}$, which we do not repeat here involves an integral over $v_y$ and can be performed in a manner similar to Eq. \ref{eq:uyalpha}.  

\begin{figure}
	\includegraphics[width=0.6\textwidth]{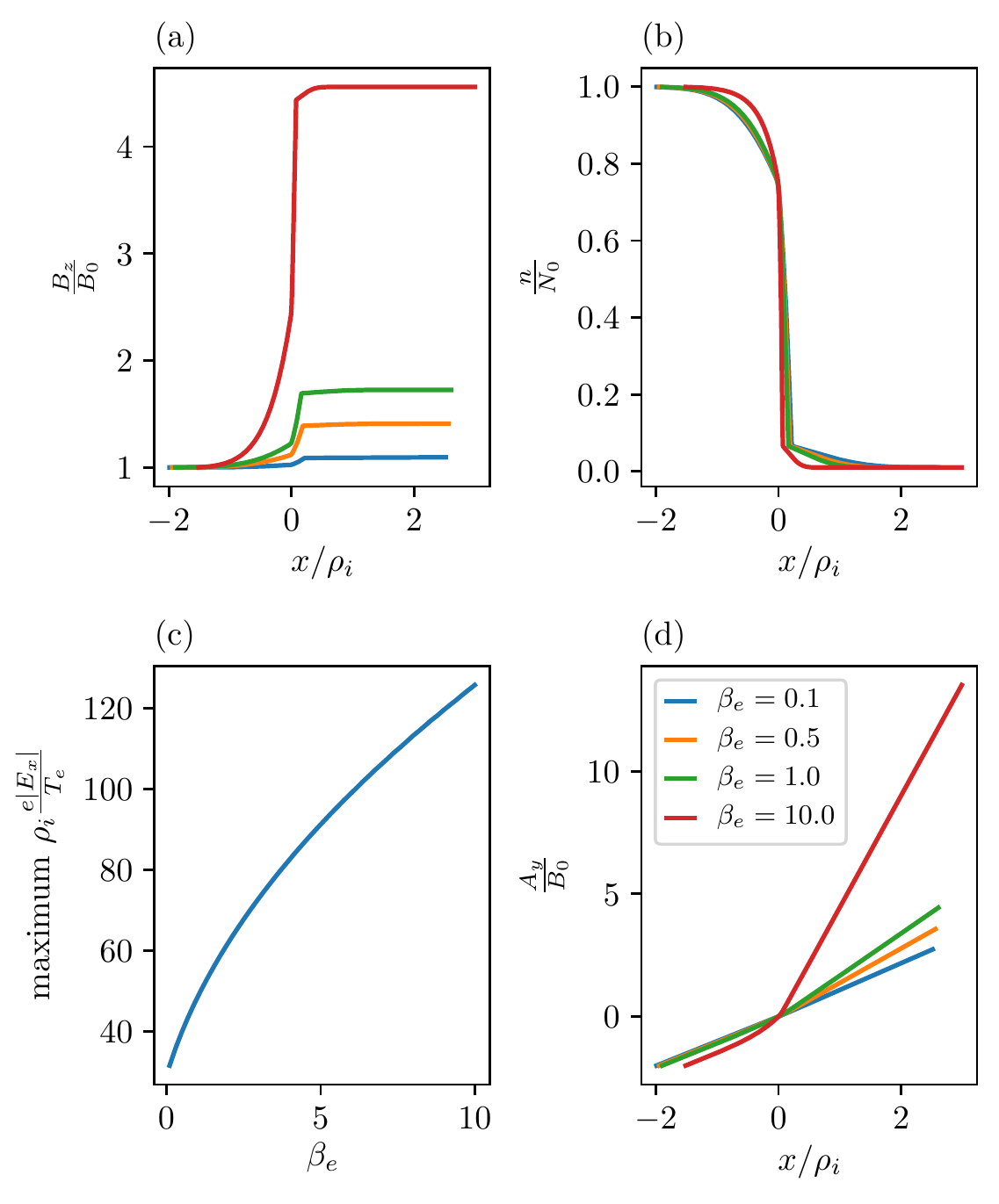}
	\caption{\label{fig:fig10} Electromagnetic effects on equilibrium.  (a) Magnetic field for different values of $\beta_{e}$.  (b) Density.  (c) Maximum electric field seen over the layer as as function of $\beta_{e}$.  (d) Vector potential as a function of position.  The legend in panel (d) refers to panels (a),(b), and (c).  The parameters are as follows $a_{1i,e}=0,0$, $a_{2i,e}=0.2,0.2$, $R_{i,e}=1.0,1.0$, $S_{i,e}=0.01,0.01$, $T_e/T_i=1.0$, and $m_i/m_e=1836.0$.}
\end{figure}

\subsubsection{Electromagnetic correction to the equilibrium distribution function}

Fig. \ref{fig:fig10} shows the electromagnetic effects on the static background structure.  To illustrate the difference we choose the input parameters to be the same as in Fig. \ref{fig:Fig2} but we increase $\beta_{e}$.  As seen from panels (a) and (c) the electric and magnetic fields increase with $\beta_{e}$.  Panel (b) indicates that the density gradient steepens with increasing $\beta_{e}$, which explains the increase in the electric field.  Panel (d) shows that as long as $\beta_{e}$ is less than unity the electromagnetic effects on static structures are minimal.  Hence, the use of the simpler electrostatic model of Section 2.1.1 to understand the static background features is sufficient.  However, in dipolarization fronts higher $\beta_{e}$ is typical.  Ganguli \textit{et~al.} \cite{Ganguli:2018vf} and Fletcher \textit{et~al.} \cite{Fletcher:2019kq} have analyzed the MMS data in detail and illustrated the difference between the electrostatic and electromagnetic models for a specific observation.

\subsubsection{Effects of magnetic field curvature:  Generation of parallel electric field}
\label{sec:vpequi_curv}

\begin{figure}[t]
\includegraphics[width=0.5\textwidth]{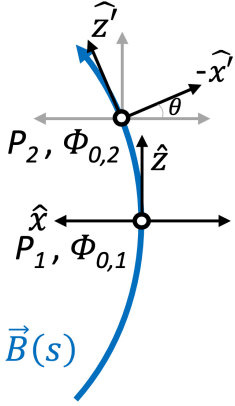}
\caption{\label{fig:fig11}Geometry along the magnetic field line of a DF.  In a typical DF the variation of plasma parameters across the magnetic field is stronger than the variation along the magnetic field which reduces the problem to 1D.  Since the plasma parameters ($T$,$|B|$) are different at the two points the electrostatic potentials assumes different values, which leads to a potential difference ($\Phi_{0,2} -\Phi_{0,1}$) along the magnetic field causing the parallel electric field.  }
\end{figure}

In the above discussion of the equilibrium structure of a DF we considered the stronger variation normal to the magnetic field and ignored the slower variation along the field.  For a typical DF the transverse electric field is strongest at a particular point; for example marked $P_1$ in Fig. \ref{fig:fig11}.  As we move from this point along the magnetic field, to point $P_2$, the $x$ and $z$ coordinates rotate by an angle $\theta$  as indicated in Fig. \ref{fig:fig11}.  Since the local values of the magnetic field, temperature, density, \textit{etc.} are different at positions $P_1$ and $P_2$ along the magnetic field, the electrostatic potential will vary, giving rise to an electric field along the magnetic field direction proportional to the potential difference between the two positions, $\Phi_{02}-\Phi_{01}$. Since $\Phi_0\simeq \Phi_0 (B(s))$, the parallel electric field is $E_{\|} (s)\equiv -\partial\Phi_0 (B(s))/\partial s=(x/L_{\|})E_x (x)$.  Fig. 3c of \cite{Ganguli:2018vf} shows that $E_{\|}$ peaks in the electron layer and varies in $x$ for a typical DF. Non-thermal plasma particles subjected to $E_{\|}$ will be accelerated along the magnetic field to form inhomogeneous beams or flows. The generation of the beam along the field line by this process provides the physical basis for a non-reconnection origin of the observed beams and its causal connection to the global compression.

Existence of $E_{\|}$ indicates that the off-diagonal terms of the pressure tensor, $\mathbf{P}_{\alpha}=m_{\alpha}\int (\mathbf{v}-\mathbf{u})(\mathbf{v} -\mathbf{u})f_{0\alpha}d^3\mathbf{v}$, are non-zero and are necessary to balance it in equilibrium, \textit{i.e.},
\begin{equation}          
                   en(x)E_{\|}=-(\mathbf{\nabla}\cdot\mathbf{P}_{\alpha}(x))
                   \cdot\mathbf{s}
                   =-(\partial_x p_{xx} \mathbf{b}_x
                   +\partial_x p_{xz} \mathbf{b}_z ), 
\end{equation}
where $\mathbf{b}_x=\sin(\theta)$ and $\mathbf{b}_{z}=\cos(\theta)$, and to leading order $\partial/\partial y=\partial/\partial z\rightarrow 0$ because the spatial variation is strongest in the $x$ direction at a given location along the magnetic field.  These equilibrium features along the magnetic field can also be important to the dynamics of the compressed plasma layers and affect the measurable quantities such as spectral character of the emissions and particle energization.  This is discussed in sections 3.3.4 and 4.3. 

\subsection{Vlasov-Maxwell System:  Field reversed geometry in the magnetotail}
\label{subsec:frequi}

While the electromagnetic effects of compression are important in DFs, especially when the plasma $\beta$ is large, electromagnetic effects are essential for the magnetic field reversal geometry and current sheets.  Current sheets are important in magnetic fusion experiments and magnetospheric, solar, and astrophysical dynamics because the reversed magnetic field geometry can lead to magnetic reconnection and thus a large-scale reconfiguration of the system.  The formation of the current sheet is the result of a global compression with opposing magnetic fields and the resulting nonlinear reconnection is often further driven by compression of a large fluid scale current sheet down to kinetic scales \cite{Schindler:1993,Sitnov:2006,Nakamura:2002,Artemyev:2019}.  We now extend the above equilibrium boundary layer methodology to the case of a current sheet with magnetic field reversal \cite{Crabtree:2020} to investigate the effects of an inhomogeneous ambipolar electric field resulting from global compression that cannot be transformed away.  Traditionally the field reversed case has been addressed by the Harris equilibrium \cite{Harris:1962cw} which is restrictive because it is a specialized distribution designed to produce density and potential gradients such that there is no net electric field by using a transformation to a uniform velocity frame (described below).  As a result, this distribution is inflexible and unable to account for the observed spatially localized structures such as embedded \cite{McComas:1986,Sergeev:1993,Sanny:1994} and bifurcated current sheets \cite{Hoshino:1996,Asano:2004,Runov:2004,Schindler:2008} that develop during active periods when the plasmasheet thins due to large scale compression causing the current sheet to structure.  We remove this inflexibility by constructing a solution to the Vlasov equation that is a generalization of the Harris equilibrium \cite{Harris:1962cw} with the inclusion of a non-uniform guiding-center distribution $Q_{\alpha}(x_{g\alpha})$, 
\begin{equation}
\label{eq:f0_cs}
f_{0\alpha}(x,\vec{v}) = \frac{N_{0\alpha}}{\left(\pi v_{t\alpha}^2\right)^{3/2}} Q_{\alpha}(x_{g\alpha})
	\exp\left(-\frac{E_{\alpha} - U_{\alpha} p_y+ \frac{1}{2}m_{\alpha}U_{\alpha}^2}{T_{\alpha}}\right)
\end{equation}
where the definitions of the various quantities are as before.  For $Q_{\alpha}\rightarrow 1$ Eq. \ref{eq:f0_cs} reduces to the Harris distribution while for $U_{\alpha}\rightarrow 0$ it reduces to the compressed layer distribution discussed in Secs 2.1 and 2.2.  The inclusion of the inhomogeneous guiding center distribution allows the Harris equilibrium the freedom to develop inhomogeneous structures, such as localized current sheets, as a response to external compression.  As in Secs 2.1 and 2.2, we specify only the global compression level through the choice of $x_{g,1,2\alpha}$ (or equivalently $a_{1,2}$) and allow the system to develop the density, flows, current, and temperature structures self-consistently.  Then we can compute the density of each species 
\begin{equation}
\begin{split}
n_{\alpha} &= \int d^3\mathbf{v}  f_{0\alpha}(x,\mathbf{v})
\\	
&= N_{0\alpha} \exp\left(-\frac{q_{\alpha}\phi}{T_{\alpha}}
	- \frac{U_{\alpha}m_{\alpha}\Omega_{\alpha} a }{T_{\alpha}}\right) 
	I_{\alpha}(a)
\end{split}
\end{equation}
where 
\begin{equation}
	I_{\alpha}(a) =  \frac{1}{\left(\pi v_{t\alpha}^2\right)^{1/2}} \int dv_y  Q_{\alpha}\left(a+\frac{v_y}{\Omega_{\alpha}}\right)
	\exp\left(-\frac{(v_y-U_{\alpha})^2 }{v_{t\alpha}^2}\right)
\end{equation}
As in the Harris equilibrium \cite{Harris:1962cw} we choose $U_e/v_{te}=-U_i/v_{ti}(\rho_e/\rho_i)$  by transforming to the frame where this is satisfied, and use quasi-neutrality to solve for the electrostatic potential.  Interestingly, the potential does not depend on $U_\alpha$ and has a similar form to the cases considered for the plasma sheet-lobe interface and for the dipolarization front,
\begin{equation}
\label{eq:quasineutrality}
	\frac{e\phi}{T_{e}} =
	\frac{1}{1+\frac{T_e}{T_i}}
	\log\left( \frac{N_{0i}I_{i}(a) } {N_{0e} I_{e}(a)  }	\right).
\end{equation}
 In the Harris equilibrium the choice of transformation to a uniformly drifting frame is typically made so that quasi-neutrality may be satisfied without an electrostatic potential.  This choice corresponds to a uniform drift where the inhomogeneity in the $\mathbf{E}\times\mathbf{B}$ drift is balanced by the inhomogeneity in the diamagnetic drift so that this transformation can be done globally.  While the mathematical simplicity and elegance of the transformation is appealing, it constrains the system from developing substructures as the current sheet thins due to global compression.  Introduction of the guiding center distribution, $Q_{\alpha}$, relaxes this constraint and allows for nonuniform flows to develop in response to global compression.  Nevertheless the transformation still can be made to simplify the expressions.
 
Next, we calculate the current density using the second moment as,
\begin{equation}
	j_{y\alpha} = q_{\alpha} \int d v_y\, v_y f_{0\alpha} = q_{\alpha}
	N_{0\alpha} v_{ta} \exp\left(-\frac{q_{\alpha}\phi}{T_{\alpha}}
	- \frac{U_{\alpha}m_{\alpha}\Omega_{\alpha} a }{T_{\alpha}}\right) 
	J_{\alpha}(a)
\end{equation}  
where 
\begin{equation}
	J_{\alpha}(a) =  \frac{1}{(\pi v_{t\alpha}^2)^{1/2}} \int dv_y  \,\frac{v_y}{v_{t\alpha}}  Q_{\alpha}\left(a+\frac{v_y}{\Omega_{\alpha}}\right)
	\exp\left(-\frac{(v_y-U_{\alpha})^2 }{v_{t\alpha}^2}\right).
\end{equation}
Considering a single ion species and electrons we can write down from Ampere's law the equation,
\begin{equation}
\begin{split}
\label{eq:ampereslaw}
	\rho_{i0}\frac{d^2 a}{d x^2} &=  \beta_i \left[
	\exp\left(-\frac{e\phi}{T_{i}}
	\right) 
	J_{i}(a(x))
	\right. \\ 
	&\left.
	-
	\frac{N_{0e} v_{te}}{N_{0i} v_{ti}} \exp\left(\frac{e\phi}{T_{e}}
	\right) 
	J_{e}(a(x))
	\right]	\exp\left(
	- \frac{U_{i} 2a(x) }{v_{ti}\rho_{i0}}\right) 
\end{split}
\end{equation}
where $\beta_i=8\pi N_{0i} T_i/B_0^2$, $\rho_{i0}=v_{ti}/\Omega_{i0}$, and $\Omega_{i0}=|e|B_0/(m_i c)$.  $B_0$ is a reference magnetic field value, which in the following, takes the value of the magnetic field in the asymptotic limit away from the layer for $Q_{\alpha}=1$ in the Harris limit.  Unlike the potential, the density and current depend on $U_{\alpha}$.  We note that Eq. \ref{eq:ampereslaw} has the form of an equation of motion, where $x$ is the time-variable and $a$ is the position like variable.  With the solution of Eq. \ref{eq:ampereslaw} (using Eq. \ref{eq:quasineutrality}) the equilibrium is fully specified.  In the limit of constant guiding center distribution, $\phi=0$, $N_{0i}=N_{0e}$, $J_i=U_{i}/v_{ti}$ and $J_e=U_{e}/v_{te}$, and Ampere's law becomes
\begin{equation}
	\frac{d^2 a}{dx^2} = \frac{\beta_i}{L_H}\left[1+\frac{T_e}{T_i}\right]\exp\left(-\frac{2 a(x)}{L_H}\right)
\end{equation}
where $L_H=\rho_{i0} v_{ti}/U_i$ is the single scale size associated with the Harris equilibrium \cite{Harris:1962cw}.  Eq.(23) has solutions $a(x) =  L_H \log(\cosh(x/L))+L_H/2\log(\beta_i+\beta_e)$.  This is the usual Harris sheet vector potential \cite{Harris:1962cw}.  Because the Harris sheet has only one length scale, $L_H$, it is unable to develop substructures in response to the compression.  Introduction of another scale, $L$, associated with $Q_{\alpha}$, in the generalized Harris equilibrium, Eq. (16), removes this limitation.  $L$ is dependent on the compression through the parameters, $x_{g1,2\alpha}$ as discussed in sections 2.1 and 2.2.  This makes the generalized Harris equilibrium a more accurate representation of reality.  

Using the same linear ramp functions $Q_{\alpha}(x_{g\alpha})$ as used in Secs 2.1 and 2.2 we can calculate explicity the functions $I_{\alpha}$ and $J_{\alpha}$, for the generalized Harris equilibrium
\begin{multline}
\begin{split}
	I_{\alpha}(a) &= \frac{1}{2}(R_{\alpha}+S_{\alpha})
	\\
	&+\frac{b_{\alpha}(R_{\alpha}-S_{\alpha})}{2|b_{\alpha}|(\xi_{1\alpha}-\xi_{2\alpha})}
	\left[ \frac{1}{\sqrt{\pi}}\left(e^{-\xi_{1\alpha}^2}-e^{-\xi_{2\alpha}^2}\right)
	+
	\xi_{1\alpha} \textrm{Erf}(\xi_{1\alpha}) - \xi_{2\alpha}\textrm{Erf}(\xi_{2\alpha})\right]
	\\	
	J_{\alpha}(a) &= \frac{u_{\alpha}}{2}(R_{\alpha}+S_{\alpha})
	\\&+\frac{b_{\alpha}(R_{\alpha}-S_{\alpha})}{2|b_{\alpha}|(\xi_{1\alpha}-\xi_{2\alpha})}\left[
	\frac{u_{\alpha}}{\sqrt{\pi}}\left(e^{-\xi_{1\alpha}^2}-e^{-\xi_{2\alpha}^2}\right)
	\right.\\&\qquad\qquad\qquad\qquad\left.
	-\frac{1}{2}\left(1-2u_{\alpha}\xi_{1\alpha}\right) \textrm{Erf}(\xi_{1\alpha}) 
	+\frac{1}{2}\left(1-2u_{\alpha}\xi_{2\alpha}\right) \textrm{Erf}(\xi_{2\alpha})
	\right]
\end{split}
\end{multline}
where we have normalized distances by $\rho_{i0}$ so that $a_{i\alpha}=x_{gi\alpha}/\rho_{i0}$ and we have defined $\xi_{i\alpha}=(-b_{\alpha}u_{\alpha}-a/\rho_i+a_{i\alpha})/b_{\alpha}$ where $u_{\alpha}=U_{\alpha}/v_{t\alpha}$ and $b_{\alpha}=\textrm{sign}(q_{\alpha})\rho_{\alpha}/\rho_{i0}$ is negative for electrons.

There are two general cases of the differential equation where the effects of the non-uniform flow are important.  Both are achieved by choosing $a_{1\alpha},a_{2\alpha}$ such that the guiding center distribution changes on a scale comparable to the ion gyroradius.  This leads to a current due to an ambipolar electric field drift, which corresponds to a global compression on the current sheet, in addition to the current that supports the current-sheet in the Harris equilibrium due to the drift $U_{\alpha}$ in the distribution functions.  There are two cases to consider (1) when this additional current is in the same direction as the Harris current or (2) when it is in the opposite direction to the Harris current.  In this paper, we only review the case when these currents are aligned.  For the alternative case see Crabtree \textit{et al.} \cite{Crabtree:2020}.   

\begin{figure}
	\includegraphics[width=0.6\textwidth]{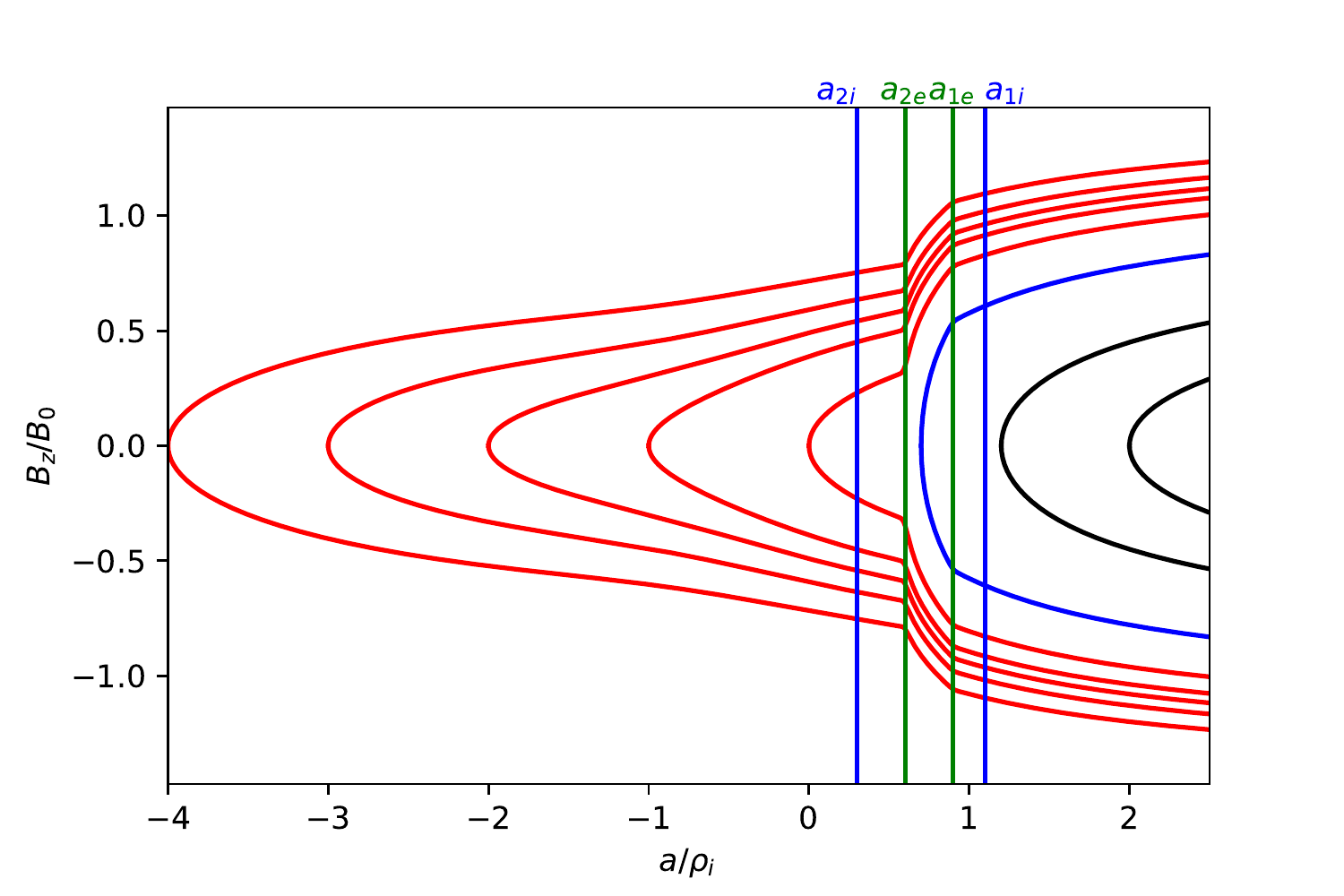}
	\caption{\label{fig:AlignedCurrentTopology}Phase plane analysis for the case when the current due to the density layer is in the same direction as the Harris current.  For this case $a_{1i,e}=1.1,0.9$, $a_{2i,e}=0.3,0.6$, $R_{i,e}=0.1,0.1$, $S_{i,e}=1.0,1.0$, $U_{i}/v_{ti}=0.2$, $T_e/T_i=1.0$, and $m_i/m_e=1836.0$.}
\end{figure}

In this case we can examine the possible categories of equilibria by examining the phase-plane analysis of Eq. \ref{eq:ampereslaw}.  We do this by solving the differential equation numerically and plotting $da/dx=B_z/B_0$ vs $a/\rho_{0}$.  In Figure \ref{fig:AlignedCurrentTopology} we show the phase-plane figure for the case when the currents are in the same direction.  In this case we find three different kinds of equilibria that are determined by the choice of initial conditions for $B_z/B_0$ and $a/\rho_{0}$.  The choice of the initial point, \textit{e.g.} the value of $a$ at $B_z=0$, is in general arbitrary.  In nature all initial values are possible.  The choice of a particular one depends on the global condition, which is beyond the purview of this model but may be obtained from a global model.  However once the initial condition is determined our model can predict the resulting sub-structures of the current sheet corresponding to the level of the global compression.  This level is represented by both the initial point and the choice of parameters $a_{1,i,e}$ and $a_{2,i,e}$ in the guiding center density function $Q_{\alpha}$.  The particular choices of the $a_{1,i,e}$ and $a_{2,i,e}$ are indicated by vertical lines in the figure.  The first type of solution (in black) is a Harris-like equilibrium because the solutions remain in the asymptotic regime of the guiding center distribution (\textit{i.e.} where $Q_{\alpha}\simeq {const.}$) so there is no significant additional current.  The second type of solution (in blue) reaches its turning point at $B_z=0$ within the guiding center distribution gradient and has solutions that are flattened in the phase plane.  The third type of solution (in red) completely traverses the gradient region and becomes elongated in the phase plane. 
 
\begin{figure}
	\includegraphics[width=0.6\textwidth]{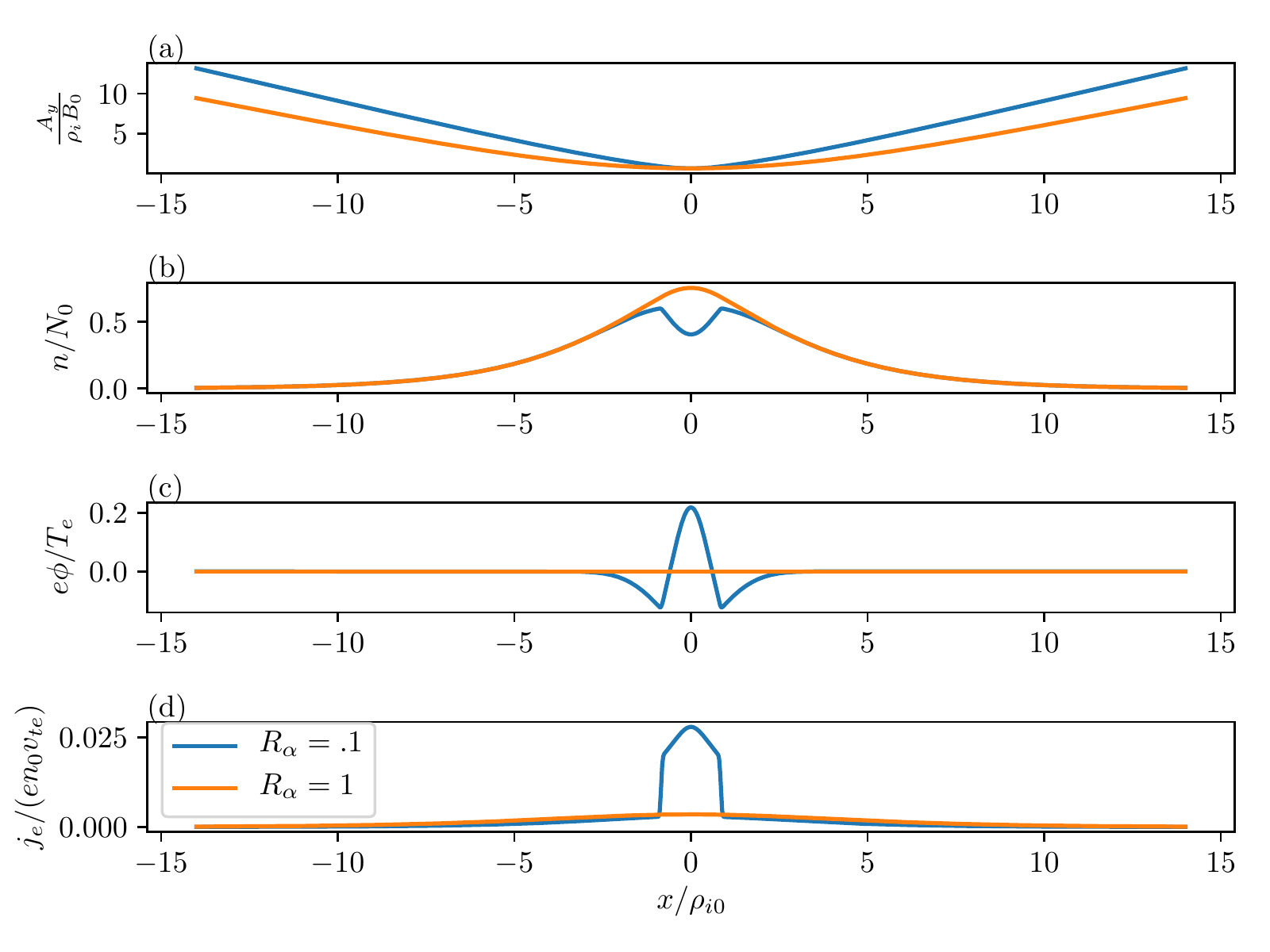}
	\caption{\label{fig:ETCS_phiandA}Embedded thin current sheet.  (a) Vector potential, $a/\rho_{i}$, (b) density, (c) potential, and (d) Electron current density across layer.  In all panels the blue curve corresponds to the case with a density gradient achieved by setting $R_{i,e}=0.1,0.1$ and the orange curve shows the Harris sheet achieved by setting $R_{i,e}=1$ and the rest of the parameters are as follows $a_{1i,e}=1.0,0.94$, $a_{2i,e}=0.3,0.56$, $S_{i,e}=1,1$, $U_{i}/v_{ti}=0.2$, $T_e/T_i=1.0$, and $m_i/m_e=1836.0$.}
\end{figure}

\begin{figure}
	\includegraphics[width=0.6\textwidth]{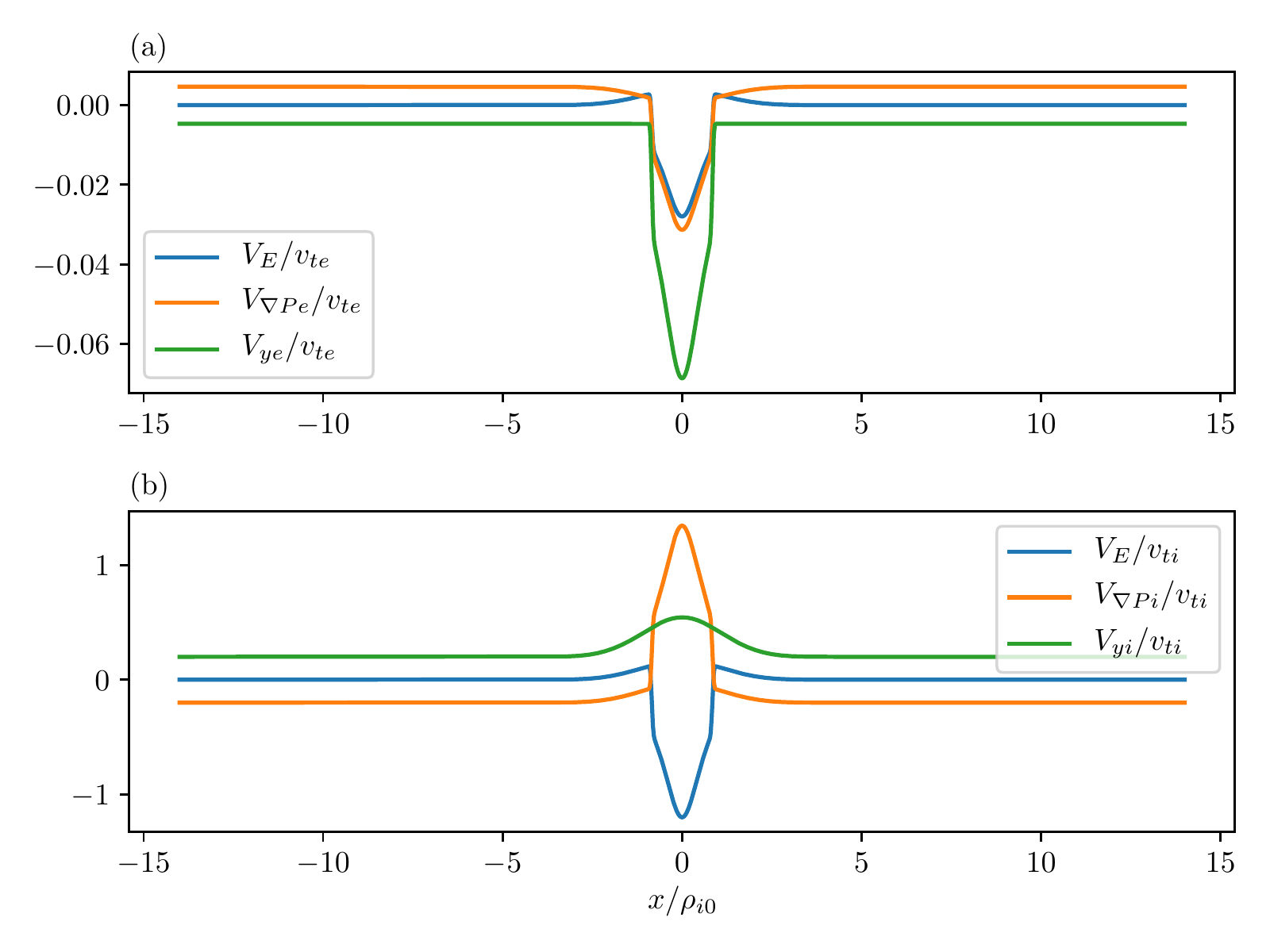}
	\caption{\label{fig:ETCS_drifts}Embedded thin current sheet.  (a) Electron drifts and the total fluid velocity across the layer normalized to the electron thermal velocity.  (b) Ion drifts and total fluid velocity normalized to the ion thermal velocity.   The parameters are the same as in Fig. \ref{fig:ETCS_phiandA}}
\end{figure}

In Figure \ref{fig:ETCS_phiandA} we show the equilibrium attributes corresponding to the blue region of curves in Figure \ref{fig:AlignedCurrentTopology}.  For reference, we added the Harris solution in orange.  The density gradient scale is comparable to the ion gyroradius and is self-consistently determined.  This generates an ambipolar electrostatic potential that cannot be transformed away (panel (c)).  The small dip in density (as opposed to a peaked density) is necessary to create the electric field in the proper direction (away from the current sheet) to generate a current that is in addition to the Harris current.  Also note that around $x = 0$, where the magnetic field vanishes and hence magnetic confinement of the particles becomes weak, the electrostatic potential peaks.  Consequently, around this point the particles can be electrostatically confined.  As a result, the  velocity profile  peaks around the null point, which is midway between the turning points of the electrostatic potential  (Fig. \ref{fig:ETCS_drifts}).  This creates an ideal situation in which the velocity gradient driven waves (Sec. 3) can originate in the vicinity of the null region and contribute to anomalous resistivity \cite{Romero:1993ip} necessary for the magnetic reconnection process. Further details are discussed in Crabtree \textit{et al.} \cite{Crabtree:2020}.  The case without a density gradient, \textit{i.e.} the Harris case, is shown in orange in the figure and correspondingly has no electrostatic potential.  In panel (d) we show that the current density across the layer consists of a thin central current sheet, of scale size $\sim L$, due to the electron Hall current, embedded in a broader current sheet of scale size $\sim L_H$ due to the bulk drifting component of the distribution function (the $U_{\alpha}$ drift).  This solution resembles an embedded thin current sheet which are commonly observed \textit{in situ} by spacecraft \cite{McComas:1986,Sergeev:1993,Sanny:1994}.  In Figure \ref{fig:BCS_drifts} we show the individual drift components.  The electrons have a small gyro-orbit  compared to the electric field scale size and thus have a standard $E\times B$ drift in the ambipolar electric field.  The ions have a larger orbit and thus the orbit averaged electric field sampled is smaller, thus the total flow of the ions is reduced.  This is the source of the additional current.        

The existence and the magnitude of the electrostatic potential around the magnetic null (Fig. 13c) leads to another interesting question, \textit{i.e.}, how does the electrostatic potential affect the individual particle orbits around the magnetic null?  For the 1D equilibria considered here, the particle orbits are all integrable and the details of how the figure eight orbits \cite{Speiser:1965} are modified by the electric field are discussed in Crabtree \textit{et al.} \cite{Crabtree:2020}.  An open question remains with the addition of a $B_x$ (north-south component in our coordinates) so that the magnetic field becomes approximately parabolic.  Will the orbits still be chaotic near the null-sheet as they are in the case without an electric field \cite{Chen:1986}?  If so, how does the electrostatic potential affect the extent of the region over which they are chaotic?  How does the electrostatic potential affect the onset condition for chaos if chaotic orbits can still survive?  These questions remain to be debated and answered in the future.

Current sheet thinning, which is the result of a global compression, is often observed in the magnetotail just prior to the onset of reconnection \cite{Schindler:1993,Sitnov:2006,Nakamura:2002,Artemyev:2019}.  With a thin embedded current sheet there are narrow layers of electron flow with large flow shear which can drive many kinds of instabilities, that would not exist in a standard Harris equilibrium.  These shear-flow driven instabilities (discussed in Sec. 3) can provide a source of anomalous resistivity for the onset of magnetic reconnection.  Lower-hybrid drift instabilities (LHDI) have been extensively studied in Harris sheets \cite{Huba:1980,Huba:1983,Daughton:1999ex,Tummel:2014} because of their potential to provide a source of anomalous resistivity, however, these studies were done in a Harris equilibrium where the LHDI is confined away from the magnetic null because LHDI favors strong magnetic field and strong density gradients.  With compression we expect current sheets to develop kinetic scale features as shown here, and also observed in the \textit{in situ} data,  such that the source of the instability can be closer to the magnetic field reversal region and thus can play a significant role in reconnection.  This is a topic for further investigation.       
 
\begin{figure}
	\includegraphics[width=0.6\textwidth]{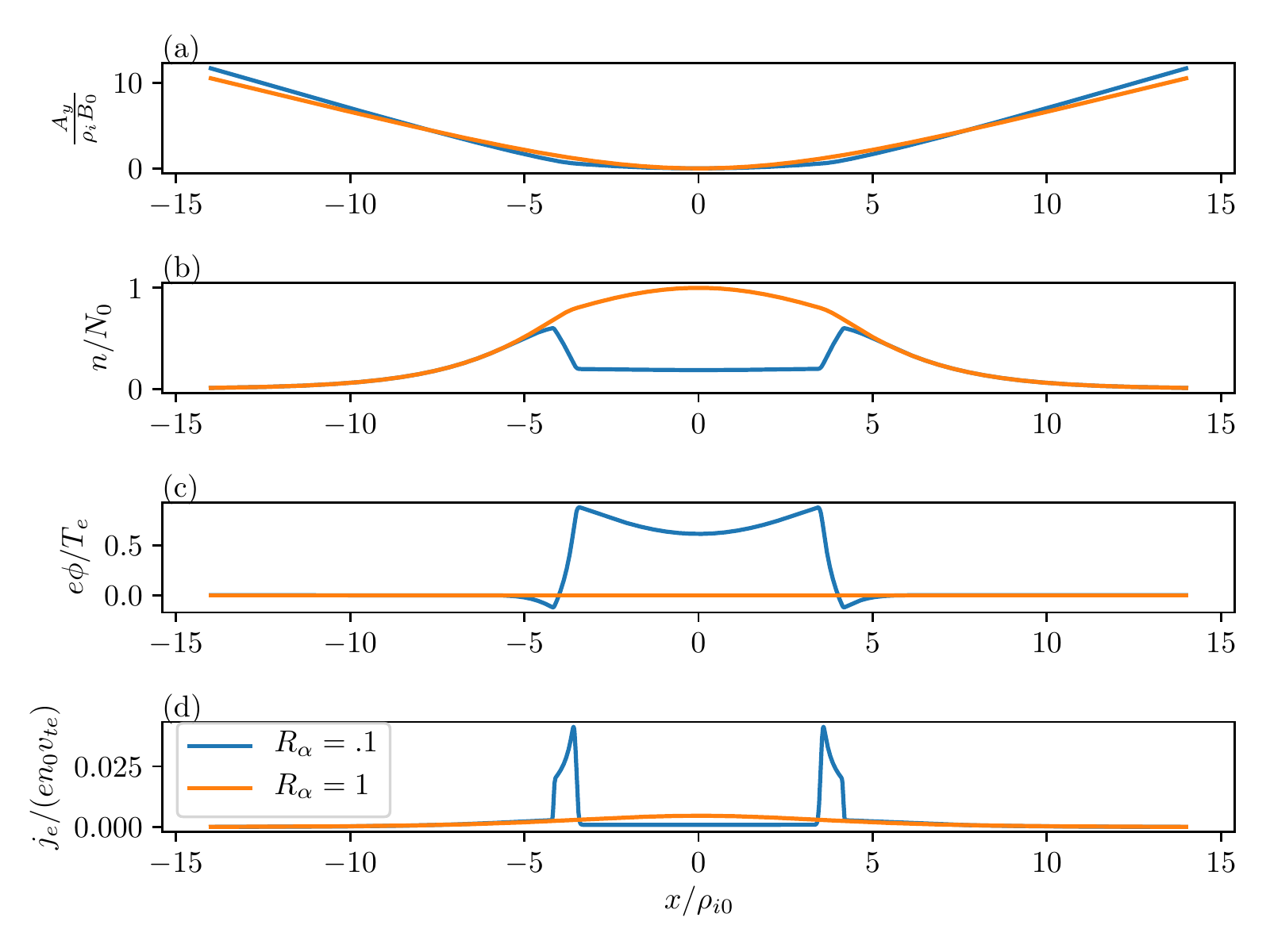}
	\caption{\label{fig:BCS_phiandA}Bifurcated current sheet.  (a) Vector potential, $a/\rho_{i}$, (b) density, (c) potential, and (d) Electron current density across layer. (c) Electron current density.  In all panels the blue curve corresponds to the case with a density gradient achieved by setting $R_{i,e}=0.1,0.1$ and the orange curve shows the Harris sheet achieved by setting $R_{i,e}=1$.  For both cases the the solution curve for the vector potential was chosen by selecting $A_{0}=0$ the rest of the parameters are as follows $a_{1i,e}=1.0,0.94$, $a_{2i,e}=0.3,0.56$, $S_{i,e}=1,1$, $U_{i}/v_{ti}=0.2$, $T_e/T_i=1.0$, and $m_i/m_e=1836.0$.}
\end{figure}

\begin{figure}
	\includegraphics[width=0.6\textwidth]{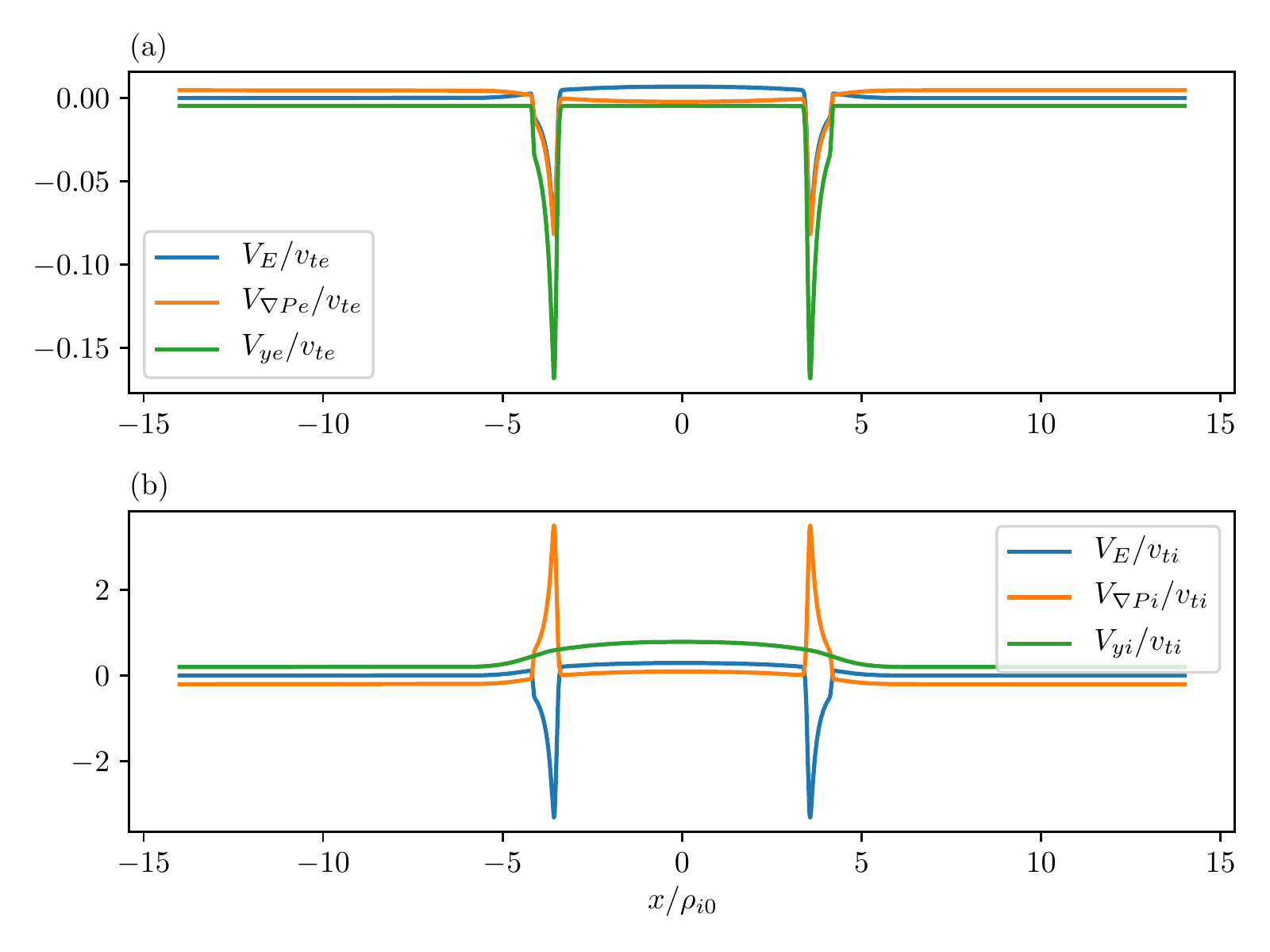}
	\caption{\label{fig:BCS_drifts}Bifurcated current sheet.  (a) Electron drifts and the total fluid velocity across the layer normalized to the electron thermal velocity.  (b) Ion drifts and total fluid velocity normalized to the ion thermal velocity.   The parameters are the same as in Fig. \ref{fig:BCS_phiandA}}
\end{figure}

In Figure \ref{fig:BCS_phiandA} we show the vector potential in panel (a), the density in panel (b), the electrostatic potential in panel (c) and the electron current density in panel (d) as a function of the distance across the layer where the magnetic field reversal is located at $x=0$.  The orange curves correspond to the Harris sheet solution with no ambipolar electric field and the blue curves correspond to the new generalized Harris solution.  This solution corresponds to the class of red curves in Fig. \ref{fig:AlignedCurrentTopology} where we chose $a=0$ at the field reversal.  Figure \ref{fig:BCS_phiandA} shows that near the guiding center gradient on either side of the field reversal there is a strong electron Hall current that is stronger than the current of scale size $L_H$ supported by the uniformly drifting component of the distribution function (\textit{i.e.} the current due to $U_{\alpha}$) but in the same direction.  In Figure \ref{fig:BCS_drifts} we show the electron drifts (in panel a) and ion drifts in panel (b) as well as the total fluid velocities.  We see that the $E\times B$ drift of the electrons (panel a) is in the same direction as the diamagnetic drift in the layer which leads to a strong net sheared flow of electrons.  Whereas with the ions (panel b) they are in opposite directions.   This figure shows that the electrons experience a significant $E\times B$ drift but the ions do not because narrow electric fields exist on scales a fraction of the ion gyroradius.  

The current sheet solution shown in Figures \ref{fig:BCS_phiandA} and \ref{fig:BCS_drifts} resemble a bifurcated current sheet that have been commonly observed by spacecraft in the magnetotail.  Such bifurcated current sheets have also been observed in 1D particle in cell simulations \cite{Schindler:1993}.  In these simulations the starting point was a Harris equilibrium and then the layer was compressed by applying time-dependent in-flows at the boundaries (in $x$ in our coordinate system).  A steady state was reached in the simulation after compression that resembled the bifurcated equilibrium shown here in Figure \ref{fig:BCS_phiandA}d.  Thus, there are simulation studies showing that by further compressing a Harris current sheet one can develop ambipolar electric fields which drive an electron current and form a bifurcated current sheet that are consistent with the Vlasov  equilibrium solutions discussed here.    

As in Secs 2.1 and 2.2, we find that even in the field reversed magnetic field geometry as the plasma is compressed an electrostatic potential is self-consistently generated.  This introduces plasma flows that are highly sheared.  As we study in Sec. 3 below, such sheared flows have a natural tendency to relax through emissions that ultimately leads to a new reconfigured steady state.  Further details of the current sheet behavior during active periods and its importance to the magnetic reconnection process is discussed in Crabtree \textit{et~al.} \cite{Crabtree:2020}.

\section{Plasma Response to Compression}
\label{sec:lineartheory}

From Sec. 2 we can conclude that in collisionless environment plasma compression generates an ambipolar electric field across the magnetic field when the layer width becomes less than an ion gyrodiameter.  It also describes some natural examples of plasma compression but this can also happen in laboratory devices.  The amplitude and gradient of the ambipolar field is proportional to the intensity of the compression, which creates the pressure gradient that forms in the layer. It is therefore reasonable to identify the transverse ambipolar electric field as a surrogate for the compression for practical purposes.  It is interesting that the electric field is a better surrogate for the compression than the pressure gradient because, as we discussed in Sec. 2.1, density and temperature gradients could combine to reduce the pressure gradient in the layer but still lead to intense electric fields as the scale size of the layer reduces with increasing compression.  With this identification it becomes possible to quantitatively address the plasma response to compression by studying the variety of linear and nonlinear processes that are triggered by the transverse electric field.  

At the kinetic level the collective behavior in plasma is sensitive to the individual particle orbits.  The particle orbits are affected by the electric field gradient, which develops self-consistently as a result of the compression.  The orbit distortion could be quite substantial and can affect the character of the waves emitted and their nonlinear evolution as well as saturation properties.  Hence, we review the particle orbit modifications due to inhomogeneous transverse electric field.

\subsection{Particle orbit modification due to localized transverse electric field}
\label{subsec:particle_orbits}

In a uniform magnetic field the charged particle orbit modification to the gyro-motion introduced by a uniform transverse electric field is a uniform $\mathbf{E}\times\mathbf{B}$ drift and this electric field can be transformed away in the moving frame.  Since the $\mathbf{E}\times\mathbf{B}$ drift is mass and charge independent, both the electron and ion drifts are identical, which implies that there is no net transverse current.  This is no longer true for an inhomogeneous electric field and has implications for plasma fluctuations.  In realistic plasmas, both in nature and the laboratory, the transverse electric field encountered is inhomogeneous.  For example, we found in Sec. \ref{sec:Equilibrium} the ambipolar electric field that arises self-consistently due to plasma compression is highly nonuniform.  Therefore, we analyze the modifications to particle orbits that such electric field inhomogeneity introduces. 

Consider a uniform magnetic field, $\mathbf{B}_0$, in the z-direction and an inhomogeneous electric field, $\mathbf{E}_0(x)$, in the $x$-direction.  The energy per mass for a charged particle in this field configuration is $K(x)=v_x^2/2+v_y^2/2+e\Phi_{0}(x)/m$, where $\Phi_{0}(x)$ is the external electrostatic potential, \textit{i.e.}, $E_{0}=-d\Phi_{0}(x)/dx$.  The equations of motion for a charged particle in the $x$- and $y$-directions are,
\begin{equation}
\label{eq:porbit_vx}
\dot{v}_x = \Omega v_y - \Omega V_{E}(x),
\end{equation}
\begin{equation}
\label{eq:porbit_vy}
\dot{v}_y = -\Omega v_x
\end{equation}
where $V_{E}=-cE_0(x)/B$ is the $\mathbf{E}_0(x)\times\mathbf{B}$ drift and dots imply time derivative. Integrating Eq. \ref{eq:porbit_vy} we obtain a constant of motion $X_g=x+v_y/\Omega$, which is the guiding center position when the electric field is absent.  Expressing  $v_y=\Omega(X_g-x)$and using it in a Hamiltonian formulation we get,
\begin{equation}
	H(x)=\frac{v_x^2}{2}
	+\frac{\Omega^2}{2} (X_g-x)^2
	+e\Phi_{0}(x)/m
	=
	v_x^2/2 + G(x)
\end{equation}
Minimizing the pseudo potential $G(x)$ at $x=\xi$,
\begin{equation}
	\left.\frac{dG}{dx}\right|_{x=\xi}
	= -\Omega^2(X_g-\xi)
	+ \frac{e}{m}\left.\frac{d\Phi_{0}(x)}{dx}\right|_{x=\xi} = 0
\end{equation}
we obtain the guiding center position $\xi=x+(v_y-V_E(\xi))/\Omega$, when an electric field is present.  For an inhomogeneous electric field this expression is an implicit function for $\xi$.    These definitions help understand the modification to the $\mathbf{E}\times\mathbf{B}$  drift due to the inhomogeneity in the electric field.

 At steady state the time-averaged $y$-drift can be obtained from Eq. \ref{eq:porbit_vx}, \textit{i.e.}, $\langle\dot{v}_{x}\rangle=0=\Omega\langle v_{y}\rangle -\Omega \langle V_{E}(x)\rangle$.  Expanding around the guiding center position and retaining terms up to $O(1/L^{2})$, where $L$ is the scale size of the transverse electric field, the time averaged $y$-drift is, 
\begin{equation}
\label{eq:vyavg_expansion}
	\langle v_y\rangle = \langle V_E(x)\rangle
	=
	V_E(\xi) + \langle(x-\xi)^2\rangle V_E''(\xi)/2 + ...
\end{equation}
The first order term, $\langle(x-\xi)\rangle$, is oscillatory and vanishes on time averaging and  $\langle v_y\rangle$ is time independent.  Thus, in general  $v_y=u_y+\langle v_y\rangle$, where $u_y$ is the oscillatory component of the velocity in the y-direction.  Using the definition of the guiding center, $x-\xi=-(v_y-V_E(\xi))/\Omega$, in Eq. \ref{eq:vyavg_expansion} we can express $\langle v_y\rangle$  as,
\begin{equation}
\label{eq:vyavg_expansion2}
\langle v_y\rangle = 
V_E(\xi) + 
\frac{V_E''(\xi)\langle u_y\rangle^2}{2\Omega^2 \eta(\xi)}
+ O(V''^2)
\end{equation}
where  $\eta(\xi)=1+(dV_E(\xi)/d\xi)/\Omega$.  The parameter $\eta$ is a comparison of  the influences of the electric and magnetic fields on particle orbits.  It is also a measure of the velocity shear strength, and hence of the plasma compression.  $\eta -1$ is the ratio of the shear frequency, $\omega_s = dV_E/dx$, and the gyrofrequency, $\Omega$.  In the limit  $\omega_s \rightarrow -\Omega$ the particle orbits become ballistic as in a field free environment.  In the limit $\omega_s \gg \Omega$ the particles execute trapped orbits in the electrostatic potential and the electric field dominates.  In between the particles respond to both electric and magnetic fields.  Because of spatial variability there may be regions where  each of these effects could be pronounced.  This makes the typical particle orbits much different from the ideal gyro-orbits in a magnetic field, which can affect the collective plasma dynamics.  Besides the usual $\mathbf{E}\times\mathbf{B}$ drift represented by the first term in the right hand side of Eq. \ref{eq:vyavg_expansion2}, there is also a mass dependent second order term.  While there is no transverse current in the zeroth order, a second order current arises due to electric field curvature, which is proportional to the magnitude of the compression.  This is an important modification to the mean or bulk plasma transverse flow, which is a fluid property.   We shall see in Sec. \ref{subsec:vlasov_stability} that this term is an important contributor to plasma collective effects and hence cannot be ignored with respect to the order unity term in Eq. \ref{eq:vyavg_expansion2}.

There is another important kinetic effect due to the electric field inhomogeneity that affects the individual particle orbits.  To understand this we cast the equation of motion in the guiding center frame \cite{Ganguli:1988hh},
\begin{equation}
	\label{eq:vx_gc}
	\dot{v}_x = \eta(\xi)\Omega u_y 
	+ \frac{V_E''(\xi)}{2\Omega}\left(\langle u_y^2\rangle - u_y^2\right),
\end{equation}
\begin{equation}
	\label{eq:uy_gc}
	\dot{u}_y -\Omega v_x
\end{equation}
Taking the time derivative and multiplying by $\dot{u}_y$, Eq. \ref{eq:uy_gc} becomes $\dot{u}_y\ddot{u}_y = -\Omega \dot{u}_y \dot{v}_x$.  Substituting $\dot{v}_x$ from Eq. \ref{eq:vx_gc} yields another constant of motion,
\begin{equation}
w_{\perp}^2 = v_x^2 + \eta(\xi)u_y^2 
- \frac{V_{E}''(\xi)}{\Omega^2}\left(\frac{u_y^3}{3}-\langle u_y^2\rangle u_y\right),	
\end{equation} 
which reduces to the perpendicular velocity for uniform electric case when $L\rightarrow\infty$.  Using this and solving Eqs. \ref{eq:vx_gc} and \ref{eq:uy_gc} for the particle velocities and orbits we get,
\begin{equation}
\label{eq:vx_gc2}
	v_x = w_{\perp} \sin(\sqrt{\eta(\xi)}\Omega\tau + \varphi)
	- 
	\frac{V_{E}''(\xi)w_{\perp}^2}{6\eta(\xi)^{3/2}\Omega^2}\sin(2\sqrt{\eta(\xi)}\Omega\tau + 2\varphi),
\end{equation}
\begin{equation}
\label{eq:uy_gc2}
	u_y = \frac{w_{\perp}}{\sqrt{\eta(\xi)}}
	\cos(\sqrt{\eta(\xi)}\Omega\tau + \varphi)
	- \frac{V_{E}''(\xi)w_{\perp}^2}{12\eta(\xi)^{2}\Omega^2}\cos(2\sqrt{\eta(\xi)}\Omega\tau + 2\varphi),
\end{equation}
From Eq. \ref{eq:uy_gc2} $\langle u_y^2\rangle = w_{\perp}^2/(2\eta)+O(V_E''^2)$ can be calculated so that $\langle v_y\rangle$ (Eq. \ref{eq:vyavg_expansion2}) becomes,
\begin{equation}
	\langle v_y \rangle = 
	V_E(\xi) +
	\frac{V_E''(\xi) w_{\perp}^2}{4\Omega^2\eta^2(\xi)} 
	+ O(V_{E}''^2)
\end{equation}
Integrating the velocities particle orbits are,
\begin{equation}
\begin{split}
\label{eq:final_orbitsx}
	x-x_0 &= - \frac{w_{\perp}}{\sqrt{\eta(\xi)}\Omega}
	\left[\cos(\sqrt{\eta(\xi)}\Omega\tau + \varphi) - \cos(\varphi)\right] +
	\\
	&\frac{V_E''(\xi)w_{\perp}^2}{12\eta(\xi)^2\Omega^3}
	\left[\cos(2\sqrt{\eta(\xi)}\Omega\tau + 2\varphi) - \cos(2\varphi)\right]
\end{split}
\end{equation}
\begin{equation}
\begin{split}
\label{eq:final_orbitsy}
	y-y_0 &=  \frac{w_{\perp}}{\eta(\xi)\Omega}
	\left[\sin(\sqrt{\eta(\xi)}\Omega\tau + \varphi) - \sin(\varphi)\right] -
	\\
	&\frac{V_E''(\xi)w_{\perp}^2}{24\eta(\xi)^{5/2}\Omega^3}
	\left[\sin(2\sqrt{\eta(\xi)}\Omega\tau + 2\varphi) - \sin(2\varphi)\right]
	+\left\langle v_y\right\rangle \tau
\end{split}
\end{equation}
A major departure from the uniform electric field case is an effective renormalization of the gyrofrequency.   To leading order in the field gradient $\Omega\rightarrow\bar{\Omega}=\sqrt{\eta}\Omega$.  Hence, even the oscillatory part of the particle orbits is dependent on the electric field gradient and the effective gyrofrequency becomes spatially dependent even when the magnetic field is uniform.

Depending on the magnitude and sign of the electric field gradient, $\eta$ can be positive or negative.  This has implications for particle orbits.  Consider a weak electric field gradient, \textit{i.e.}, $\rho/L<1$ where $\rho$ is the particle gyroradius, and $\eta>0$.   To leading order in the gradient the equation of motion may be simplified to $\ddot{v}_x=-\eta(x)\Omega^2v_x+O(V_E'')$, which shows that the particle orbit is either oscillatory or divergent depending on the sign of $\eta(x)$.  Depending on the magnitude of the gradient, the effective gyroradius, $\bar{\rho}=v_t/\bar{\Omega}$, can be larger or smaller compared to the uniform electric field case for which  $\eta=1$.  This will be reflected in the averaged equilibrium quantities as larger or smaller temperatures and affect plasma distribution functions, as we shall discuss in detail in Sec. \ref{subsec:analytical_f}.  While the $\eta\rightarrow0$ limit leads to weak magnetization with large gyroradius resulting in weak magnetic confinement of the particles,   $\eta\rightarrow\infty$ leads to strong magnetization, which effectively is electrostatic confinement of the particles.  This property may be especially consequential to the chaotic orbits \cite{Chen:1992} in the neighborhood of the null sheet in the magnetic field reversed geometry in the earth’s magnetotail when there is guiding magnetic field normal to the current sheet.  As discussed in Sec. \ref{subsec:frequi}, an electrostatic potential self-consistently develops around the null sheet that has not been considered in the studies of the chaotic particle orbits in this region.

In the weak gradient limit, the higher-order derivatives of the electric field are not important but they become critical for stronger gradients when $\eta<0$.  For $\eta < 0$ the equation of motion becomes $\ddot{v}_x=|\eta(x)|\Omega^2 v_x+O(V''_E)$ indicating that the restoring nature of the force becomes divergent and the particle accelerates along the electric field.  Gavrishchaka \cite{Gavrishchaka:1996phd} studied the strong gradient limit. He showed that for strong gradients, multiple guiding centers can arise and the particles do not accelerate indefinitely unless the electric field is linear, which is a pathological case. Higher order derivatives prevent indefinite linear acceleration, which results in modified orbits that are no longer the ideal gyromotion.  Effectively, the particle acquires a larger gyroradius around a new guiding center.  As shown in Sec. 2, this can have major implications to the equilibrium properties when $\eta_i$ becomes small and negative in the narrow layers with $\rho_i > L > \rho_e$.

When the scale size of localization reduces much below the gyroradius the gyro-averaged electric field experienced by the particle reduces until a limit is reached below which the electric field becomes negligible \cite{Gavrishchaka:1996phd}.  Consequently, the particle $\mathbf{E}\times\mathbf{B}$ motion is drastically reduced if not eliminated.  In plasmas this can lead to an interesting regime when $\rho_i \gg L > \rho_e$ in which the ions do not experience the $\mathbf{E}\times\mathbf{B}$ drift but the electrons do.  For short time scale processes, such that $\Omega_i\ll\omega < \Omega_e$, the ions effectively behave as an unmagnetized fluid while the electrons remain magnetized.   This gives rise to a Hall current even in a collisonless uniform plasma.  In plasmas undergoing compression, or relaxing from it, the scale size of the electric field varies in time, which affects the particle orbits differently at different stages of compression or relaxation.  These changes in particle orbits affect the collective dynamics resulting in the observed spectral characteristic that includes broadband emission as we discuss in Sec. \ref{subsec:vlasov_stability}.

\subsection{Analytical distribution function}
\label{subsec:analytical_f}

To understand the ramifications of the orbit modification discussed in Sec. \ref{subsec:particle_orbits} on plasma collective effects it is necessary to develop a kinetic formalism to analyze the stability of plasmas including localized DC electric fields.  For doing so we must obtain a representative zeroth order distribution function appropriate for the initial equilibrium state characterized by a homogeneous magnetic field and an inhomogeneous electric field in the transverse direction.  In Sec. \ref{sec:Equilibrium} we found such a distribution function for arbitrary magnitude of the compression but it is a solution that uses special functions and does not lend itself transparently to perturbative analysis of the stability properties, which is ideal for a general understanding of the plasma response to localized electric fields.  In this Section we construct an analytical distribution function for weak shear, \textit{i.e.} for $\rho/L<1$  and $\eta>0$, using the constants of motion $H(x)$ and the guiding center position $\xi$, which will then be perturbed in Secs \ref{subsec:vlasov_stability} to understand the stability of the Vlasov equilibrium state of a compressed plasma.  Consider the equilibrium distribution function introduced by Ganguli \textit{et~al.} \cite{Ganguli:1988hh},
\begin{equation}
	\label{eq:analytical_f0}
	f_{0}(H(x),\xi)=\frac{N}{\sqrt{\eta(\xi)}}
	g(\xi) e^{-\beta_t H(x)} e^{-\beta_t H_{\|}(\xi)},
\end{equation}
\begin{equation}
	g(\xi)=e^{\beta_t\left[\frac{e}{m}\Phi_{0}(\xi) + \frac{V_{E}^2}{2}\right]},
\end{equation}
where $N=n_{0}(\beta_t/(2\pi))^{3/2}$, $\beta_t=1/v_t^2$, $H_{\|}(\xi)=(v_z-V_{\|}(\xi))^2/2$, $v_{t}=\sqrt{T/m}$ is the thermal velocity, and $V_{\|}(\xi)$ is an inhomogeneous drift along the magnetic field.  In constructing the distribution function two requirements are imposed: 1) the velocity integrated distribution function should produce a constant density so that a static electric field generated in a quasi-neutral plasma without a significant density gradient can be studied.  However, a density gradient, as prevalent in the compressed layers discussed in Sec. \ref{sec:Equilibrium} , can be included through $n_{0}(\xi)$ when necessary, and 2) although any function constructed out of constants of the motion is a Vlasov solution, the particular choice must reduce to the fluid limit when the temperature $T\rightarrow 0$ .  The importance of the later will become apparent in Sec \ref{subsec:vlasov_stability}.

In the weak compression limit when $\epsilon=\rho/L<1$ and for $V_{\|}(\xi)=0$  the distribution function can be simplified.  Using $v_y=u_y+\langle v_y\rangle$ in the argument of distribution function Eq. \ref{eq:analytical_f0} and expanding the argument around the guiding center position it can be simplified to,
\begin{equation}
	\label{eq:analytical_f0_simplified}
	f_{0}\simeq \frac{n_{0}}{\sqrt{\eta(x)}(2\pi v_{t}^{2})^{3/2}}
	e^{-(v_x^2 + (v_y-V_{E}(x))^2/\eta(x) + v_z^2)/(2 v_{t}^2)}
	+ O(\epsilon)
\end{equation}
where terms up to  $O(V_{E}')$ are retained.  For a uniform electric field, \textit{i.e.}, $V_{E}'=0$, $\eta=1$ and $w_{\perp}^2=v_x^2+(v_y-V_E^0)^2$.   Eq. \ref{eq:analytical_f0_simplified} reduces to a Maxwellian distribution with $v_{y}$ shifted by a constant $V_{E}^{0}$  velocity.  Since the  $\mathbf{E}\times\mathbf{B}$ drift is identical for both electrons and ions in collisionless plasma there is no relative drift between the species to feed energy to waves and hence the distribution is stable.  This shows that global compression results in a deviation from a Maxwellian distribution through the velocity gradient, which is a source of free energy for waves. In a collisionless environment compression triggers a relaxation mechanism to reach a steady state through the emission of waves and hence by dissipating the velocity gradient.  The dependence of the distribution function on the spatial gradient of the velocity through the parameter $\eta$ and its asymmetric appearance in the distribution function is noteworthy.  It shows that the temperature in the y direction is preferentially affected by the localized electric field across the magnetic field in the x direction, which introduces an asymmetry and breaks the gyrotropy of the distribution function.  This results in a difference in the temperature in the x and y directions orthogonal to the magnetic field \cite{Ganguli:2018vf}.

In the following sections we will analyze how the electric field gradient can excite broadband waves that can relax the gradients and hence the compression.

Transforming to the cylindrical coordinates ( $w_{\perp}$, $\varphi$, $v_z$ ) by using the Jacobian, 
\begin{equation}
	|J|= \frac{\sqrt{\eta}w_{\perp}}{
	1+\frac{V''_{E} w_{\perp} \cos(\varphi)}{2\eta^{3/2}\Omega^2}}
\end{equation}
the velocity integrals can be performed to obtain $n_{0}(x)=n_{0}(1+O(\epsilon^2))$\cite{Ganguli:1988hh}.  This shows that a large localized static electric field can be maintained in a quasi-neutral plasma across the magnetic field with negligible density gradient, as is observed in the earth’s auroral region  \cite{Mozer1977}.

\subsection{Stability of the Vlasov equilibrium}
\label{subsec:vlasov_stability}

Electric fields encountered in both laboratory and natural plasmas are nonuniform, albeit with varying degree of nonuniformity.  For example, in Sec. \ref{sec:Equilibrium} we showed that the ambipolar electric field that develops self-consistently in a compressed plasma is highly nonuniform.  In Sec \ref{subsec:analytical_f} we established that such electric fields make the equilibrium distribution function non-Maxwellian and therefore introduces a source of free energy for waves.  In collisionless plasmas these waves are a natural response to compression since they relax the shear in the electric field so that a steady state can be achieved.  Due to the strong spatial variability across the magnetic field the plane wave or WKB approximations will break down. Also, some of the modes due to transverse flows discussed below are essentially nonlocal in nature with no local limit.  Hence, the analysis of these waves must be treated as an eigenvalue problem.  Their dispersion relation is usually a differential or an integral equation.  In the following we highlight the key aspects of the derivation of the eigenvalue condition and refer the readers to \cite{Ganguli:1988hh} for details.  

Linearizing the analytical equilibrium distribution given in Eq. \ref{eq:analytical_f0} with a nonuniform density,   $N(\xi)$, we get $f(x,\mathbf{v},t)=f_{0}(x,\mathbf{v})+f_{1}(x,\mathbf{v},t)$.  Since the inhomogeneity is in the $x$-direction the fluctuating quantities, \textit{e.g.}, the electrostatic potential, is periodic in $y$- and $z$- directions but localized in the $x$-direction, \textit{i.e.}, $\phi(r',t)=\exp[-i(\omega t'-k_y y'-k_z z')]\phi(x')$ where $\phi(x')=\int dk'_x\exp(ik_x x')\phi_{k}(k'_x)$.  Then, the perturbed density fluctuation may be obtained as $n_{1}(x)=\int d^3\mathbf{v}\,f_{1}(x,\mathbf{v})$.  Using the orbits given in Eqs. \ref{eq:final_orbitsx} – \ref{eq:final_orbitsy} it can be shown that, 
\begin{equation}
	n_{1}(k_x) = 
	-\frac{e\beta_t}{2\pi m} \iiint dx\,d^{3}\mathbf{v} dk'_{x}\,
	\phi(k'_{x}) f_{0}(\xi,w_{\perp})
	\left[
	e^{i(k'_{x}-k_{x})x} - e^{i(k'_{x}-k_{x})\bar{\xi}}F
	\right],
\end{equation}
\begin{equation}
\label{eq:Fdef}
	F=(\omega-k_y V_g)\sum_{l,l’,m,m’}
	\frac{ J_{l'}(\sigma') J_{m'}(\hat{\sigma}') J_{l}(\sigma) J_{m}(\hat{\sigma}) 
	}{
	\omega -(l'-2m')\bar{\Omega} - k_{y}\langle v_{y}\rangle
	}
	e^{i\left\{ 2(m-m')-(l-l') \right\}\varphi}
	e^{i\left\{l\delta -l'\delta' - m\hat{\delta} + m'\hat{\delta}'\right\}},
\end{equation}
where $J_{m}(\sigma)$ are Bessel functions, $\sigma'=k_{\perp}' w_{\perp}/\Omega$, $k_{\perp}'^{2}=k_{x}'^2/\eta + k_{y}^2/\eta^2$, $\delta'=\tan^{-1}(k'_{x}\sqrt{\eta}/k_{y})$, $\hat{\sigma}'=\hat{k}_{\perp}'\hat{w}_{\perp}/(12\Omega)$, $\hat{w}_{\perp}= V''_E w_{\perp}^2/\Omega^2$, $\hat{\delta}'=\tan^{-1}(2k'_{x}\sqrt{\eta}/k_{y})$, $\hat{k}_{\perp}'^2=k_{x}'^2/\eta^4 + k_{y}^2/4\eta^5$, $\bar{\Omega}=\sqrt{\eta}\Omega$, and $\bar{\xi}=x+u_{y}/\Omega$.  $V_g$ is the bulk fluid drift in the plasma and is given by
\begin{equation}
	V_{g}(\xi) = \frac{1}{\eta(\xi)\Omega \beta_{t}}
	\frac{1}{f_{0}} \frac{\partial f_{0}}{\partial \xi}
	= V_{E}(\xi) - 
	\frac{V_{E}''(\xi)\rho^2}{2\eta^2(\xi)} 
	- \frac{\epsilon_n \rho \Omega}{\eta},
\end{equation}
so that,
\begin{equation}
	\omega-k_{y}V_{g}(\xi) = \omega - k_{y}V_{E}(\xi)
	+ \frac{k_{y} V_{E}''(\xi) \rho^2}{2\eta^2(\xi)}
	- \frac{k_{y} \epsilon_{n} \rho \Omega}{\eta}
	\equiv \omega_{1}+\omega_{2}-\omega^{*},
\end{equation}
where $\omega_{1}=\omega-k_{y}V_{E}(\xi)$ is the local Doppler shifted frequency, $\omega_{2}=k_{y} V_{E}''(\xi)\rho^2/2\eta^2$ is a frequency that is introduced due to the second derivative, \textit{i.e.} the curvature, of the electric field, $\omega^{*}=k_{y}\rho\epsilon_{n}\Omega/\eta$ is the diamagnetic drift frequency, and $\epsilon_{n}=\rho/L_{n}$ where the density gradient scale size $L_{n}=[(dn/dx)/n]^{-1}$.
 
A number of noteworthy features arise compared to the uniform electric field case.  Unlike the trivial case when a global Doppler shift is appropriate, in the nonuniform case a local doppler shift arises and no global transformation can eliminate this spatially dependent shift.  Because of the spatial inhomogeneity the plane wave assumption in the direction of the inhomogeneity is no longer possible.  Higher harmonics of quantized eigenstates are possible, which can broaden the frequency and wave vector bandwidth of the emissions.  The transverse electric field becomes an irreducible feature defining the bulk plasma and affects its dielectric properties including the normal modes of the system.  New time scales, represented by the frequencies $\omega_{1}$ and $\omega_{2}$, are introduced.  A resonance with the bulk plasma flow arises that can affect the fluid (macro) stability.  Landau and cyclotron resonances with individual particles are affected through orbit modifications altering the kinetic (micro) stability of the plasma.   Consequently, the transverse electric field can affect both the real and imaginary parts of the dispersion relation and therefore affect both the real and imaginary parts of the frequency of oscillations.  This can vastly alter the known waves that characterize a plasma with uniform magnetic field and their nonlinear behavior.  Under certain conditions the transverse electric field can suppress some waves while in others waves can be reinforced \cite{Gavrishchaka:1996}.  In addition, an entirely new class of oscillation becomes possible due to an inhomogeneity in the wave energy density introduced by the variable Doppler shift \cite{Ganguli:1985a}.

Quasi neutrality, \textit{i.e.}, $\sum_{\alpha}\int dk_{x}\exp(ik_{x}x)n_{1\alpha}(k_{x})=0$, gives the general dispersion condition for the waves, in the electrostatic approximation, which is an integral equation and cumbersome to solve.  However, for weak gradients, \textit{i.e.} $\rho/L<1$, $\eta\sim 1$, and $k_{x}\simeq k_{x}'$, some simplifications are possible.  For example, $\hat{\sigma}'\propto (\rho/L)^2\ll 1$ so we may use $J_{0}(\hat{\sigma})\sim J_{0}(\hat{\sigma}')\sim 1$ and ignore terms higher than $m=m'=0$.  Furthermore, $k_{x}\simeq k_{x}'$ implies $\sigma'\simeq\sigma$ and $\delta'\simeq\delta$.  In the $O(\rho/L)^2$ term in the denominator of Eq. \ref{eq:Fdef} we may replace $w_{\perp}^2$, that appears in $\langle v_{y}\rangle$, by $2v_{t}^2$.  This simplifies $F$ considerably to,
\begin{equation}
\label{eq:Fsimplified}
	F=(\omega_1+\omega_2-\omega^*)
	\sum_{l',l} \frac{ J_{l'}(\sigma') J_{l}(\sigma)
	}{\omega_1-\omega_2-l'\Omega
	}
	e^{\left[
	i(l'-l)\varphi + il\delta - il'\delta' 
	\right]}.
\end{equation}
It is interesting to note that the electric field curvature related frequency, $\omega_{2}$, that appears in the numerator of Eq. \ref{eq:Fsimplified} originates from the fluid plasma flow, while the one in the denominator originates from the individual particle orbit due to its kinetic behavior and will be absent in the fluid framework.  With these simplifications and transforming coordinates from Cartesian, $(x,v_x,v_y,v_z)$, to cylindrical, $(\xi,w_{\perp},\varphi,v_z)$, the velocity integrals can be readily performed to obtain the density fluctuations,
\begin{multline}
n_{1}(x) = \frac{e\beta_t}{2\pi m}
\int dk_{x} \exp(ik_{x}x) \iint d\xi dk_x' \, \phi(k'_x)
\exp[i(k_x'-k_x)\xi]n_{0}(\xi) \times
\\
\left\{
1 + \sum_{l}\left(
\frac{\omega_{1}+\omega_{2}-\omega^*}{\sqrt{2}|k_{\|}|v_{t}}
\right) Z\left(
\frac{\omega_{1}-\omega_{2}-l\Omega}{\sqrt{2}|k_{\|}|v_{t}}
\right) \Gamma_{l}(\bar{b})
\right\}	
\end{multline}
where $Z(\zeta)=(\pi)^{-1/2}\int_{-\infty}^{\infty} dt\,\exp(-t^2)/(t-\zeta)$ is the plasma dispersion function, $\Gamma_{n}(\bar{b})=\exp(-\bar{b})I_{n}(\bar{b})$, $\bar{b}=(k_{\perp}\rho)^2$, and $I_{n}(\bar{b})$ is the modified Bessel function.  The weak gradient condition allows the expansion $\Gamma_{l}(\bar{b})=\Gamma_{l}(b)-\Gamma'_{l}(b)\rho^2k_{x}^2+O((\rho k_{x})^4)$, where $b=(k_{y}\rho)^2$ so that the remaining integrals can be easily performed to obtain,
\begin{multline}
\label{eq:fluc_density}
	n_{1}(x) = - \frac{\omega_{p}^2}{4\pi v_{t}^2 q}\left[
	-\sum_{n}\left(\frac{\omega_{1}+\omega_{2}-\omega^*}{\sqrt{2}|k_{\|}|v_{t}}
	\right) 
	Z\left(
\frac{\omega_{1}-\omega_{2}-n\Omega}{\sqrt{2}|k_{\|}|v_{t}}
\right) \frac{d\Gamma_{n}(b)}{db} \rho^2 \frac{d^2}{dx^2}
	\right.
	\\
	\left.
	+ 1 + \sum_n \left(\frac{\omega_{1}+\omega_{2}-\omega^*}{\sqrt{2}|k_{\|}|v_{t}}
	\right) 
	Z\left(
\frac{\omega_{1}-\omega_{2}-n\Omega}{\sqrt{2}|k_{\|}|v_{t}}
\right)\Gamma_{n}(b)
	\right] \phi(x)
\end{multline}
which, in conjunction with the quasi-neutrality condition or the Poisson equation, provides the electrostatic dispersion eigenvalue condition in the form of a second order differential equation.  Using the fluid model for ions the derivation has also been generalized to the electromagnetic regime \cite{Penano:1999prl,Penano:2000,Penano:2002}.

\subsubsection{Low frequency limit:  fully magnetized ions and electrons}
\label{subsec:linear_low_freq_limit}

We first consider the linear plasma response to a weak compression where the electric field scale size $L>\rho_{i}$.  As discussed in Sec. \ref{subsec:particle_orbits}, in this case both ions and electrons experience identical electric field magnitude since on average they sample the electric field throughout their gyro-motion.  Hence, to the zeroth order, their $\mathbf{E}\times\mathbf{B}$ drift will be identical.  Under this condition the fluctuating density for both the ions and the electrons is given by Eq. \ref{eq:fluc_density} with respective mass and charge, which leads to the electrostatic dispersion relation under the quasi neutrality condition, $\sum_{\alpha} q_{\alpha} n_{1\alpha}=0$ .  Ignoring terms of the order of $(m_{e}/m_{i})^2$ and considering low frequency waves, $\omega_{1}<\omega_{LH}=\omega_{pi}/(1+\omega_{pe}^2/\Omega_e^2)^{1/2}$, where $\omega_{LH}$ is the lower-hybrid frequency, the $n=0$ cyclotron harmonic term for the electrons is sufficient.  Then the  eigenvalue condition is, 
\begin{equation}
	\label{eq:low_freq_eigenval}
	\left[
	\rho_{i}^2 A(x)\frac{d^2}{dx^2} + Q(x)\right] \phi(x) 
	+ O(\epsilon^3) = 0,
\end{equation}
where
\begin{equation}
	\label{eq:A_def}
	A(x) = -\sum_{n} \left(\frac{\omega_{1}+\omega_{2}-\omega^*}{\sqrt{2}|k_{\|}|v_{ti}}
	\right) 
	Z\left(
\frac{\omega_{1}-\omega_{2}-n\Omega_i}{\sqrt{2}|k_{\|}|v_{ti}}
\right)\frac{d\Gamma_{n}(b)}{db},
\end{equation}
\begin{multline}
	\label{eq:Q_def}
Q(x)= 1 + \sum_n \left(\frac{\omega_{1}+\omega_{2}-\omega^*}{\sqrt{2}|k_{\|}|v_{ti}}
	\right) 
	Z\left(
\frac{\omega_{1}-\omega_{2}-n\Omega_{i}}{\sqrt{2}|k_{\|}|v_{ti}}
\right)\Gamma_{n}(b)
	\\
	+\tau\left[
	 1 + \left(\frac{\omega_{1}+\omega_{2}/\tau\mu-\omega^*/\tau}{\sqrt{2}|k_{\|}|v_{te}}
	\right) 
	Z\left(
\frac{\omega_{1}-\omega_{2}/\tau\mu}{\sqrt{2}|k_{\|}|v_{te}}
\right)\right],
\end{multline}
and $\tau=T_{i}/T_{e}$, $\mu=m_{i}/m_{e}$.  There are two branches of oscillations driven by the electric field in this equilibrium configuration \cite{Ganguli:1988hh}.  These branches do not require a density gradient so in the following analysis we set $\omega^{*}=0$.

\begin{figure}
	\includegraphics[width=\textwidth]{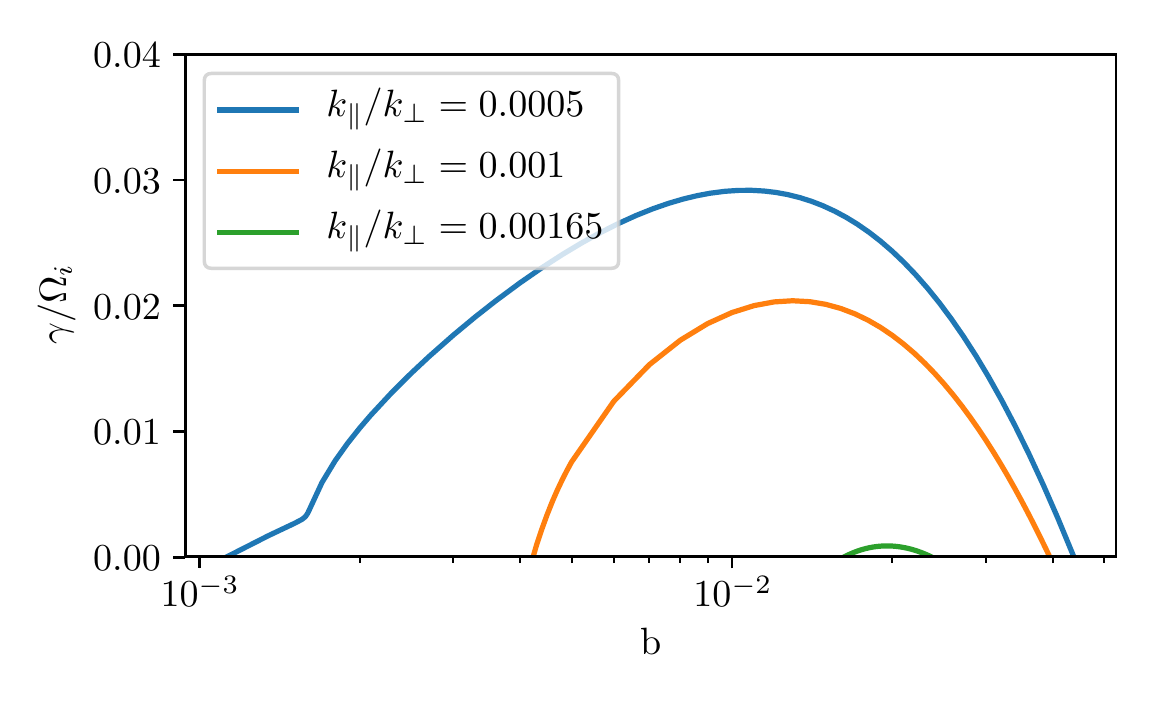}
	\caption{\label{fig:KinKHmodes}Kinetic solutions (for three values of $k_{\|}/k_\perp$) showing that the KH modes are strongly Landau damped.  The parameters used are $\epsilon =\rho_{i}/L =.1$, $\tau = T_i/T_e=5$, $\bar{V}_E =V_{0E}/v_{ti}=2$, $\mu=m_i/m_e=1837$, and no density gradient.}
\end{figure}

%
%
%

\subsubsubsection{Kelvin-Helmholtz Instability Branch}

For low frequencies, such that $\omega_{1}\ll n\Omega_{i}$, the $n=0,\pm 1$ terms for the ions are sufficient in Eq. \ref{eq:low_freq_eigenval}.  This gives the kinetic generalization of the dispersion relation for the Kelvin-Helmholtz (KH) modes.  Kinetic solutions of Eq. \ref{eq:low_freq_eigenval} in this limit with  $E(x)=E_0 \textrm{sech}^2(x/L)$, $L = 10 \rho_i$, shown in Fig. \ref{fig:KinKHmodes}, indicate that the KH mode is strongly Landau damped.

The KH instability is the quintessential shear flow driven instability invoked in innumerable applications in the fluid phenomenology both in space and laboratory plasmas.  It is extensively invoked in large-scale fluid models in space plasmas.  If long wavelengths, \textit{i.e.}, $k_{\|}\rightarrow 0$, or cold plasma, \textit{i.e.}, $T\rightarrow 0$, can be realized then this may be justified.  But caution must be exercised since, as evident from Fig. \ref{fig:KinKHmodes}, the KH mode is highly sensitive to Landau damping especially for $T_{i}\geq T_{e}$, which is usually the case in the magnetosphere and the $T\rightarrow 0$ assumption is not realistic.   Also, because of the inhomogeneous magnetic field structure in the region, which may introduce geometrical constraints, very long wavelengths necessary to avoid Landau damping may not be possible.  Even for long parallel wavelength, $k_{\|}\rightarrow 0$, such that the parallel phase speed of the waves is larger than the ion and electron thermal speeds the KH modes can be damped by finite Larmor radius (FLR) effects if the perpendicular wavelengths are sufficiently short, which is likely in the thin compressed layers.  In this case $A(x)$ and $Q(x)$ reduces to,
\begin{equation}
	A(x) = \left(\frac{\omega_{1}+\omega_{2}}{\omega_{1}-\omega_{2}}\right)
	\Gamma'_{0}(b) + 
	\left(\frac{\omega_{1}^2-\omega_{2}^2}{(\omega_{1}-\omega_{2})^2-\Omega_{i}^2}\right)
	2\Gamma'_{1}(b)
\end{equation}
\begin{equation}
	Q(x) =1-\left(\frac{\omega_{1}+\omega_{2}}{\omega_{1}-\omega_{2}}\right)
	\Gamma_{0}(b) + 
	\left(\frac{\omega_{1}^2-\omega_{2}^2}{(\omega_{1}-\omega_{2})^2-\Omega_{i}^2}\right)
	2\Gamma_{1}(b)
\end{equation}
The Bessel functions diminish the magnitude of the source term for the KH modes, which is proportional to $\omega_{2}$ as will become clear in Eq. \ref{eq:classical_KH}.  If the perpendicular wavelength is also sufficiently long such that $b=(k_{y}\rho_{i})^2\ll 1$, then  $\Gamma_{0}(b)\sim 1-b$, $\Gamma_{0}'(b)\sim -1$, $\Gamma_{1}(b)\sim b/2$, and $\Gamma_{1}'(b)\sim 1/2$.  With these values and in the low frequency limit, $\Omega_{i}>\omega_{1}>\omega_{2}$, the order unity terms in $Q(x)$ cancel out making the second order terms proportional to $(\rho_{i}/L)^2$ as the leading order in the eigenvalue condition, which then yields the classical fluid KH mode equation \cite{Raleigh:1896,Drazin:1966},
\begin{equation}
\label{eq:classical_KH}
\left[
\frac{d^2}{dx^2} - k_{y}^2 + \frac{k_y V_{E}''(x)}{\omega-k_{y}V_{E}(x)}
\right]\phi(x) = 0	
\end{equation}
 
In producing the fluid limit the frequency $\omega_{2}$ in the numerator of $Q(x)$, which originates from the fluid plasma property, combines in equal part with the one in the denominator, which originates from the kinetic plasma property, to constitute the source term proportional to $V_{E}''$ that feeds the KH instability.  

Another kinetic effect is gyro-averaging. As a result, the fluid flow due to the $E\times B$ drift and its derivatives become smaller as the scale size of the velocity shear becomes comparable or less than an ion gyroradius.  This reduces the curvature of the flow and hence lowers the KH source term (see Fig. \ref{fig:ebarprofiles}).  This shows that the kinetic effects are deeply entrenched in the KH mechanism, which can modify the source term substantially.  The kinetic effects can be strong enough to stabilize the instability in a large portion of the parameter space allowed to it within the fluid framework thereby limiting its applicability.  In addition Keskinen \textit{et~al.} \cite{Keskinen:1988} and Satyanarayana \textit{et~al.} \cite{Satyanarayana1987} have shown that a density gradient has a stabilizing effect on the KH modes.

It is important to realize that in the fluid limit all the order unity terms exactly cancel each other in Eq. \ref{eq:low_freq_eigenval}, making the otherwise negligible second-order terms responsible for KH instability as leading terms. This is critical to the recovery of the KH eigenvalue condition in the fluid limit, implying that the KH limit is sensitive to the choice of the initial distribution function.  A number of different initial distribution functions are possible and were tried but only the particular one described by Eq. \ref{eq:analytical_f0} yielded the classical KH eigenvalue condition in the fluid limit \cite{Ganguli:1988hh,Ganguli:1997}. Since many distribution functions are possible but not all of them lead to the KH modes, the robustness of the KH instability in warm plasma becomes questionable in comparison to the Inhomogeneous Energy Density Driven Instability (IEDDI) discussed below, which does not depend on a particular choice and therefore may be more ubiquitous. 

%
%
%

\subsubsubsection{Inhomogeneous Energy Density Driven Instability Branch}

\begin{figure}
	\includegraphics[width=\textwidth]{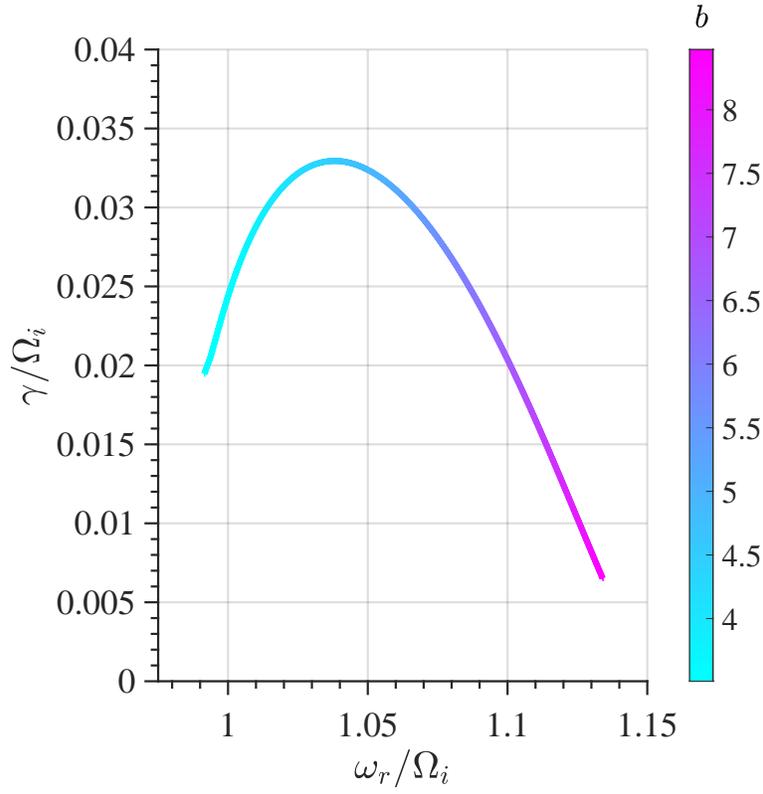}
	\caption{\label{fig:IEDDI_modes} Growth rate vs frequency for IEDDI instability as a function of $b=(k_y\rho_i)^{1/2}$ in color.  For these calculations $k_{\|}/k_{\perp}=0.011$, $\epsilon=\rho_i/L=.3$, $a=1.87$, $\tau=5$, $\mu=1837$, and no density gradient.}
\end{figure}

The above discussion on the Kelvin-Helmholtz limit also implies that in the kinetic regime for shorter wavelengths such that the wave phase speed is larger than or of the order of the ion thermal velocity but smaller than the electron thermal velocity, \textit{i.e.} $v_{te}>(\omega_{1}-n\Omega_{i})/k_{\|}\geq v_{ti}$, and $\omega_{1}\sim n\Omega_{i}$ the second order terms in $A(x)$ and $Q(x)$ may be neglected with respect to the order unity terms.  This regime leads to a different branch of oscillations arising due to the inhomogeneity in the wave energy density introduced by the velocity shear \cite{Ganguli:1985a}.   Unlike the KH instability the IEDDI can be enhanced by a density gradient \cite{Ganguli:1988eb,Liu:2018,Ilyasov:2015}.  Fig. \ref{fig:IEDDI_modes} shows the typical linear spectrum of the IEDDI.  The background electric field profile used is $E(x)=E_0 \textrm{sech}^2(x/L)$ with $L=3.3\rho_i$, $\tau=5$, $V_E/v_{ti}=.1$, and $k_{\|}=0.011 k_{\perp}$.  The spectrum remains relatively unaffected for an electric field with a top hat profile (Fig. \ref{fig:ebarprofiles}), although the growth rates reduce as the field profile becomes smoother.  This is because the IEDDI does not depend on the local value of a specific derivative of the electric field like the KH mode.

\begin{figure}
	\includegraphics[width=\textwidth]{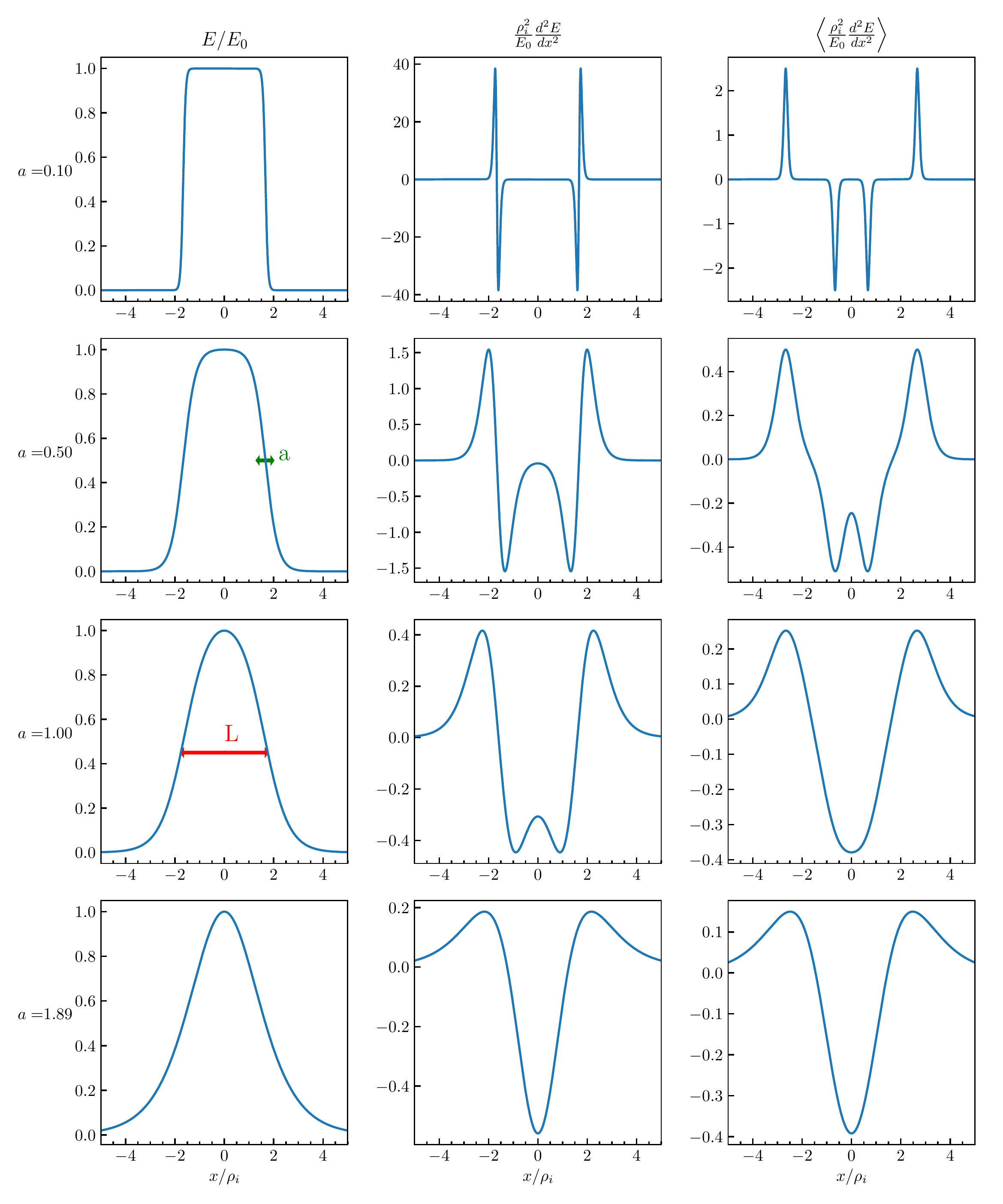}
	\caption{\label{fig:ebarprofiles} The left column shows profiles of the electric field for different values of $a$ with $\epsilon=0.3$.  The middle column shows the profile of the second derivative normalized to the ion thermal gyroradius.  The column on the right indicates the gyro-averaged second derivative. }
\end{figure}

To understand the general characteristics of the two (KH and IEDDI) branches of oscillations we have considered a generic electric field profile,
\begin{equation}
\label{eq:2scaleE}
	E(x) = \frac{E_0}{A \sinh^2(x/a)+1}
\end{equation}
where $A=1/\sinh^2(x_0/a)$, $x_0=L/2$, $\epsilon=\rho_i/L$.  At $x=x_0$ the value of $E(x)$ reduces to $E_0/2$.  For $a=x_0/\sinh^{-1}(1)$, $A=1$ and $E(x)=E_0\textrm{sech}^2(x/a)$.  For $a\rightarrow 0$ the profile becomes a top-hat profile.  This profile is characterized by two scale lengths, $L$ and $a$.  In the natural environment, especially under compression, the static electric fields are likely to be generated with multiscale profiles.  This also becomes apparent from our equilibrium studies in Sec. 2.  In Eq. \ref{eq:2scaleE} while $L$ determines the overall extent of the localization of the electric field, $a$ determines its local gradient.  For $A\rightarrow 1$, the scale lengths $a$ and $L$ become comparable.  The first column of Fig. \ref{fig:ebarprofiles} shows the transition of the electric field profile in Eq. \ref{eq:2scaleE} from a top hat to a smooth $\textrm{sech}^2(x)$ as a function of increasing $a$.  The second and the third columns of Fig. \ref{fig:ebarprofiles} show the second derivative and the gyro-averaged second derivative of the electric field.  For $a\rightarrow 0$ the gyro-averaged second derivative of the electric field becomes smaller compared to the un-averaged, indicating that the source of the KH modes become weaker due the kinetic effect of gyro-averaging as $a$ decreases.  This has a stabilizing effect on the KH mode (see eq. (56) below).  On the other hand, electric field profiles with smaller a favors the IEDDI mechanism as it primarily depends on the localized nature of the electric field rather than the local value of any specific derivative (see Eq. (58)) below.  The gyro-averaging effect becomes more prominent as the external compression increases and the scale sizes shrink compared to the ion gyroradius.  (For the KH instability in neutral fluids there is no gyro-averaging, since the particles are not charged, and this stabilizing effect does not exist in a neutral medium.) 

As discussed in Sec 3.3, the general eigenvalue condition for the IEDDI is an integral equation.  For weaker shear it may be approximated to a second order differential equation.  The numerical solution for the truncated IEDDI eigenvalue condition is easier in the $a\rightarrow 0$ limit when the electric field profile is top hat like.  It becomes difficult as the profile becomes smoother with increasing $a$.  The potential, $Q(x)/A(x)$, of the second order differential equation, Eq. (50), becomes stiff and there are a number of roots in close vicinity of each other.  This poses considerable difficulty in tracking the IEDDI roots by solving the differential equation.  Potential barriers develop that obstruct the energy flux away from the negative energy density region created by the localized electric field that is necessary for the IEDDI (as elucidated in Eq. 61).  This may partly be because of the truncation of the integral equation to second order.  Ganguli \textit{et al.} \cite{Ganguli:1988hh} had to use a small density gradient in order to circumvent this difficulty to obtain the roots. 

Thus, unlike the KH modes, the solution to the eigenvalue problem, Eq. \ref{eq:low_freq_eigenval}, with the potential $Q/A$ given by Eqs. \ref{eq:A_def} and \ref{eq:Q_def} for the IEDDI is not trivial.  As $x\rightarrow\infty$, Eq. \ref{eq:low_freq_eigenval} has two asymptotic solutions: one that is exponentially growing and the other exponentially decaying.  The decaying one is the physical solution, but the growing one can easily contaminate numerical solutions.  Furthermore, $Q/A$ has poles scattered around the complex plane that can also make finding precise eigenvalues difficult.

The effects of the exponentially growing solution can be minimized significantly by using the Riccati transform.  This technique was recently applied to tearing instabilities and an explanation of how and why the method works was provided \cite{Finn:2020}.  In this method the potential is transformed using $u =\phi'/\phi$, where the prime denotes an $x$ derivative.  This gives the transformed equation
\begin{equation}
\frac{du}{dx} = -\frac{Q}{A} - u^2\label{eq:riccati}
\end{equation}
which has asymptotic solutions
\begin{equation}
u(x\rightarrow\infty) = \pm i \sqrt{\frac{Q_\infty}{A_\infty}}
\end{equation}
where the $+/-$ refers to growing/decaying solutions.  Therefore, the decaying solution may be chosen at $x\rightarrow\infty$ and integrated backwards, using Eq. \ref{eq:riccati}, towards $x=0$.  For modes with even parity in $\phi(x)$, \textit{i.e.} $\phi'(0)=0$, $u(x)$ should be zero at $x=0$.  A complex root finder (\textit{e.g.} Muller's method or Newton's method) finds the appropriate eigenvalue, $\omega$, that leads to $u(0)=0$.  A close guess for an appropriate $\omega$ is still necessary for the root finder to converge reliably. 

The spiky nature of $Q/A$ can introduce further difficulty, but as long as the poles do not lie exactly on the real axis, a standard numerical integrator that controls accuracy will be sufficient.  In the case that the poles are on the real axis (\textit{e.g.} both the real and imaginary parts of $\phi$ are zero simultaneously), a numerical integrator based on Pad\'e approximations is useful \cite{Fornberg:2011}.  These numerical techniques allow robust solutions to be found without the need to add any density gradient (as was needed in 1988 \cite{Ganguli:1988hh}).

Both the KH and IEDDI branches and their applications have been extensively studied in the literature and are not repeated here.  Instead, below we review the physical mechanisms that are responsible for the two branches of oscillations.

%
%
%

\subsubsubsection{Physical Origin of the Kelvin-Helmholtz Instability}

Although both the branches mentioned above are sustained by the velocity gradient, they rely on different mechanisms for drawing the free energy from it.  This is best understood by analyzing the energy balance conditions.  For the KH modes the energy quadrature can be derived as \cite{Ganguli:1997},
\begin{equation}
	\label{eq:kh_energy_quad}
	\frac{\partial}{\partial t}
	\int dx\left[
	\frac{|E_{1}|^2}{8\pi} +
	\frac{n_{0}m_{i}}{2}\frac{|cE_{1}|^2}{B^2} + 
	\frac{n_{0}m_{i}}{2}|x_1|^2 V_{E}V_{E}''(x)
	\right] =0,
\end{equation}
where $E_{1}=-ik_{y}\phi$, $x_{1}=v_{1x}/(\omega_r-k_{y}V_E(x))$, $v_{1x}=-cE_{1y}/B$, and $\mathbf{E}_{1}$ and $\mathbf{v}_{1}$ are the fluctuating electric field and velocity  The first two terms of Eq. \ref{eq:kh_energy_quad} are due to the fluctuating wave electric field. The first term represents the electrostatic wave energy density in vacuum, the second term is the wave-induced kinetic energy of the ions.  The energy balance condition in Eq. \ref{eq:kh_energy_quad} indicates that reduction in the equilibrium flow energy, \textit{i.e.}, $(\langle V_{E}(x+x_1)\rangle^2 - V_{E}^2(x))=|x_{1}|^2 V_{E}(x)V_{E}''(x)+O((1/L)^3)$, at a given position $x$, which occurs due to time averaging by the waves, is available as the free energy necessary for the growth of the KH instability. The time averaging removes the first derivative and therefore the free energy is proportional to the second derivative of the dc electric field.  Consequently, to leading order the KH instability is explicitly dependent on the second derivative, \textit{i.e.}, the curvature, of the electric field. This condition may be a limiting factor to the viability of the KH instability compared to its sister instability, the IEDDI, which does not depend on any particular velocity derivative as we discuss next.  

%
%
%

\subsubsubsection{Physical Origin of the IEDDI}

When both the electrons and the ions are cold fluids it leads to the classical KH description as shown above. The ions play the crucial role while the electrons simply provide a charge neutralizing background.   But for $k_{\|}\neq 0$, $T_{e}\neq 0$ and for waves with $\omega_{1}\sim n\Omega_{i}$ the electron response can be adiabatic, \textit{i.e.}, $v_{te}>(\omega_1-n\Omega_{i})/k_{\|}\geq v_{ti}$.  In this limit ignoring the $(\rho_{i}/L)^2$ terms in $A(x)$ and $Q(x)$ we obtain the eigenvalue condition for the IEDDI branch. To understand the physics of this branch of oscillations we may assume the ion response to be fluid so that $b\ll 1$  and the eigenvalue condition for the IEDDI reduces to,
\begin{equation}
	\left[
	\frac{d^2}{d\bar{x}^2} -
	\bar{k}_{y}^2 + \left(\frac{\omega_{1}}{\Omega_{i}}\right)^2 - 1
	\right]\phi = 0
\end{equation}
where $\bar{x}=x/\rho_{s}$, $\rho_{s}=c_{s}/\Omega_{i}$, $c_{s}=T_{e}/m_{i}$, and $\bar{k}_{y}=k_{y}\rho_{s}$.  Following the procedure outlined in Ganguli \cite{Ganguli:1997} we obtain the condition,
\begin{equation}
	\label{eq:IEDDI_cond}
	S+\frac{2}{\Omega_{i}^2}
	\int_{-\infty}^{\infty} d\bar{x}\,
	\gamma(\omega_{r}-k_{y}V_{E}(x))|\phi|^2 = 0,
\end{equation}
where $S=(\phi^{*}\phi'-\phi\phi'^{*})/2i$ is the flux and is a positive real number, $\gamma$ is the growth rate for the IEDDI, $\phi^{*}$ is the complex conjugate of $\phi$, and the primes indicate spatial derivatives.  In order for Eq. \ref{eq:IEDDI_cond} to be valid the second term must be negative which implies that the product $\gamma(\omega_{r}-k_{y}V_{E})<0$ in at least a finite interval of space, since other factors are positive definite. Therefore, the necessary condition for IEDDI growth, \textit{i.e.}, $\gamma>0$, is that the Doppler shifted frequency $(\omega_r-k_{y}V_{E})$ be negative in some region of space. 

To understand the physical consequences of $(\omega_{r}-k_{y}V_{E})<0$ that can lead to wave growth consider the ion-cyclotron waves. The homogeneous electrostatic dispersion relation for the ion cyclotron waves is \cite{Drummond:1962}, 
\begin{equation}
	D(\omega) = 1 + \tau - \Gamma_{0}(b)
	- \sum_{n>0}\frac{2\omega^2}{\omega^2-n^2\Omega_{i}^2}\Gamma_{n}(b)=0
\end{equation}
The wave energy density is given by,
\begin{equation}
	U\propto \omega \frac{\partial D}{\partial \omega}
	=\omega\left(
	\sum_{n>0} \frac{4\omega n^2 \Omega_{i}^2\Gamma_{n}(b)}{
	(\omega^2-n^2\Omega_{i}^2)^2}
	\right) \equiv \omega^2 \Xi(\omega)
\end{equation}
Clearly, the ion cyclotron waves are positive energy density waves.   However, introduction of a uniform electric field in the $x$ direction initiates an $\mathbf{E}\times\mathbf{B}$ drift in the $y$-direction and consequently there is a Doppler shift in the dynamical frequency, \textit{i.e.}, $\omega\rightarrow\omega_{1}$. The energy density in the presence of the Doppler shift is, $U_{I}\propto \omega\omega_{1}\Xi(\omega)$, which can be negative provided $\omega\omega_{1}<0$. 

\begin{figure}
	\includegraphics[width=\textwidth]{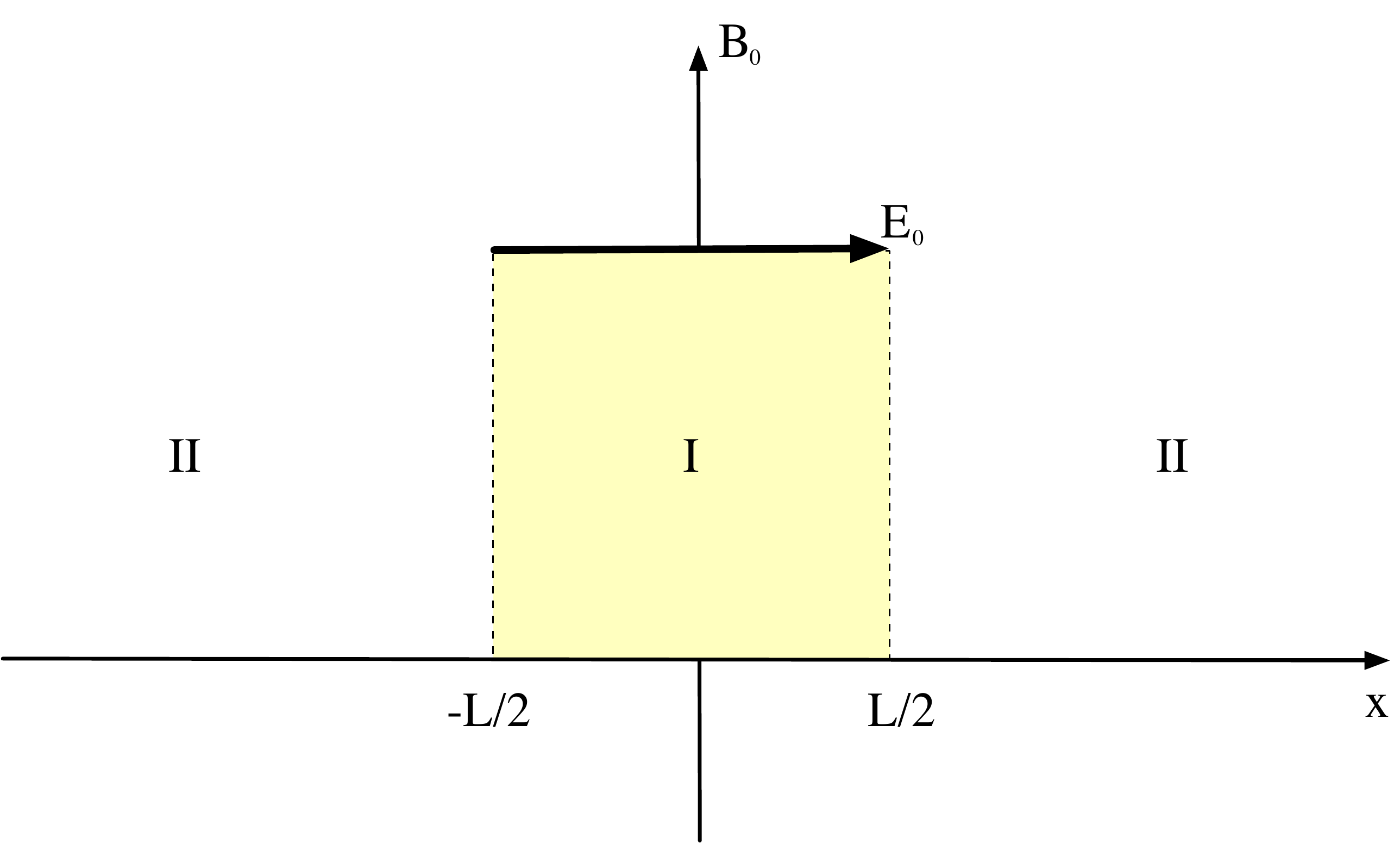}
	\caption{\label{fig:IDDIgeom}Geometry of Inhomogeneous Energy Density Driven Instability (IEDDI).}
\end{figure}

Now consider the simplest example of an inhomogeneous electric field geometry given by a piece-wise continuous configuration as shown in Fig. \ref{fig:IDDIgeom} in which a uniform electric field is localized in the region-I of extent $L$.  It is clear that because of the localized nature of the $\mathbf{E}\times\mathbf{B}$ drift in region-I, the energy density in region-I can become negative provided the Doppler shifted frequency $\omega_{1}<0$, while it remains positive in region-II. A nonlocal wave packet can couple these two regions so that a flow of energy from region-I into region-II will enable the wave to grow.  In region-I it is a negative energy wave while it is positive energy wave in region-II.  The situation is complementary to the two-stream instability.   In that case there are two waves one of positive energy density and the other of negative energy density at every location and their coupling in velocity space leads to the instability.  In the IEDDI case there is only one wave but two regions, one in which the wave energy density is negative and positive in the other.  The coupling of these two regions in the configuration space by a wave packet leads to the instability \cite{Ganguli:1985a}.  

This simple idea may be quantified further using the wave-kinetic framework. The growth of the wave in region-I implies a loss of energy from that region.  By conservation of energy, this must be the result of convection of energy into region-II in the absence of local sources or sinks.  The rate of growth of the total energy deficit in region-I is proportional to the growth rate of the wave, the wave energy density $U_I$ in region-I, and the volume of region-I given by the extent in the x-direction of region-I times a unit area $A_{\perp}$ in the plane perpendicular to $x$. The rate of energy convection through $A_{\perp}$ is $V_{g}U_{II}$, where $V_{g}$ is the group velocity in the $x$-direction and $U_{II}$ is the wave energy density in region-II, which is positive since the electric field is absent in this region. We can then write the power balance condition as
\begin{equation}
\gamma U_{I} L A_{\perp} = - V_{g} U_{II} A_{\perp},	
\end{equation}
which implies that the growth rate of the IEDDI is $\gamma\propto-U_{II}/U_{I}$.  Consequently, if $U_{I}$ is negative then the growth rate is positive showing that the growth of the wave can be sustained by convection of energy into region-II from region-I. On the other hand, if $U_{I}$ is positive then the convection of energy out of region-I would lead to a negative growth rate and, therefore, to damping of the waves.  This shows that if the wave energy density is sufficiently inhomogeneous to change its sign over a small distance then it can support wave growth.  This is in contrast to the KH mechanism in which there is an exchange of energy between the medium and the wave via local plasma flow gradient (Eq. \ref{eq:kh_energy_quad}).  In the IEDDI mechanism such an exchange is not necessary. Instead, as described in Eq. \ref{eq:IEDDI_cond}, the IEDDI is dependent on energy transport from one region to another such that the sign of energy density changes. 

In addition to the driving mechanism described above, dissipative mechanisms are also present in a realistic system.  If the energy gained from the dc electric field is larger than the energy dissipated the wave can exhibit a net growth. It is important to note that this phenomenon is not restricted to a resonant group of particles in velocity space. The only requirement is that $(\omega_r-k_y V_E)<0$  in a localized region.  Thus, the bulk plasma in this region can participate, which results in a broadband frequency spectrum.  

Although we used the ion cyclotron waves as a specific example, the IEDD mechanism described here can affect other waves in the system and therefore represents a genre of instabilities in plasmas that contains a localized electric field.  This makes the transverse electric field a unique source of free energy.

%
%
%

\subsubsubsection{Magnetron Analogy of the IEDDI.}

\begin{figure}
	\includegraphics{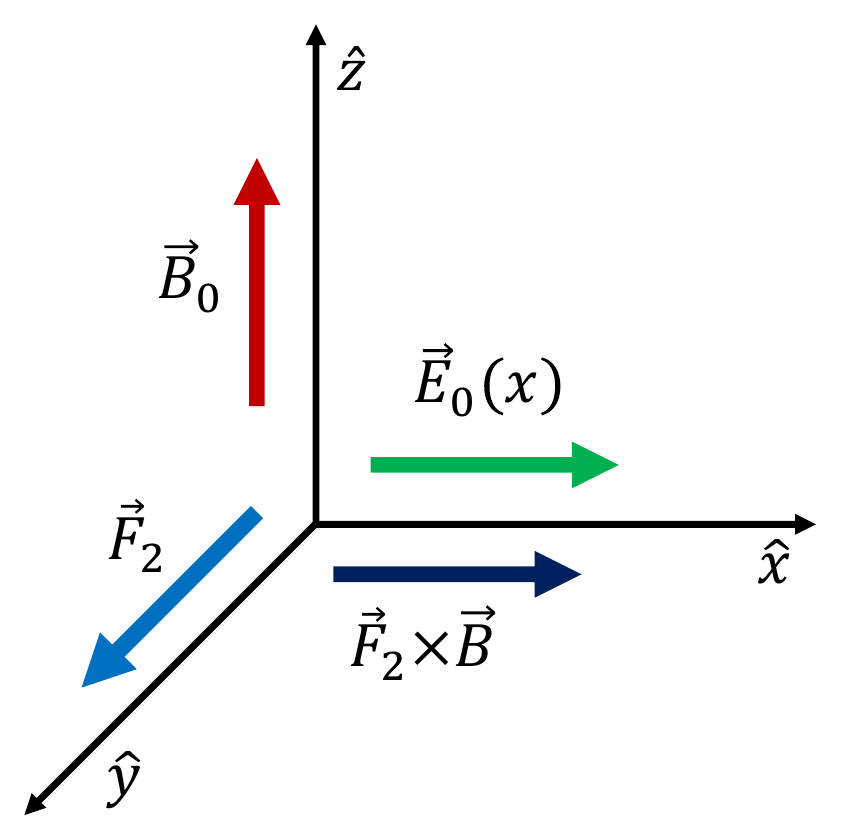}
	\caption{\label{fig:fig13a} Geometry of the ponderomotive force and nonlinear particle drift.}
\end{figure}

\begin{figure}
	\includegraphics[width=\textwidth]{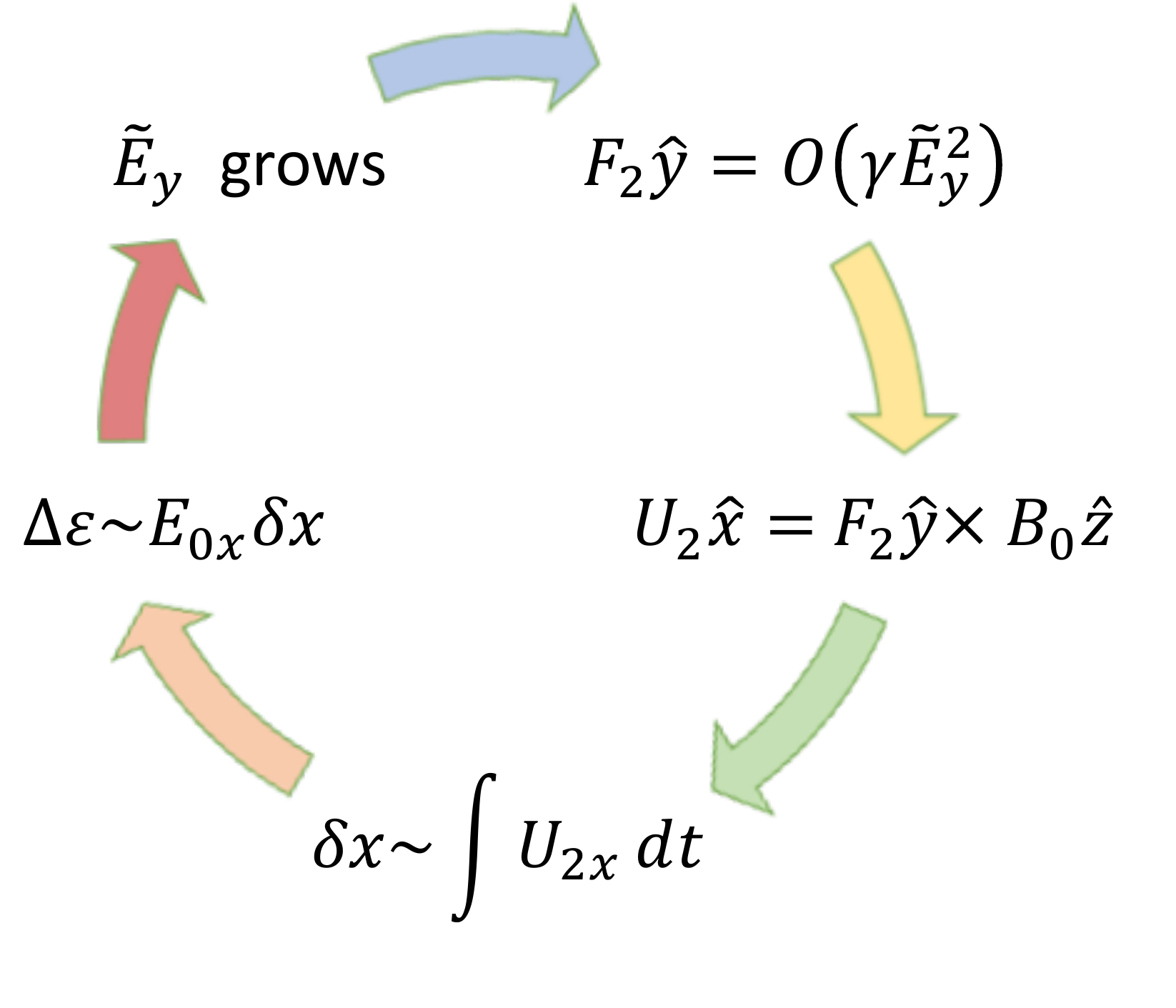}
	\caption{\label{fig:13f} Positive feedback loop for IEDDI instability.}
\end{figure}

A nonlinear description of the wave–particle interaction responsible for IEDDI was given by Palmadesso \textit{et al.} \cite{Palmadesso:1986}. It was shown that the fluctuating wave electric field $\mathbf{E}_{1}$ leads to an average secular (ponderomotive) force $F_{2y}\sim O(\gamma E_{1y}^2)$ in the $y$-direction (see Fig. \ref{fig:fig13a}). This leads to a $\mathbf{F}_{2y}\times\mathbf{B}$ drift in the $x$-direction, which in the small gyroradius limit is $u_{2x}\propto -\gamma(\omega-k_{y}V_{E})^{-3}E_{1y}^2$, leading to a shift in the particle position in the $x$-direction given by $\delta x=\int u_{2x}d\tau\sim E_{1y}^2$.  As there is dc electric field $E_{0}(x)$ in the $x$ direction there is a potential energy gain given by $E_{0}(x)\delta x$ if $(\omega_{r}-k_{y}V_{E})<0$.  Since the particle motion is perpendicular to $F_{2y}$ there can be no net increase in the particle energy. Thus, the energy gained by the particles by falling in the potential of the dc electric field in the $x$-direction is lost to the waves in the $y$-direction. Consequently, $E_{1y}$ grows and $F_{2y}$ is further enhanced, which closes a positive feedback loop as shown in Fig. (\ref{fig:13f}).  This leads to the instability in a way similar to a magnetron.  

The second order ion drift in the direction of the electric field constitutes a polarization current that reduces the magnitude of the external electric field.  Such polarization current was observed in the Particle-in-Cell (PIC) simulation of the IEDDI by Nishikawa \cite{Nishikawa:1988}.

%
\subsubsection{Intermediate frequency limit:  partially magnetized ions and fully magnetized electrons}
\label{subsec:intermediate_f}

As compression increases the self-consistent electric field becomes more intense and narrower in scale size.  In the intermediate compression regime the scale size is narrower than an ion gyroradius but larger than an electron gyroradius, \textit{i.e.}, $\rho_{i}>L>\rho_{e}$.  As discussed in Sec. \ref{subsec:particle_orbits}, the ions in this regime do not experience the electric field over their entire gyro-orbit.  Consequently, the ions experience a lower gyro-averaged electric field than the electrons.  For sufficiently localized electric field the ions experience vanishingly small electric field.  In this regime for intermediate frequencies and short wavelengths, \textit{i.e.}, $\Omega_{i}<\omega<\Omega_{e}$ and $k_{y}\rho_{i}>1>k_{y}\rho_{e}$, the ions behave as an unmagnetized plasma species but the electrons are magnetized.  The cyclotron harmonics for the ions can be integrated to rigorously show their unmagnetized character \cite{Ganguli:1988eb}.  Since the wave frequency is much smaller than the electron cyclotron frequency it will suffice to consider only the $n=0$ cyclotron harmonic term for the electrons.  Also, for simplicity, we assume that the velocity shear that the electrons experience is small enough so that we may use $\eta=1$ for the electrons.  The ions do not experience a Doppler shift so the phase speed of the waves can remain larger than the thermal velocity, which allows the assumption of fluid ions in which the density perturbation is given by \cite{Ganguli:1988eb},
\begin{equation}
	\label{eq:ni_intermed}
	n_{1i}(x)=\frac{1}{4\pi q_{i}}
	\frac{\omega_{pi}^2}{\omega^2}\left(
	k_y^2 + k_{\|}^2 - \frac{d^2}{dx^2}\right)\phi(x)
\end{equation}
However the electrons experience a spatially varying Doppler shift. The phase speed of the waves can become comparable to the electron thermal velocity at some locations.  So for generality we use the kinetic response for the electron, which leads to their density perturbation 
\begin{multline}
	\label{eq:ne_intermed}
n_{1e}(x) = -\frac{\omega_{pe}^2}{4\pi v_{te}^2 q_{e}}
\left[
-\left(\frac{\omega_{1}+\omega_{2e}-\omega^*}{\sqrt{2}|k_{\|}|v_{te}}
	\right) 
	Z\left(
\frac{\omega_{1}-\omega_{2e}}{\sqrt{2}|k_{\|}|v_{te}}
\right)\frac{d\Gamma_{n}(b_e)}{db}\rho_{e}^2 \frac{d^2}{dx^2}
\right.\\\left.
+1+\left(\frac{\omega_{1}+\omega_{2e}-\omega^*}{\sqrt{2}|k_{\|}|v_{te}}
	\right) 
	Z\left(
\frac{\omega_{1}-\omega_{2e}}{\sqrt{2}|k_{\|}|v_{te}}
\right)\Gamma_{0}(b_e)
\right]\phi(x).	
\end{multline}
Combining Eqs. \ref{eq:ni_intermed} and \ref{eq:ne_intermed} with the Poisson equation we get the general eigenvalue condition of the EIH instability in the kinetic limit that includes the electron diamagnetic drift.
\begin{multline}
\label{eq:intermed_gen}
0=\frac{d^2	\phi}{dx^2}+
\\
\left\{\frac{
\left(1-\frac{\omega_{pi}^2}{\omega^2}\right)
(k_y^2+k_{\|}^2)
-\frac{\omega_{pe}^2}{v_{te}^2}\left[
1+\left(\frac{\omega_{1}+\omega_{2e}-\omega^*}{\sqrt{2}|k_{\|}|v_{te}}
	\right) 
	Z\left(
\frac{\omega_{1}-\omega_{2e}}{\sqrt{2}|k_{\|}|v_{te}}
\right)\Gamma_{0}(b_e)\right]
}{
1-\frac{\omega_{pi}^2}{\omega^2} + \frac{\omega_{pe}^2}{\Omega_{e}^2}
\left(\frac{\omega_{1}+\omega_{2e}-\omega^*}{\sqrt{2}|k_{\|}|v_{te}}
	\right) 
	Z\left(
\frac{\omega_{1}-\omega_{2e}}{\sqrt{2}|k_{\|}|v_{te}}
\right)\frac{d\Gamma_{n}(b_e)}{db}
}
\right\}\phi(x)
\end{multline}
In the long wavelength ($k_{\|}\rightarrow0$, $k_y\rightarrow0$) limit Eq. \ref{eq:intermed_gen} reduces to,
\begin{equation}
	\label{eq:intermed_eig}
	\frac{d^2\phi}{dx^2} 
	- (k_y^2+k_{\|}^2)\phi +
	\left(\frac{\omega_{pe}^2}{\Omega_{e}^2+\omega_{pe}^2}\right)
	\frac{\omega^2}{(\omega^2-\omega_{LH}^2)}
	\left[
	\frac{k_y(V_{E}''-\Omega/L_n)}{\omega_{1}} 
	-\frac{k_{\|}^2\Omega_{e}^2}{\omega_{1}^2}
	\right]\phi(x)=0
\end{equation}
Eq. \ref{eq:intermed_eig} includes the modified two-stream instability \cite{McBride:1972}, which was not in Fletcher \textit{et al.} \cite{Fletcher:2019kq} since $k_{\|}=0$ was assumed.  The modified two-stream instability dispersion relation can be recovered if the electric field curvature and the density gradient are neglected in Eq. \ref{eq:intermed_eig}.  Including the density gradient Eq. \ref{eq:intermed_eig} represents the lower-hybrid drift instability \cite{Krall:1971}.  The lower-hybrid drift modes depend upon the density gradient and hence their growth relaxes the density gradient.  If the density gradient is ignored but $V_{E}''\not=0$ then Eq. \ref{eq:intermed_eig} reduces to the eigenvalue condition for the electron-ion hybrid (EIH) instability \cite{Ganguli:1988eb} where the free energy is obtained from the sheared electron flow through fast time averaging by the perturbations \cite{Ganguli:1988eb} similar to the KH modes discussed earlier. The growth of the EIH waves relaxes the velocity shear.  

From Eq. \ref{eq:intermed_eig} it is clear that the intermediate frequency waves depend on a double resonance $\omega\simeq \omega_{LH}\simeq k_y V_E (x)$. The spatial variation of $k_y V_E (x)$ is particularly important because at some point in $x$ the argument of the $Z$ function in Eq. \ref{eq:intermed_gen} can become of the order of unity so that Landau damping cannot be ignored unless $k_{\|}$ is sufficiently small.  Hence, the limit $k_{\|}\rightarrow 0$ where Landau damping is eliminated and both the EIH and LHD instability growth are maximized is used to determine the most likely modes that will arise in the intermediate frequency range in compressed plasma.  The modified two stream instability, whose modification due to shear flow has not been studied sufficiently in the literature, requires $k_{\|}\neq 0$.  It is included in the last term in Eq. \ref{eq:intermed_eig} but its contribution is minimal because for $k_{\|}\rightarrow 0$ the growth rate of the intermediate frequency waves is largest.

For dense plasmas of interest  $\omega_{pe}>\Omega_{e}$, so $\omega_{LH}\simeq\sqrt{(\Omega_i \Omega_e )}$ and the first factor in the third term of Eq. \ref{eq:intermed_eig} is about one.  In the $k_{\|}\rightarrow 0$ limit the eigenmode equation, Eq. \ref{eq:intermed_eig}, in dimensionless form becomes,
\begin{equation}
	\label{eq:intermed_eig_dim}
	\left\{
	\frac{d^2}{d\bar{x}^2} - \bar{k}^2
	+ \left(\frac{\bar{\omega}^2}{\bar{\omega}^2-1}\right)
	\frac{ 
	\bar{k} (\alpha_s \bar{V}''_{E}(\bar{x}) - \frac{L}{L_n})
	}{
	\bar{\omega} - \bar{k}\alpha_s \bar{V}_{E}(\bar{x})}
	\right\}\phi(\bar{x})=0,
\end{equation}
where $\bar{x}=x/L$, $\bar{\omega}=\omega/\omega_{LH}$, $\bar{k}=k_{y}L$, $\bar{V}_{E}=V_{E}/V_{0}$, $V_{0}=cE_{0}/B_{0}$, $\alpha_s=V_{0}/L\Omega_{e}$ is the shear parameter.

\begin{figure}
\includegraphics[width=\textwidth]{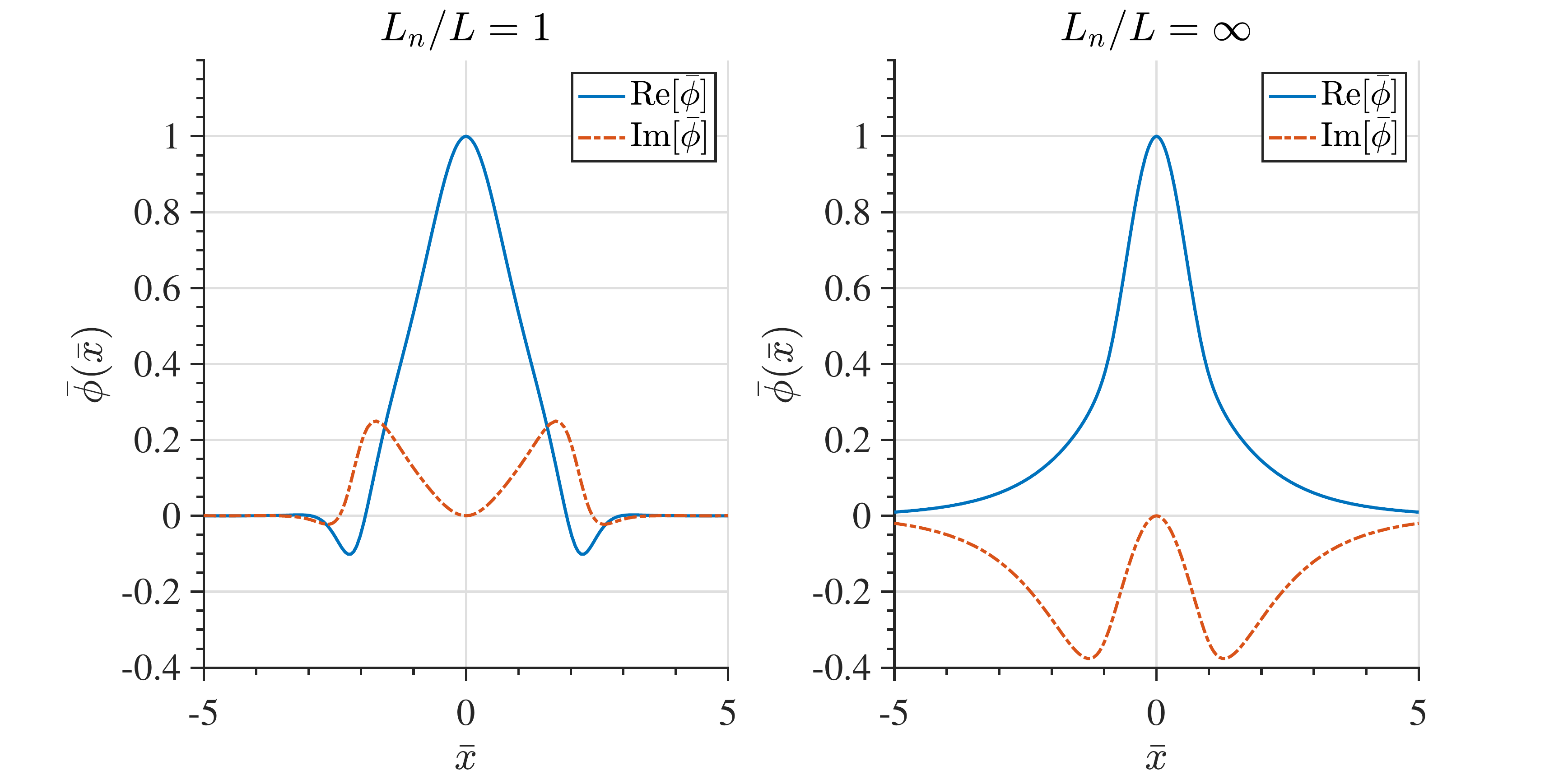}
\caption{\label{fig:EIH_Eigenfunctions}Eigenfunctions for $L_n/L=1$ and $L_n/L=\infty$ and $\alpha=1$ with $k_y L$ chosen to maximize the growth rate. }
\end{figure}

\begin{figure}
\includegraphics[width=\textwidth]{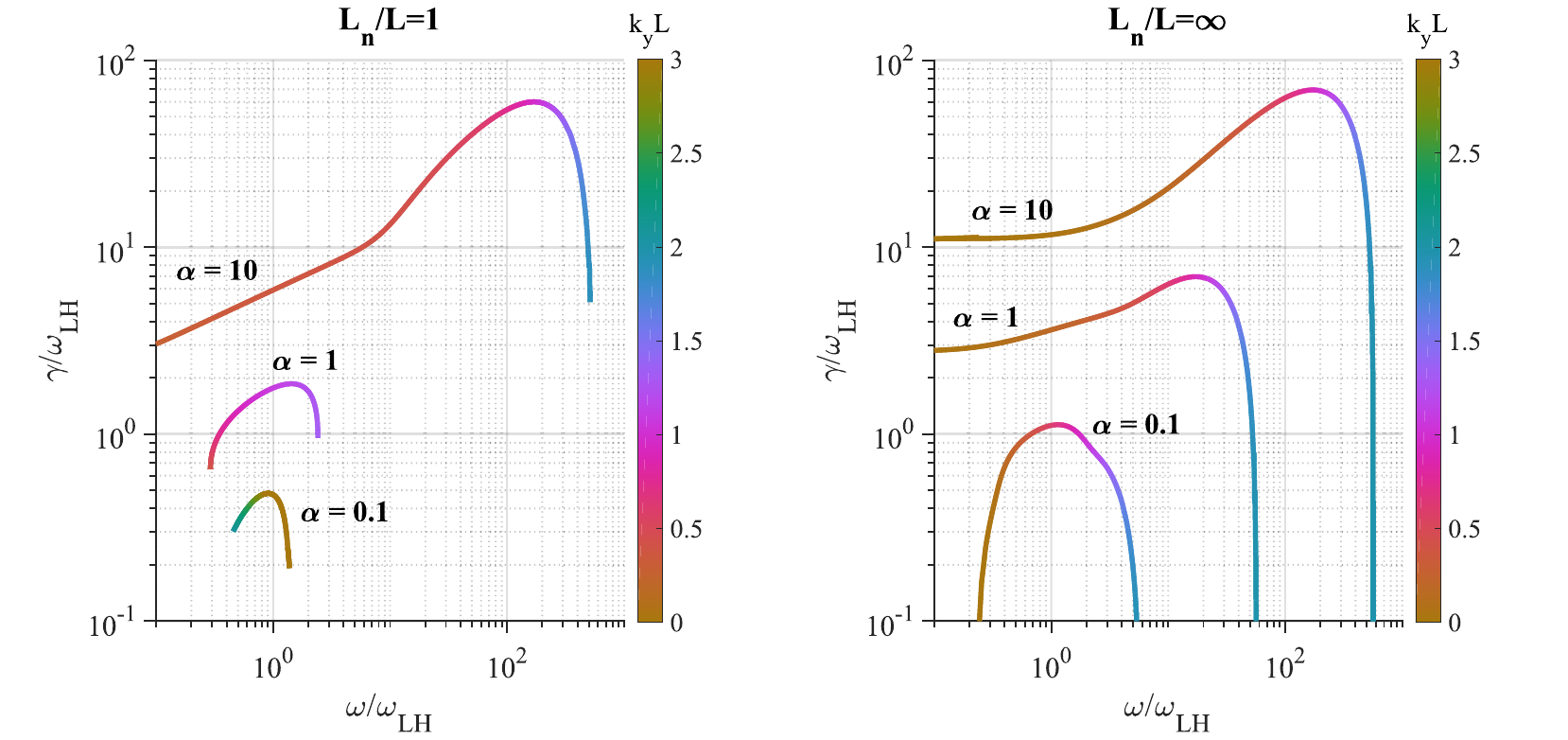}
\caption{\label{fig:EIH_Eigenvalues}Linear growth rate as a function of real frequency, colored by associated $k_y L$ value. On the left is the fluid case, where the electric field balances the density gradient. On the right is the limit of the kinetic case. Reproduced from Figure 14 of Fletcher \textit{et~al.} \cite{Fletcher:2019kq}. }
\end{figure}

Fig. \ref{fig:EIH_Eigenfunctions} shows two solutions to Eq. \ref{eq:intermed_eig_dim}  (\textit{i.e.} the real and imaginary parts of the eigenfunctions).  Fig. \ref{fig:EIH_Eigenvalues} is a plot of the linear growth rate and the real frequency obtained from solving the eigenvalue condition given in Eq \ref{eq:intermed_eig_dim}. The eigenfunctions and eigenvalues were found via a shooting method in which the large $\bar{x}$ solution goes to zero at infinity.  The density profile is $n(x)=n_0 \textrm{tanh}(x/L_n)$ and the electron flow profile by $E(x)=E_0 \textrm{sech}^2 (x/L)$ are chosen to match the self-consistent low $\beta$ \cite{Ganguli:2018vf} DF discussed in Sec. \ref{subsec:df_eq} and its parameters are based on the MMS observations.  As the shear parameter is increased, implying higher compression, the growth rate increases. The real frequency is around the lower-hybrid frequency while Doppler shifting broadens the frequency spectrum. The bandwidth increases with shear parameter.

In the two cases shown, the growth peaks for $k_y L\sim 1$. The wavelength is much longer than $\rho_e$ since $L\gg \rho_e$. As $L_n/L$ is reduced, the wavelengths become shorter and in the limit of uniform electric field ($L\rightarrow\infty$) it is well known that $k_{y}\rho_{e}\sim 1$ \cite{Krall:1971}.  Note that these discrete eigenmodes in $x$ are still continuously dependent on $k_y$.   The parallel wave vector, $k_{\|}$, is assumed to be zero.  In Sec. \ref{subsec:nl_ifw_in_sf} the nonlinear evolution of this equilibrium condition and its observable signatures are studied by PIC simulation and show that the spectral bandwidth becomes even broader nonlinearly as lower frequency waves are naturally triggered with increasing $L$.  

Since Eq. \ref{eq:intermed_eig_dim} contains both density and electric field gradients, an interesting question is which one of these is responsible for the waves?

\begin{figure}
\includegraphics[width=\textwidth]{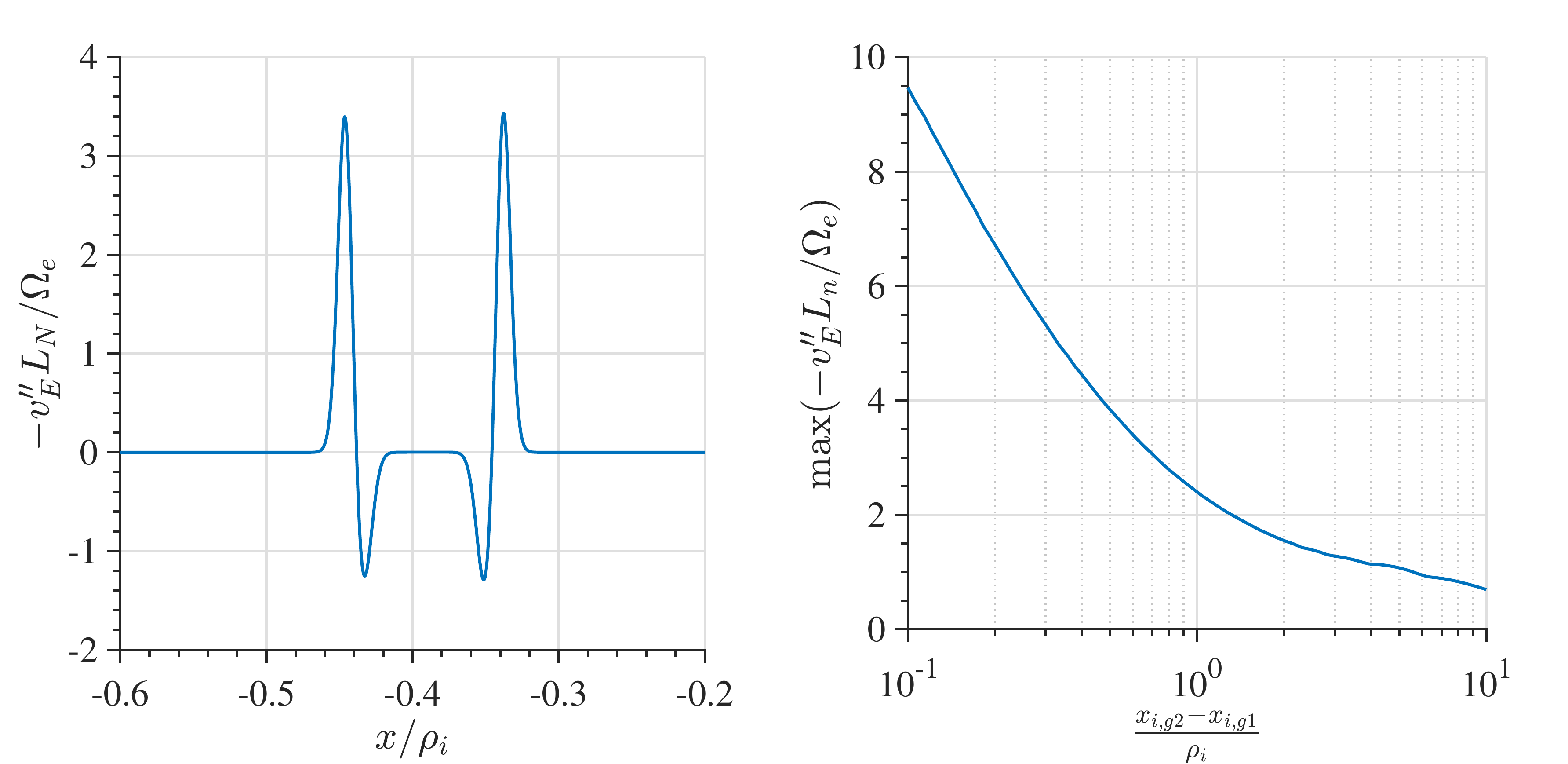}
\caption{\label{fig:EIH_DriveTerms} The ratio of the two driving terms in Equation \ref{eq:intermed_eig_dim} as a function of x (left) and as a function of layer width (right).  Reproduced from Figure 15 of Fletcher \textit{et~al.} \cite{Fletcher:2019kq}.}
\end{figure}

To answer this question Fig. \ref{fig:EIH_DriveTerms} compares the relative strength of the LHD and the EIH terms in Eq. \ref{eq:intermed_eig_dim}. The left plot shows the ratio of these EIH to LHD instability source terms for the low beta MMS parameters \cite{Ganguli:2018vf,Fletcher:2019kq}, which can be reproduced by our electrostatic equilibrium model discussed in Sec. 2.1 with $R_i=R_e=1$, $S_i=S_e=0.793$, $x_{g1e}=-0.438\rho_i$, $x_{g2e}=-0.346\rho_i$, $x_{g1i}=-0.0390\rho_i$, $x_{g2i}=0.850\rho_i$, $n_0=0.355$ cm$^{-3}$, $T_{e0}=654.62$ eV, $T_i/T_e=6.714$, and $B_0=12.55$ nT.  It shows that even for weak compression, as in the case considered, the EIH term is three times as large as the LHD term. In the stronger compression high beta case (Fig. 26), the EIH term is more than an order or magnitude larger. The right plot shows the maximum of the ratio of EIH/LHD terms as the compression is increased. This plot was made by using the same parameters as the low $\beta$ case and compressing and expanding the layer via choice of $x_{g1\alpha}$ and $x_{g2\alpha}$. Clearly, the EIH instability dominates over the LHD instability as long as the scale size of the density gradient is comparable to ion gyroradius or less.

Magnetic field gradients result in a stronger EIH instability \cite{Romero:1994}, but a weaker LHD instability \cite{Davidson:1977}. In Sec. 2.1 we showed that a gradient in the temperature can also develop, which can make the pressure gradient in the layer (and hence the diamagnetic drift) weaker, but not significantly affect the ambipolar electric field. This also favors the EIH instability over the LHD instability. Thus, the EIH mechanism will dominate wave generation and hence the nonlinear evolution in a compressed plasma system in the intermediate frequency range.

\begin{figure}
	\includegraphics[width=\textwidth]{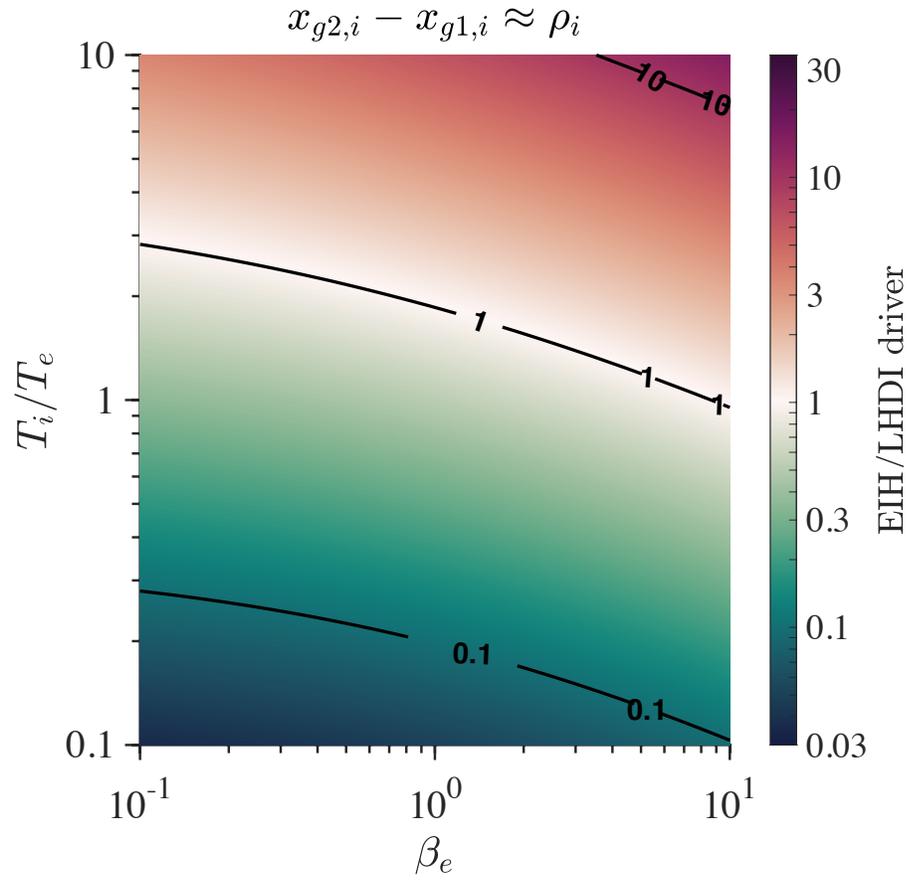}
	\caption{\label{fig:termratio_beta}EIH to LHD instability growth term ratio vs temperature ratio and $\beta_e$ for the equilibrium model where the width of the transition layer is equal to the ion thermal gryoradius.  Red indicates that the EIH drive term dominates.}
\end{figure}

In general, the self-consistent generation of an ambipolar electric field is unavoidable in warm plasmas with a density gradient scale size comparable to or less than the ion gyroradius. This raises an interesting question: How ubiquitous in nature is the classical LHD instability?  To examine this we generalize the Fig. 25 results to include electromagnetic effects in the equilibrium condition and compare the relative strengths of the two drivers of the electrostatic instability in Eq. 66: 1) $\alpha_s \bar{V}_E''(\bar{x})$, which is the shear-driven EIH instability, and 2) $-L/L_n$, which is the density gradient-driven LHD instability.  By using the electromagnetic equilibrium model of section 2.2 we can investigate the magnitude of these two driving terms.  In general we find that for $T_i/T_e>1$, which is typical in space plasmas (particularly in the magnetotail), we find that the EIH instability drive dominates.  In the opposite limit $T_e/T_i<1$, which is typical in laboratory plasmas, LHD tends to dominate.  Fig. \ref{fig:termratio_beta} shows the same ratio of terms for different values of $\beta_e$.  As $\beta_e$ increases, the EIH term also becomes more dominant because the ambipolar electric field intensifies with $\beta$ as shown in Figure 10.  For typical conditions in the magnetotail (high $\beta_e$ and $T_i/T_e$), the EIH term is greater than the LHD term.  The dominance of the EIH over LHD wave becomes further evident in the nonlinear analysis in Sec. \ref{sec:nonlinear}.

%
%
%

\subsubsection{Higher frequency transverse flow shear driven modes}
\label{subsec:hf_due_to_transverse_shear}

As compression increases further so that $\rho_{i}\gg L\geq \rho_{e}$ then even higher frequency modes with $\omega_{1}\leq\Omega_{e}$ are possible.  For these modes the ions do not play any important role other than charge neutralizing background and they may be ignored.  The dispersion relations will become similar to the KH and IEDDI discussed in Sec. \ref{subsec:linear_low_freq_limit} but for the electron species.  By symmetry for $\omega_{1}<\Omega_{e}$ the electron KH modes can be recovered and for $\omega_{1}\sim n\Omega_{e}$ the electron IEDDI can be recovered.

%
%
%

\subsubsection{Stability of the Vlasov Equilibrium Including $V_{\|}(x)$}
\label{subsec:stability_due_to_par_shear}

In sections \ref{subsec:linear_low_freq_limit} through \ref{subsec:hf_due_to_transverse_shear} we discussed the waves that are driven by the shear in transverse flows.  However, as discussed in Sec. \ref{sec:vpequi_curv}, large-scale magnetic field curvature can lead to a potential difference along the magnetic field.   This originates because the global compression is strongest at a particular point and decreases away from it and hence the transverse electrostatic potential generated by compression also decreases proportionately away from this point along the magnetic field.  The potential difference along the field line results in a magnetic field aligned electric field as sketched in Fig. \ref{fig:fig11}.  Non-thermal particles can be accelerated by the parallel electric field to form a beam along the magnetic field direction, with a transverse spatial gradient, \textit{i.e.}, $dV_{\|}/dx$.  The gradient in the parallel flow is also a source for free energy.  This has been established both theoretically \cite{DAngelo1965,Lakhina:1987,Gavrishchaka:1998,Gavrishchaka:2000,Ganguli:2002} and through laboratory experiments \cite{DAngelo1965,Agrimson:2001,Agrimson:2002,Teodorescu:2002a,Teodorescu:2002b}.  Like its transverse counterpart the spatial gradient in the parallel flow can also support a hierarchy of oscillations.   Below we summarize the physical origin of these waves.  

\begin{figure}
	\includegraphics{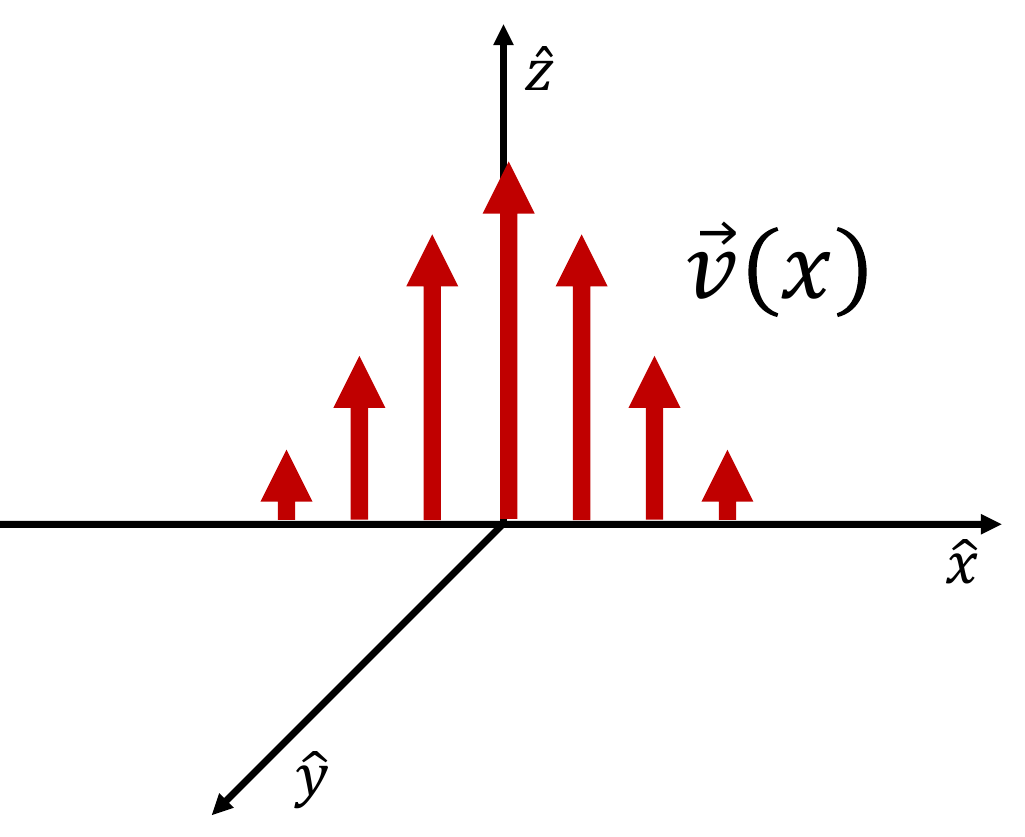}
	\caption{\label{fig:parallel_shear_geom} Geometry for parallel shear flow.}
\end{figure}

Consider a uniform magnetic field in the $z$ direction with a transverse gradient in the flow along the magnetic field ($dV_{\|}/dx$).  The background plasma condition is sketched in Fig. \ref{fig:parallel_shear_geom}.  Unlike the transverse flow shear, the parallel flow shear does not affect the particle gyro-motion, which simplifies the analysis considerably.  For simplicity consider a locally linear flow, \textit{i.e.}, $V_{\|,\alpha}(x)=V_{\|,\alpha}+(dV_{\|,\alpha}/dx)x$ where $V_{\|,\alpha}$ and $dV_{\|,\alpha}/dx$ are constants and $\alpha$ represents the species and let $dV_{\|,e}/dx=dV_{\|,i}/dx\equiv dV_{\|}/dx$.  Transforming to the ion frame (\textit{i.e.}, $V_{\|,i}=0$) so that $V_{\|,e}\equiv V_{\|}$ represents the relative electron-ion parallel drift.  Although a nonlocal eigenvalue condition is desirable, a local limit exists for the parallel flow shear driven modes.  

First consider the general dispersion relation for waves with $\omega\ll\Omega_{e}$ so that only the $n = 0$ cyclotron harmonic of the electrons is sufficient.  For this condition the dispersion relation is \cite{Ganguli:2002},
\begin{equation}
	\label{eq:62t}
	1 + \sum_{n} \Gamma_{n}(b) F_{ni} + \tau(1+F_{0e})=0,
\end{equation}
\begin{multline}
	F_{ni} = 
	\left(\frac{\omega}{\sqrt{2}|k_{\|}|v_{ti}}\right)
	Z\left(\frac{\omega-n\Omega_{i}}{\sqrt{2}|k_{\|}|v_{ti}}\right)
	\\
	-\frac{k_{y}}{k_{\|}}\frac{1}{\Omega_{i}}\frac{dV_{\|}}{dx}
	\left[
	1+ \left(\frac{\omega-n\Omega_{i}}{\sqrt{2}|k_{\|}|v_{ti}}\right)
	Z\left(\frac{\omega-n\Omega_{i}}{\sqrt{2}|k_{\|}|v_{ti}}\right)
	\right]
\end{multline}
\begin{multline}
	F_{0e}=
		\left(\frac{\omega-k_{\|}V_{\|}}{\sqrt{2}|k_{\|}|v_{te}}\right)
	Z\left(\frac{\omega-k_{\|}V_{\|} }{\sqrt{2}|k_{\|}|v_{te}}\right)
	\\
	+\frac{k_{y}}{k_{\|}}\frac{1}{\mu\Omega_{i}}\frac{dV_{\|}}{dx}
	\left[
	1+ \left(\frac{\omega-k_{\|}V_{\|}}{\sqrt{2}|k_{\|}|v_{te}}\right)
	Z\left(\frac{\omega-k_{\|}V_{\|}}{\sqrt{2}|k_{\|}|v_{te}}\right)
	\right]
\end{multline}
In the absence of shear (i.e., $dV_{\|}/dx\equiv V_{\|}'=0$), the dispersion relation reduces to the case of a homogeneous flow as discussed by Drummond and Rosenbluth \cite{Drummond:1962} and applied to space plasmas by Kindel and Kennel \cite{Kindel:1971}. 

%
%
%

\subsubsubsection{Low Frequency Limit: Sub-Cyclotron Frequency Waves.}
We first discusss low (sub-cyclotron) frequency ion-acoustic waves for which only the $n=0$ cyclotron harmonic term for the ions is sufficient.  For long wavelength, \textit{i.e.} $b\ll 1$, $\Gamma_{0}(b)\sim 1$, and Eq. \ref{eq:62t} simplifies to,
\begin{equation}
\label{eq:65t}
	\sigma^2 + \tau\hat{\sigma}^2 +\sigma^2 \xi_{0}Z(\xi_{0}) 
	+ \tau \hat{\sigma}^2\xi_{e}Z(\xi_e)=0
\end{equation}
where $\xi_{0}=\omega/(\sqrt{2}|k_{\|}|v_{ti})$, $\xi_{e}=(\omega-k_{\|}V_{\|})/(\sqrt{2}|k_{\|}|v_{te})$, $\sigma^2=(1-k_yV_{\|}'/k_{\|}\Omega_{i})$, $\hat{\sigma}^2=1+k_yV_{\|}'/(k_{\|}\Omega_{i}\mu)$.  Assuming the ions to be fluid ($\xi_{0}\gg1$) and electrons to be Boltzmann ($\xi_{e}\ll1$) and equating the real part of Eq. \ref{eq:65t} to zero we get,
\begin{equation}
	\label{eq:66t}
	\omega=k_{z}c_{s}\sigma/\hat{\sigma}\sim k_{z}c_{s}\sigma 
\end{equation}
where $\hat{\sigma}\sim 1$ is used since $\mu\gg1$.  In the absence of shear ($dV_{\|}/dx=0$, \textit{i.e.} $\sigma^2=1$) the classical ion acoustic limit is recovered.  If $\sigma^2<0$ then Eq. \ref{eq:66t} reduces to the dispersion relation for the D'Angelo instability \cite{DAngelo1965} for which the real frequency $\omega_{r}=0$ in the drifting ion frame.  The D'Angelo instability has been the subject of numerous space and laboratory applications  \cite{Catto:1973,Huba:1981,Gary:1981}.  
	
\begin{figure}
	\includegraphics[width=\textwidth]{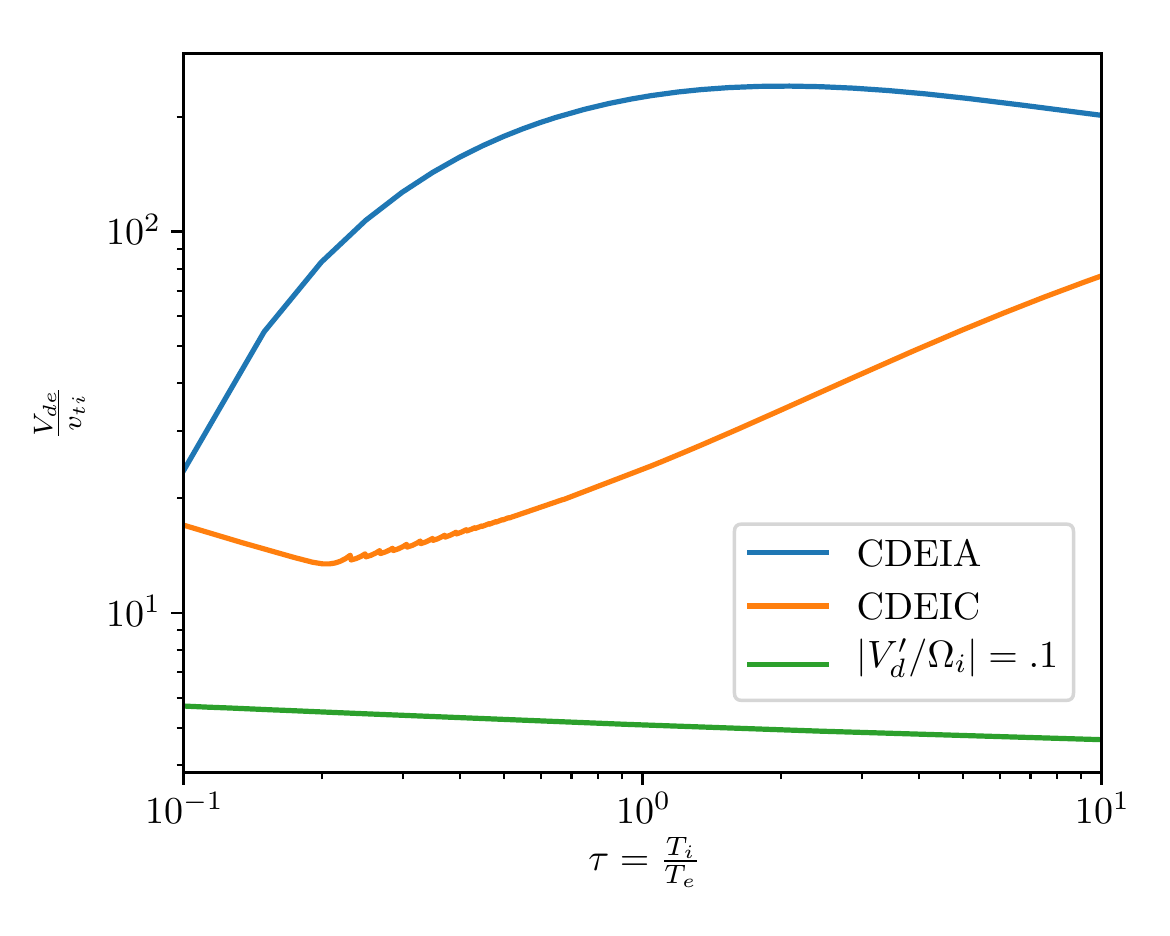}
	\caption{\label{fig:criticaldrift} Critical Drift vs temperature ratio.  Blue curve is for the classical current driven electrostatic ion acoustic mode (CDEIA).  Orange curve is for the shear modified ion acoustic-instability.}
\end{figure}

The $\sigma^2>1$ regime was addressed by Gavrishchaka \textit{et al.} \cite{Gavrishchaka:1998}.  In this regime Eq. \ref{eq:65t} indicates that it is possible to obtain a shear modified ion-acoustic (SMIA) wave with interesting properties.  Eq. \ref{eq:66t} indicates that shear can increase the parallel phase speed ($\omega_{r}/k_{\|}$) of the ion acoustic mode by the factor $\sigma$.  For a large enough $\sigma$ the phase speed can be sufficiently increased so that ion Landau damping is reduced or eliminated.  Consequently, a much lower threshold for the ion acoustic mode can be realized even for $T_{i}>T_{e}$.  The growth rate expression for the SMIA instability is given by Gavrishchaka \textit{et al.}, \cite{Gavrishchaka:1998},
\begin{equation}
\label{eq:67t}
	\frac{\gamma}{|k_{\|}|v_{ti}}
	= \sqrt{\frac{\pi}{8}}
	\frac{\sigma^2}{\tau^2}\left[
	\frac{\tau^{3/2}}{\mu^{1/2}}
	\left(\frac{V_{\|}}{\sigma c_s}-1\right)
	-\sigma^2\exp(-\sigma^2/2\tau)\right]
\end{equation}
The classical ion-acoustic wave growth rate is recovered for $\sigma^2=1$.   From Eq. \ref{eq:67t} it is clear that $\sigma$ can rapidly lower the ion Landau damping as seen from the exponential dependence of the second term in the bracket.  The critical drift is obtained from Eq. \ref{eq:67t} by setting the growth rate to zero and minimizing over the propagation angle ($k_{\|}/k_y$) as is plotted in Fig. \ref{fig:criticaldrift} with $dV_{\|}/dx=0.1\Omega_{i}$.  It is found that even a small shear can reduce the critical drift for the ion acoustic instability by orders of magnitude and put it below that of the classical ion cyclotron wave \cite{Drummond:1962} for a wide range of $\tau=T_{i}/T_{e}$ but the shear modified ion acoustics waves propagate more obliquely than their classical counterpart.  This is a major departure from the conclusion of Kindel and Kennel \cite{Kindel:1971}; that among the waves driven by a field aligned current in the earth’s ionosphere the current driven ion cyclotron instability has the lowest threshold.  Kindel and Kennel's conclusion had extensively guided the interpretation of \textit{in-situ} data for a long time until Gavrishchaka \textit{et~al.} reexamined the data \cite{Gavrishchaka:1999} with shear modified instabilities in mind.

%
%
%

\subsubsubsection{Low Frequency Limit: Ion Cyclotron Frequency Waves.}

To study the ion cyclotron frequency regime we return to Eq. \ref{eq:62t} but relax the constraints of low frequency and long wavelength used to study the shear modified ion acoustic waves.  We first examine how a gradient in the parallel plasma flow affects the threshold condition for ion cyclotron waves by analyzing the expression for critical relative drift for the ion cyclotron waves in small and large shear limits.   For the marginal stability condition ($\gamma=0$) the imaginary part of the dispersion relation, Eq. \ref{eq:62t}, is set equal to zero, \textit{i.e.},
\begin{equation}
\label{eq:68t}
	\sum_{n}\Gamma_{n}
	\left[
	\left(\xi_{0} - \frac{k_{y}V_{\|}'}{k_{\|}\Omega_{i}}\xi_{n}
	\right) \textrm{Im}Z(\xi_{n})
	\right]
	+
	\tau\left(1+ \frac{k_{y}V_{\|}'}{k_{\|}\Omega_{i}\mu}
	\right)\xi_{e} \textrm{Im}Z(\xi_{e})=0
\end{equation}
where $\xi_{n}=(\omega-n\Omega_{i})/(\sqrt{2}|k_{\|}|v_{ti})$.

Dividing Eq. \ref{eq:68t} throughout by $\xi_{0}$ and considering the electrons to be adiabatic, \textit{i.e.}, $\xi_{e}\ll 1$, we get,
\begin{multline}
\label{eq:69t}
\sum_{n}\Gamma_{n}\left[
\left(1-\frac{k_{y}V_{\|}'}{k_{\|}\Omega_{i}}
\left(1-\frac{n\Omega_i}{\omega_r}\right)
\right)
\exp\left\{
-\left(\frac{\omega_r-n\Omega_{i}}{\sqrt{2}|k_{\|}|v_{ti}}\right)^2
\right\}
\right]
\\
+
\frac{\tau^{3/2}}{\mu^{1/2}}\left(
1+\frac{k_y V_{\|}'}{k_{\|}\Omega_{i}\mu}\right)
\left(\frac{\omega_{r}-k_{\|}V_{\|c}}{\omega_r}\right)=0
\end{multline}
Under ordinary conditions $(k_y/k_{\|})(dV_{\|}/dx)/\Omega_{i}\ll\mu$, which implies that the shear in the electron flow is not as critical as it is in the ion flow and can be ignored.   Since only a specific resonant cyclotron harmonic term dominates, Eq. \ref{eq:69t} can be simplified by considering only that resonant term in the summation to obtain an expression for the critical relative drift,
\begin{equation}
\label{eq:70t}
	V_{\|c}=\frac{\omega_{r}}{k_{\|}}\left[
	1 + \Gamma_{n}(b)\frac{\mu^{1/2}}{\tau^{3/2}}
	\left\{1 - 
	\frac{k_{y}V_{\|}'}{k_{\|}\Omega_{i}}\left(1-\frac{n\Omega_{i}}{\omega_{r}}\right)
	\right\}
	\exp\left(
	-\frac{(\omega_r-n\Omega_{i})^2}{2k_{\|}^2 v_{ti}^2}
	\right)
	\right]
\end{equation}
For no shear, $V_{\|}'=0$, the critical drift reduces to,
\begin{equation}
\label{eq:71t}
	V_{\|c}=\frac{\omega_{r}}{k_{\|}}\left[
	1 + \Gamma_{n}(b)\frac{\mu^{1/2}}{\tau^{3/2}}
	\exp\left(
	-\frac{(\omega_r-n\Omega_{i})^2}{2k_{\|}^2 v_{ti}^2}
	\right)
	\right]	
\end{equation}
This is the critical drift for the homogeneous current driven ion cyclotron instability (CDICI) \cite{Drummond:1962}.  Since the relative sign between the two terms within the bracket is positive and each term is positive definite, the critical drift is always greater than the wave phase speed and increases for higher harmonics since $\omega_{r}\sim n \Omega_{i}$.

From Eq. \ref{eq:70t} it may appear that for small but non-negligible and positive values of $(k_{y}V_{\|}'/k_{\|}\Omega_{i})(1-n\Omega_{i}/\omega_{r})$ there can be a substantial reduction in the critical drift for the current driven ion cyclotron instability because of reduction in the ion cyclotron damping.  However, this is not possible and can be understood by rewriting Eq. \ref{eq:70t} as,
\begin{equation}
	\label{eq:72t}
	\frac{V_{\|c}}{V^{0}_{\|c}}
	=
	1 - \left(
	1 - \frac{(\omega_r/k_{\|})}{V_{\|c}^{0}} \right)
	\left(\frac{k_{y}V_{\|}'}{k_{\|}\Omega_{i}}\right)
	\left(1-\frac{n\Omega_{i}}{\omega_{r}}\right)
\end{equation}
where the second term represents the correction to the critical drift for the current driven ion cyclotron instability due to shear.  A necessary condition for the CDICI is that $V_{\|}>\omega_{r}/k_{\|}$.  For a given magnitude of $|dV_{\|}/dx|/\Omega_{i}\ll1$, it is clear from Eq. \ref{eq:72t} that the shear correction is small unless the ratio $k_{y}/k_{\|}$ can be made large.  However, as $k_y$ increases, the real frequency of the wave approaches harmonics of the ion cyclotron frequency and consequently $(1-n\Omega_{i}/\omega_{r})$ becomes small which makes the shear correction small.  Alternately, when $k_z$ decreases the wave phase speed increases and the condition $V_{\|}>\omega_{r}/k_{\|}$ is violated.  Thus, for realistic (small to moderate) values of the shear magnitude, the reduction in the threshold current for the current driven ion cyclotron instability by a gradient in the ion parallel flow is minimal at best.  This is unlike the current driven ion acoustic mode case as discussed in the previous section.

Although shear is ineffective in reducing the threshold current for the ion cyclotron instability, it allows for a novel method to extract free energy from the spatial gradient of the ion flow, which does not involve a resonance of parallel phase speed with the relative drift speed.  To illustrate this we return to Eq. \ref{eq:70t} and consider the limit $(k_{y}V_{\|}'/k_{\|}\Omega_{i})(1-n\Omega_{i}/\omega_{r})\gg 1$, in which Eq. \ref{eq:70t} reduces to,	
\begin{equation}
	\label{eq:73t}
	V_{\|c} = \frac{\omega_{r}}{k_{z}}\left[
	1 - \Gamma_{n}(b)\frac{\mu^{1/2}}{\tau^{3/2}}\left\{
	\frac{k_{y}V_{\|}'}{k_{\|}\Omega_{i}}\left(1 - \frac{n\Omega_{i}}{\omega_{r}}\right)
	\right\}
	\exp\left(-\frac{(\omega_{r}-n\Omega_{i})^2}{2k_{\|}^{2}v_{ti}^2}\right)
	\right],
\end{equation}
For $\omega_r > n\Omega_i$ each term of Eq. \ref{eq:73t} is still positive but the relative sign between them is now negative, which allows for $V_{\|c}=0$.  In this regime the ion flow gradient can support ion cyclotron waves.  This can be understood by examining the relevant terms in the growth rate \cite{Ganguli:2002}, 
\begin{equation}
\label{eq:75t}
	\frac{\gamma}{\Omega_{i}} \propto 
	\frac{\tau^{3/2}}{\mu^{1/2}}\left(\frac{V_{\|}}{(\omega_{r}/k_{z})}-1\right)
	-\sum_{n}\Gamma_{n}\left\{
	1 - \frac{k_{y}V_{\|}'}{k_{\|}\Omega_{i}}\left(1-\frac{n\Omega_{i}}{\omega_{r}}\right)
	\right\}
	\exp\left(-\frac{(\omega_{r}-n\Omega_{i})^2}{2k_{\|}^{2}v_{ti}^2}\right),
\end{equation}
The first term in the bracket represents a balance between growth due to the relative field-aligned drift and electron Landau damping while the second term represents cyclotron damping.  Provided the drift speed exceeds the wave phase speed and the magnitude of the first term is large enough to overcome the cyclotron damping a net growth for the ion cyclotron waves can be realized.  This is the classical case where inverse electron Landau damping leads to wave growth \cite{Drummond:1962}.  For the homogeneous case (\textit{i.e.}, $dV_{\|}/dx=0$), the second term is positive definite and always leads to damping.  However, if $(k_{y}V_{\|}'/k_{z}\Omega_{i})(1-n\Omega_{i}/\omega_{r})>1$ then the sign of the cyclotron damping can be changed and the second term can provide a net growth even for $V_{\|}=0$.   This possibility for wave growth is facilitated by velocity shear via inverse cyclotron damping and favors short perpendicular and long parallel wavelengths, which makes the term proportional to shear large even when the magnitude of shear is small.  A necessary condition for ion cyclotron instability due to inverse cyclotron damping is,
\begin{equation}
	\label{eq:76t}
	\left(1-\frac{n\Omega_{i}}{\omega_{r}}\right)
	\left(\frac{k_{y}}{k_{\|}}
	\frac{dV_{\|}/dx}{\Omega_{i}}\right)=
	\left(1-\frac{n\Omega_{i}}{\omega_{r}}\right)
	\left(\frac{V_{py}}{V_{pz}} \frac{dV_{\|}/dx}{\Omega_{i}}\right)>1
\end{equation} 
where $V_{py}$ and $V_{pz}$ are ion cyclotron wave phase speeds in the $y$ and $z$ directions.

\begin{figure}
	\includegraphics[width=\textwidth]{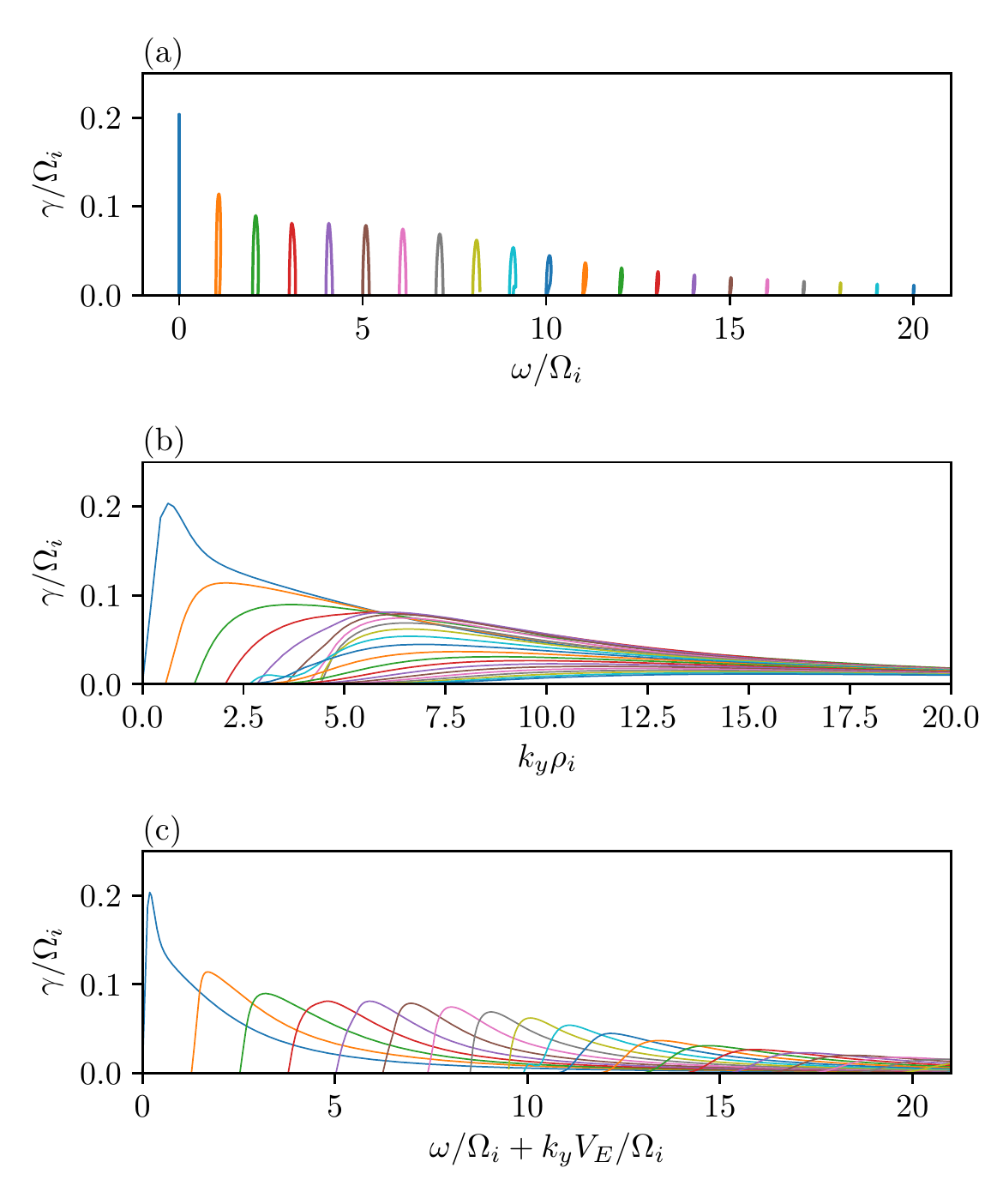}
	\caption{\label{fig:GavCycloHarm}First 20 ion cyclotron harmonics.  (a) Growth rate vs frequency, (b) Growth rate vs $k_y\rho_{i}$, (c) growth rate vs doppler shifted frequency with $V_{E}=.3 v_{ti}$.  Here $V_{de} = 0$ (current-free case), $V_{d}'=2\Omega_{H+}$, $\mu= 1837$, and $\Omega_e/\omega_{pe}=8.2$. }
\end{figure}

Another noteworthy property introduced by the ion flow gradient is in the generation of higher harmonics.  From Eq. \ref{eq:71t} we see that in the homogeneous case the n\textsuperscript{th} harmonic requires a much larger drift than the first harmonic.  However, for $\omega_{r}\sim n\Omega_{i}$ the critical shear necessary to excite the n\textsuperscript{th} harmonic of the gradient driven ion cyclotron mode, can be expressed as,
\begin{equation}
	\frac{(dV_{\|}/dx)_{c}}{\Omega_{i}}
	\sim 
	\frac{\tau^{3/2}}{\mu^{1/2}}
	\left(\frac{k_{\|}}{k_{y}}\right)
	\left(
	\frac{1+\tau-\Gamma_{0}(b)}{\Gamma_{n}^2(b)}\right),
\end{equation}
For short wavelengths, \textit{i.e.}, $b\gg1$, $\Gamma_{n}\sim 1/\sqrt{2\pi b}$  and hence, to leading order, the critical shear is independent of the harmonic number.  Consequently, a number of higher harmonics can be simultaneously generated by the shear magnitude necessary for exciting the fundamental harmonic.  This is quantitatively shown in Fig. \ref{fig:GavCycloHarm} (also in Gavrishachaka \cite{Gavrishchaka:2000}), which indicates about 20 ion cyclotron harmonics can be generated for typical ionospheric plasma parameters.  This figure also shows that when the Doppler broadening due to a transverse dc electric field is taken into account the discrete spectra around individual cyclotron harmonics overlap to form a continuous broadband spectrum such as those found in satellite observations.  This remarkable ability of velocity shear to excite multiples of ion cyclotron harmonics simultaneously via inverse cyclotron damping is similar to the ion cyclotron maser mechanism \cite{Tsang:1987} that results in broadband spectral signature.  However, important differences with the ion cyclotron maser instability exist.  The ion cyclotron maser instability is an electromagnetic non-resonant instability while we discuss the electrostatic limit of a resonant instability.  Also, in this mechanism the background magnetic field is uniform unlike the ion cyclotron maser mechanism.  

%
%
%

\subsubsubsection{High Frequency Limit.}

As discussed in the previous section multiple harmonics of the ion cyclotron frequencies can be generated by the shear in parallel flows.  In the presence of a parallel sheared flow and a transverse electric field the waves generated at the cyclotron harmonics can overlap due to Doppler shift, which can result in a broadband spectrum.  Romero \textit{et al.} \cite{Romero:1992ks} discussed the intermediate and higher frequency modes due to parallel flow shear in which ions can be assumed as an unmagnetized species but the electrons remain magnetized for waves in the frequency range $\Omega_{i}<\omega < \Omega_{e}$ and wavelengths in the range $k_{y}\rho_{i}>1>k_{y}\rho_{e}$.  

For even shorter time scales with frequencies $\omega>\Omega_{e}$ both ions and electrons behave as unmagnetized species.  Mikhailovoskii \cite{Mikhailovskii:1974} has shown that flow shear in this regime can drive modes around the plasma frequency.  

Thus, the combination of low, intermediate, and high frequency emissions that are generated by parallel velocity shear can also lead to a broadband spectral signature similar to that due to transverse velocity shear.

%
%
%

\subsubsection{Hierarchy of compression driven waves}

\begin{figure}
	\includegraphics[width=\textwidth]{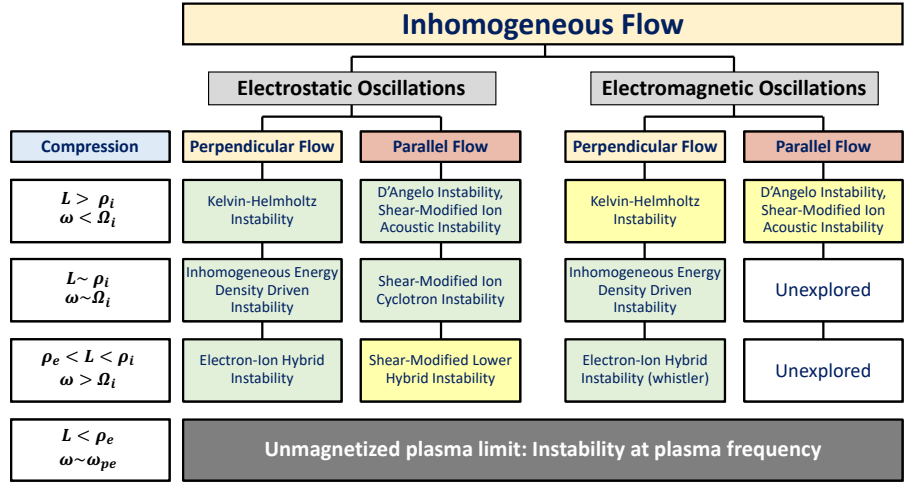}
	\caption{\label{fig:22} Hierarchy of compression driven waves as a function of the magnitude of the compression shown in the first column as shear scale size and the associated wave frequencies. Green corresponds to those cases which have been theoretically predicted and experimentally validated in the laboratory.   Yellow corresponds to the cases which have been theoretically predicted but yet to be validated in a laboratory experiment.  White indicates the cases that are expected to be there by symmetry arguments but yet to be rigorously analyzed.}
\end{figure}

Summarizing the survey of shear driven waves in sections \ref{subsec:linear_low_freq_limit} – \ref{subsec:stability_due_to_par_shear} it can be concluded that the linear response of a magnetized plasma to compression is to generate shear driven waves with frequencies and wave-vectors that scale as the compression.   In Sec. \ref{sec:Equilibrium} we showed that plasma compression self-consistently generates ambipolar electric fields that lead to sheared flows both along and across the magnetic field.   This establishes the causal connection of the shear-driven waves with plasma compression.  Cumulatively, the gradient in the parallel and perpendicular flows constitute a rich source for waves in a broad frequency and wave vector band.  In a collisionless environment their emission is necessary to relax the stress that builds up in the layer due to compression.  Fig. \ref{fig:22} schematically shows the impressive breadth of the frequency range involved with these waves starting from much below the ion cyclotron frequency and stretching to above the electron cyclotron and plasma frequencies that can be generated by a magnetized plasma system undergoing compression. 
 
In the dynamic phase the relaxing gradient can successively excite the next lower frequency wave in the hierarchy when the gradient scale size is sufficiently relaxed to turn off the higher frequency wave, or vice-versa with a steepening gradient \cite{Ganguli:1994a}.  Both relaxation and compression are longer time scale processes compared to the shear driven wave time scales.  This can result in emissions in a very broad frequency band in a quasi-static background that is usually observed in the \textit{in-situ} data.  As a proof of principle a recent laboratory experiment has demonstrated this phenomenon in a limited frequency range that was possible within the constraints of a laboratory device \cite{DuBois:2014} as we elaborate in Sec. \ref{sec:lab}.   Frequency overlap due to Doppler shift and nonlinear processes, such as scattering, vortex merging, \textit{etc.}, can smooth out the spectrum and contribute to seamless frequency broadening as typically observed by satellites.   This naturally raises a question of how these waves affect the plasma-saturated state that a satellite observes.  This is the topic of discussion in the following section.

\section{Nonlinear Evolution and Feedback of the Waves to Global Dynamics}
\label{sec:nonlinear}

We now examine how the linear fluctuations induced by the compression evolve, the dominant nonlinear processes that relax the gradients to establish a steady state, and the measurable signatures of the compression driven waves.  For this we need numerical simulations.  However, due to the huge disparity in space and time scales it is difficult to simulate the entire chain of physics in a single simulation.  Hence, we focus on limited frequency and wavelength domains in order to understand the development of the spectral signature and the steady state features in the nonlinear stage along with other nonlinear characteristics.

\subsection{Low frequency waves in transverse sheared flows}
\label{subsec:nl_lfw_in_sf}

\begin{figure}
	\includegraphics[width=0.6\textwidth]{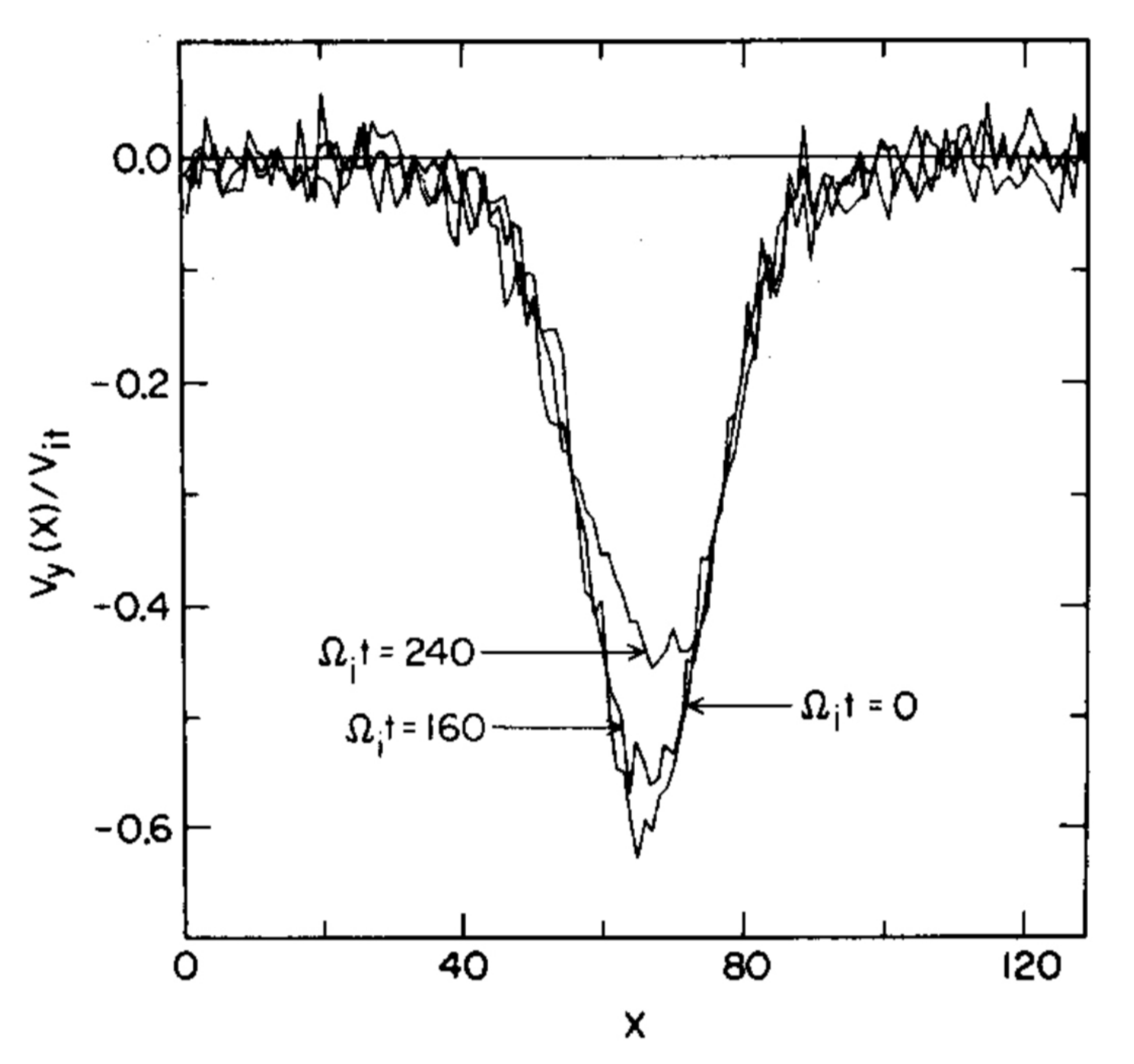}
	\caption{\label{fig:Nishikawa88_Fig8} The average ion flow velocity $v_y(x)$ at $\Omega_i t=0$, $160$, and  $240$ reproduced from Figure 5 of Nishikawa \textit{et~al.} \cite{Nishikawa:1988}. }
\end{figure}

The ion cyclotron frequency range IEDDI was first invoked to understand observations of ion cyclotron waves associated with a transverse electric field \cite{Mozer1977} in the auroral region in which the magnetic field aligned current was minimal and the background plasma density was nearly uniform.  Soon after the IEDDI mechanism was proposed \cite{Ganguli:1985a}, Pritchett \cite{Pritchett:1987} conducted a PIC simulation using the simple ‘top hat’ piecewise continuous electric field model (Fig. \ref{fig:IDDIgeom}), which was intended as a proof-of-principle calculation of the IEDDI in the initial article.  Because the electric field in the top hat model changes its value discontinuously the simulation showed immediate decay of the electric field due to gyro-averaging, which led Pritchett to conclude that the IEDDI does not exist and identified the fluctuations in the simulation as due to the KH instability.  This initiated the derivation of an appropriate equilibrium distribution function in warm plasma that includes a sheared transverse electric field and is suitable for the initial loading in a computer simulation \cite{Ganguli:1988hh} (also briefly described in Sec. \ref{subsec:analytical_f}).  This distribution function was used to obtain the general kinetic dispersion relation, which showed the existence of both the IEDDI and the KH branches in the proper parameter regimes as summarized in section 3.3.1.  Nishikawa \textit{et al.} \cite{Nishikawa:1988,Nishikawa:1990} used this equilibrium distribution function to successfully simulate the IEDDI and demonstrated that it was another branch of oscillation in magnetized plasmas with transverse electric field distinct from the KH instability.  The simulation also showed the development of a polarization current along the electric field direction that reduced the magnitude of the external electric field as the waves grew and a bursty spectrum of waves, which were consistent with the nonlinear IEDDI (ion magnetron) model of Palmadesso \textit{et al.} \cite{Palmadesso:1986}.   More important to this article, as shown in Fig. \ref{fig:Nishikawa88_Fig8} (reproduced from Nishikawa \textit{et~al.} \cite{Nishikawa:1988}), the growth of the instability relaxed the flow gradient.  This establishes that the strong transverse electric field gradients that develop as a response to plasma compression (Sec. 2) can relax through the emission of the shear driven modes discussed in section \ref{sec:lineartheory}. 

The IEDDI was later validated in laboratory experiments in NRL \cite{Amatucci:1996} and elsewhere \cite{Koepke:1994} as discussed in Sec. \ref{sec:lab}.  These laboratory experiments consistently showed that the IEDDI fluctuations have azimuthal mode number $m = 1$. Interestingly, Hojo \textit{et~al} \cite{Hojo:1995} showed that there can be no $m=1$ KH mode in a cylindrical geometry.  The KH wave growth peaks for higher m numbers in a cylindrical geometry \cite{Kent:1969,Jassby:1970,Jassby:1972} while the IEDDI growth maximizes for $m=1$ \cite{Penano:1998}.  This is an experimental confirmation that the IEDDI is distinct from the KH instability and that they form separate branches of oscillations in magnetized plasma with transverse sheared flow.  Subsequently, Pritchet \cite{Pritchett:1993} also tested the Ganguli \textit{et~al.} \cite{Ganguli:1988hh} equilibrium model and concluded that it led to more reliable results although he could not resolve the IEDDI in his simulation accurately. 

\subsection{Intermediate frequency waves in transverse sheared flows}
\label{subsec:nl_ifw_in_sf}

In the auroral region the observed velocity shear scale size is generally larger than the ion gyroradius, albeit in the saturated state.  This is the weak shear regime.  However, as we found in Sec. 2, the scale size of the velocity shear that develops in the boundary layers can be in the intermediate range, \textit{i.e.}, $\rho_i>L>\rho_e$.  Also in this region wave power around the lower hybrid frequency range has been observed.  The generation of both electrostatic and electromagnetic waves around the lower hybrid frequency by velocity gradient has been extensively studied.  Simulations \cite{Romero:1993ip} indicate that these waves produce anomalous viscosity and relax the velocity gradients to reach a steady state.  In the following sections we study the nonlinear evolution of these waves leading to formation of the steady state and the observable signatures by numerical simulation.

\subsubsection{Plasma sheet-lobe interface}

\begin{figure}
	\includegraphics[width=0.6\textwidth]{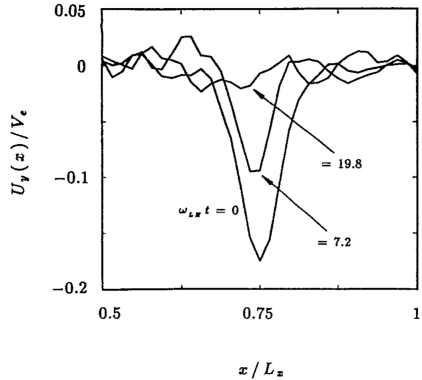}
	\caption{\label{fig:Romero1993_Fig16}  Spatial profiles of the electron cross-field flow at different times indicating the relaxation of the velocity gradient.  Reproduced Fig. 16 of Romero \textit{et~al.} \cite{Romero:1993ip}.}
\end{figure}

In understanding the behavior in the compressed plasma layer formed at the plasma sheet-lobe interface (Sec. \ref{subsec:vlasovpoisson}) Romero \textit{et~al.} \cite{Romero:1992a} used the Ganguli \textit{et~al.} \cite{Ganguli:1988hh} equilibrium (Eq. \ref{eq:analytical_f0_simplified}) for the electrons and an unmagnetized Maxwellian distribution for the ions in a 2D electrostatic PIC model to simulate the spontaneous generation of the intermediate frequency EIH waves discussed in Sec. \ref{subsec:intermediate_f}.   The localized electric field used in the simulation was in the intermediate scale length defined by $\rho_{i}>L>\rho_{e}$ and was self-consistent with the density gradient.  The simulation was motivated by the ISEE satellite observation in the plasma sheet-lobe interface as shown in Fig. \ref{fig:1}.  Spontaneous growth of the lower hybrid waves was seen in the boundary layer.  The waves nonlinearly formed vortices.  The scale size of the vortices was comparable to the velocity gradient scale size. Fig. \ref{fig:Romero1993_Fig16}, (reproduced from Romero and Ganguli \cite{Romero:1993ip}), shows that the growth of the EIH waves relaxed the velocity gradient similar to that observed in the IEDDI simulation of Nishikawa \textit{et~al.} \cite{Nishikawa:1990}.  Interestingly, the density gradient was not relaxed by the EIH instability. The difference in the two simulation was that in the Nishikawa \textit{et~al.} \cite{Nishikawa:1990} simulation of Ion cyclotron IEDDI the electric field was localized over a distance larger than $\rho_{i}$ while in the Romero and Ganguli \cite{Romero:1993ip} simulation it was localized over a smaller distance.  The inference that can be drawn from the two simulations is that if the initial compression is large such that $L<\rho_{i}$, then the growth of the lower hybrid waves could relax the velocity gradient so that $L>\rho_{i}$ at steady state.  While this saturates the lower hybrid waves, the flow shear will be in the right magnitude to trigger the lower frequency IEDDI.   When IEDDI relaxes the gradient even further so that $L\gg\rho_{i}$ then the KH modes could be triggered and so on.  This nonlinear cascade to appropriate frequencies as the background gradient scale changes is how the shear driven modes can lead to a broadband signature of the emissions that are observed in the compressed plasmas \cite{Grabbe:1984}.  In addition, the Nishikawa \textit{et al.} \cite{Nishikawa:1990} simulation showed the coalescence of smaller vortices into larger ones implying that the wavelengths become larger with time due to nonlinear vortex merging.  Thus, these lower hybrid waves have large wavelengths, roughly of the order of the shear scale length rather than an $\rho_e$ as expected due to the LHDI, as discussed in Section 3.3.2.  The spatio-temporal scales associated with the cascading frequencies are so large that it is difficult to simulate the entire bandwidth in a single simulation. 

The initial Romero \textit{et al.} simulation \cite{Romero:1992a} was followed up with more detailed studies of the nonlinear signatures of these waves, effects of magnetic field inhomogeneity on these waves, as well as their contribution to viscosity and resistivity, which provide the steady state and feedback to the larger scale dynamics \cite{Romero:1993ip,Romero:1994}.

\subsubsection{Dipolarization front}

More recently, the Romero \textit{et~al.} \cite{Romero:1993ip} simulation model was applied to the DF plasmas \cite{Fletcher:2019kq} and generalized to the electromagnetic regime \cite{Lin:2019ho}.  The plasma parameters used in the simulation \cite{Fletcher:2019kq} were $\omega_{pe}/\Omega_{e}=3.59$, $\beta_{e}=0.035$, $m_e/m_i=1/400$, and the peak of the ambipolar field consistent with the density gradient is given by $cE_0/B_0=0.32v_{te}$. The simulation time is $175/\omega_{LH}$, the spatial domain is $21\rho_i$ by $21\rho_i$ (1200 by 1200 cells), boundaries are periodic in all directions, and 537 million particles were used.

\begin{figure}
\includegraphics[width=0.8\textwidth]{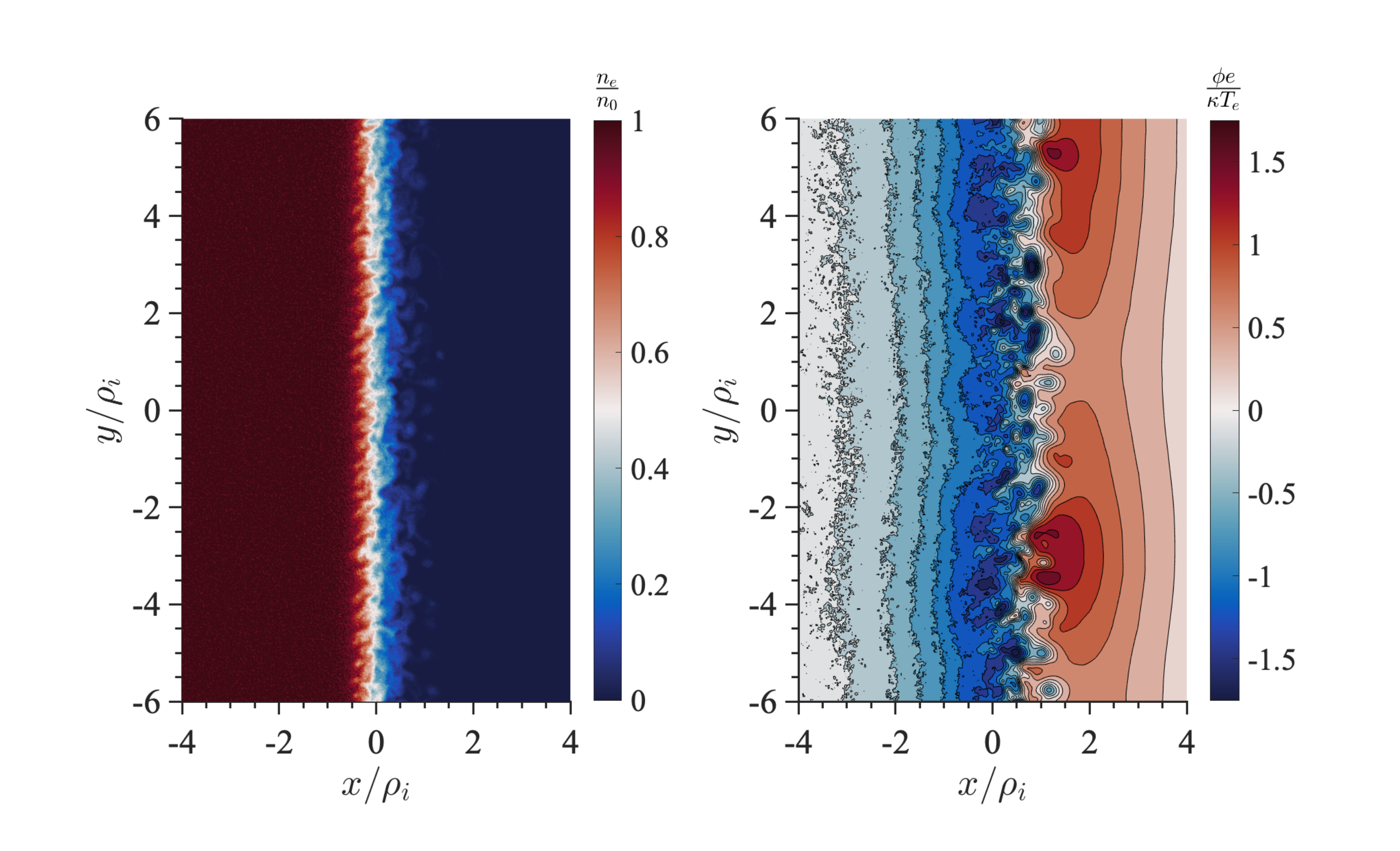}
\caption{\label{fig:DF_DensityPhi} Plasma density, $n$, (left) and electrostatic potential, $\phi$, (right) at $t\simeq28/\omega_{LH}$. Waves in the $y$ direction and vortices are both visible. Reproduced from Figure 16 of Fletcher \textit{et~al.} \cite{Fletcher:2019kq}.}
\end{figure}

Fig. \ref{fig:DF_DensityPhi} (from Fletcher \textit{et~al.} \cite{Fletcher:2019kq}) shows a snapshot of the plasma density and electrostatic potential from the simulation at $t\simeq 28/\omega_{LH}$. These images show only a part of the simulation domain in order to make features more visible.  Kinking is seen in the density.  Vortices are formed on the lower density (right) side of the layer as well; these are visible in the potential (for example, one vortex is located at ($x/\rho_i$,$y/\rho_i$)$\simeq$(1,-1.5)).  Wave activity in the $y$ direction with $k_y L\sim 1$ is apparent in both the density and the potential. The growth rate of the field energy in the simulation is consistent with the growth rate found by solving Eq. \ref{eq:intermed_eig_dim}. The mass ratio of the simulation is low in order to facilitate quick simulation but a physical mass ratio would enhance the ambipolar electric field and further drive these waves.

\begin{figure}
\includegraphics[width=0.75\textwidth]{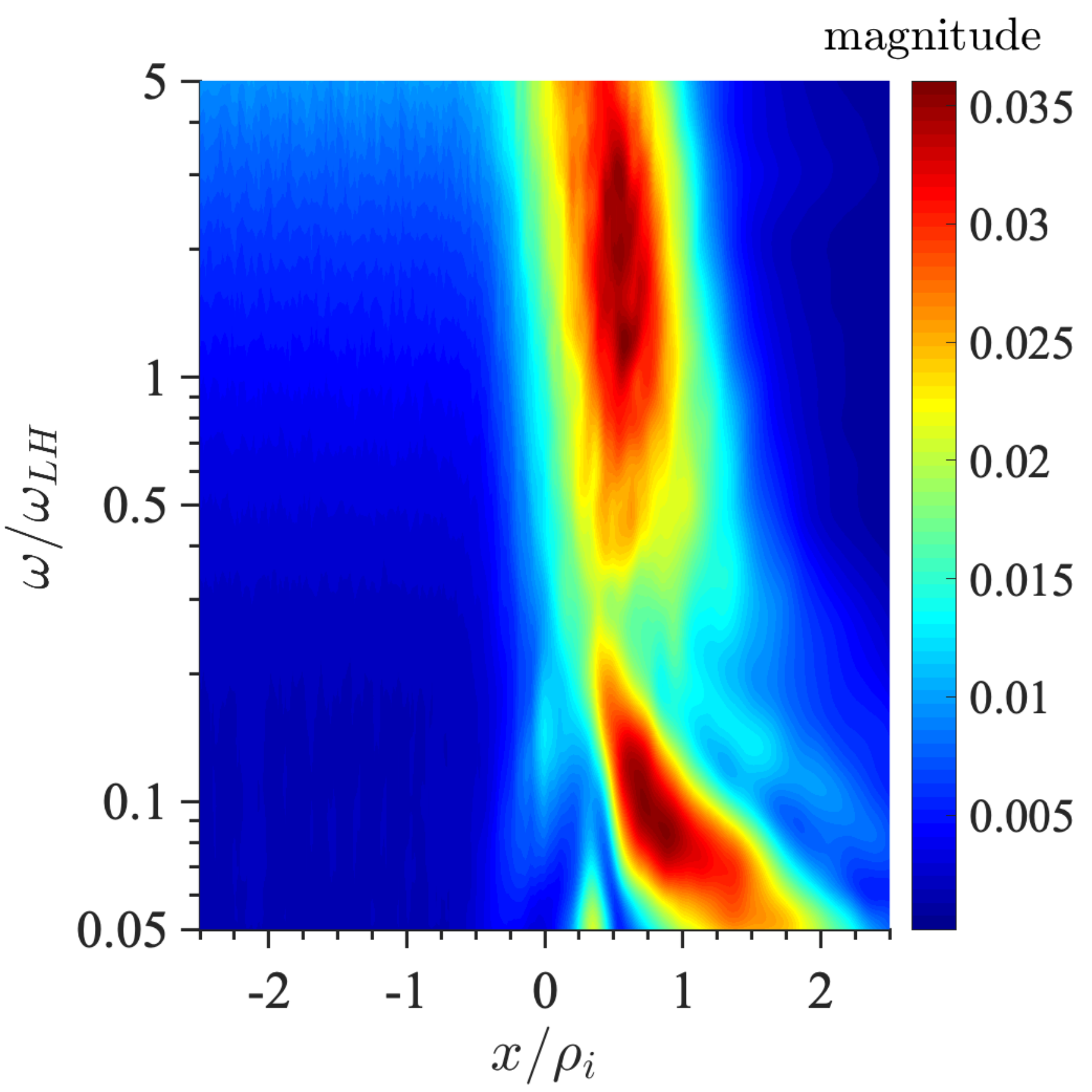}	
\caption{\label{fig:DF_Wavelet}Wavelet spectrum of the electric field as a function of position near $t\simeq 28/\omega_{LH}$. The density gradient is steepest near $x/\rho_{e}=0$.  Reproduced from Figure 17 of Fletcher \textit{et~al.} \cite{Fletcher:2019kq}.}
\end{figure}

Fig. \ref{fig:DF_Wavelet} is a wavelet spectrum as a function of $x$ position; the layer is centered near $x/\rho_{i}=0$. It is similar to what a satellite would measure if it were flying through the simulated layer or a DF would propagate past the observing satellite. There are broadband waves spread around and above the lower hybrid frequency. The lower frequency power $\omega/\omega_{LH}\simeq 0.1$ is consistent with vortices being generated and propagating away from the layer. 

\begin{figure}
\includegraphics[width=\textwidth]{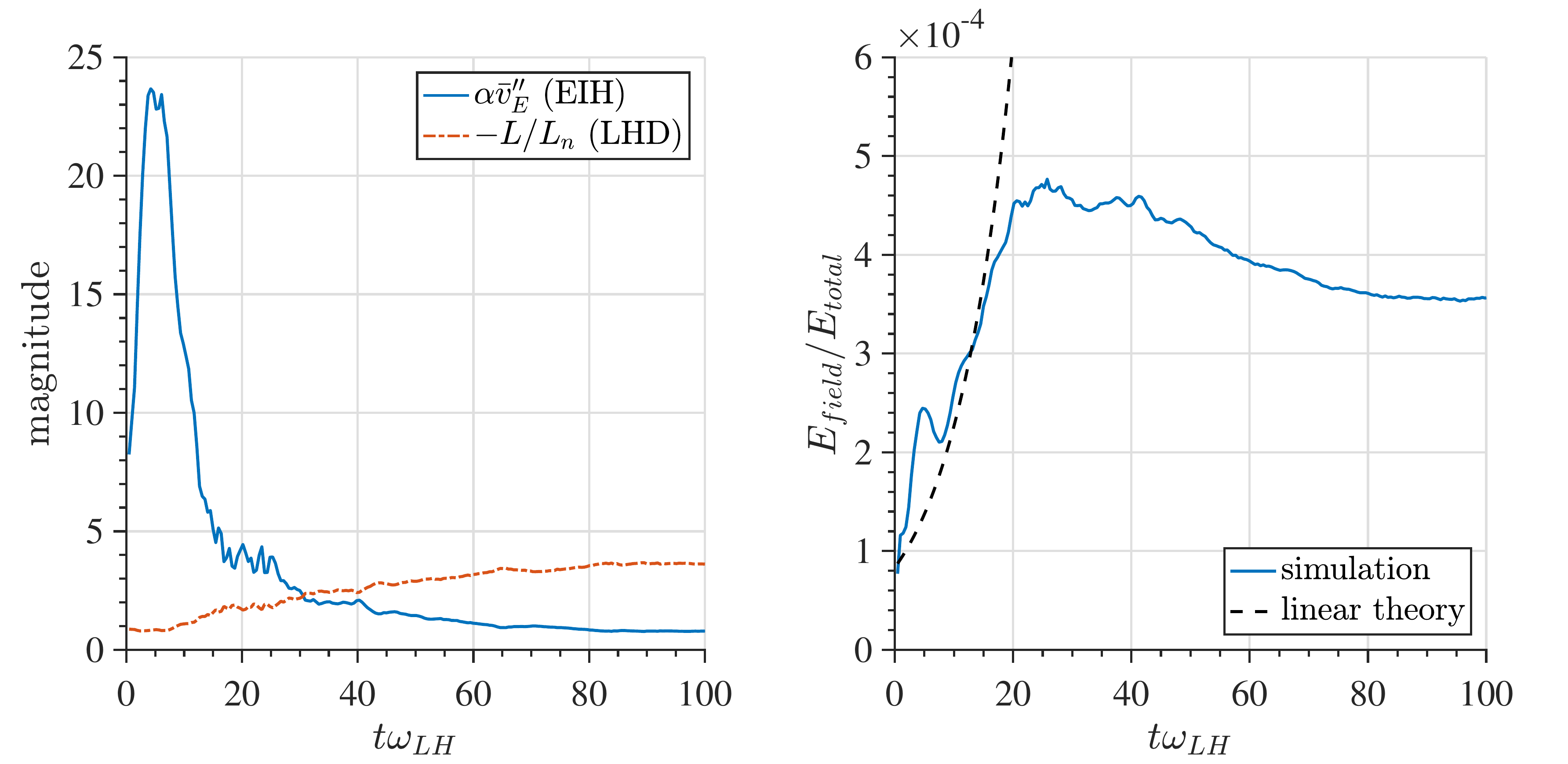}
\caption{\label{fig:DF_Drive_LinearTheory} The driving terms for the EIH instability and LHD instability (left) and the field energy fraction (right) in the simulation as a function of time.  Reproduced from Figure 18 of Fletcher \textit{et~al.} \cite{Fletcher:2019kq}.}
\end{figure}

As time passes in the simulation, the density gradient is more-or-less unaffected while the electron flow in the $y$ direction and accompanying electric field in the $x$ direction is significantly relaxed, indicating the dominance of shear-driven instability (EIH) over the density gradient-driven instability (LHD).  Fig. \ref{fig:DF_Drive_LinearTheory} shows these two separate source terms responsible for the EIH and the LHD instabilities respectively (as in the numerator of the last term Eq. \ref{eq:intermed_eig_dim}) and the field energy as a function of simulation time. Instability growth and wave emission occurs before $t=20/\omega_{LH}$. The dotted black line is the theoretical linear growth predicted by Eq. \ref{eq:intermed_eig_dim}.   During the growth phase, the EIH source term (and thus the velocity shear) is clearly falling, suggesting that the shear is the source of free energy for the waves. The simulation reaches a saturated state at $t\simeq 20/\omega_{LH}$.

\subsection{Ion cyclotron waves in parallel sheared flows}

In Secs 4.1 and 4.2, we studied the nonlinear evolution of sheared transverse flows.  We found that spontaneous generation of shear-driven waves relaxes the velocity gradient that leads to saturation.  The frequency and wavelengths of these waves scale as the shear magnitude.  Nonlinear vortex merging results in longer wavelengths.  Relaxation of stronger shear leads to weaker shear which can then drive lower frequency modes.  This cascade leads to the broadband spectrum of emissions that are often observed.  Now we examine the nonlinear behavior of parallel flow shear driven waves.

The nonlinear evolution of the parallel flow shear driven modes discussed in Section 3.3.4 was investigated with PIC simulations by Gavrishchaka \textit{et~al.} \cite{Gavrishchaka:2000}.  The simulations included full ion dynamics but used a gyrocenter approximation for the electrons.  To clearly resolve short wavelength modes 900 particles per cell were used with grid size $\Delta=\lambda_D=0.2\rho_i$ and mass ratio $\mu=1837$.  A drifting Maxwellian (H+, e-) plasma is initially loaded, with equal ion and electron temperatures.  The magnetic field is slightly tilted such that $k_{\|}/k_y=0.01$.  A parallel drift velocity $V_{\|}(x)$ is assigned to ions to obtain an inhomogeneous velocity profile.  The magnitude of the flow is initially specified and not reinforced during the simulation.  To characterize the role of spatial gradients in the flow, the relative drift between the ions and the electrons, \textit{i.e.}, field aligned current, is kept at a minimum.  Its value does not exceed $3v_{ti}$ locally while on average it is negligible.   Periodic boundary conditions are used in both $x$ and $y$ directions.  The magnitude of shear $|dV_{\|}/dx|_{max}=2\Omega_i$ is used for the simulation.  For this case the simulation box size was specified by $L_x=64\lambda_D$ and $L_y=64\lambda_D$.  

\begin{figure}
	\includegraphics[width=0.5\textwidth]{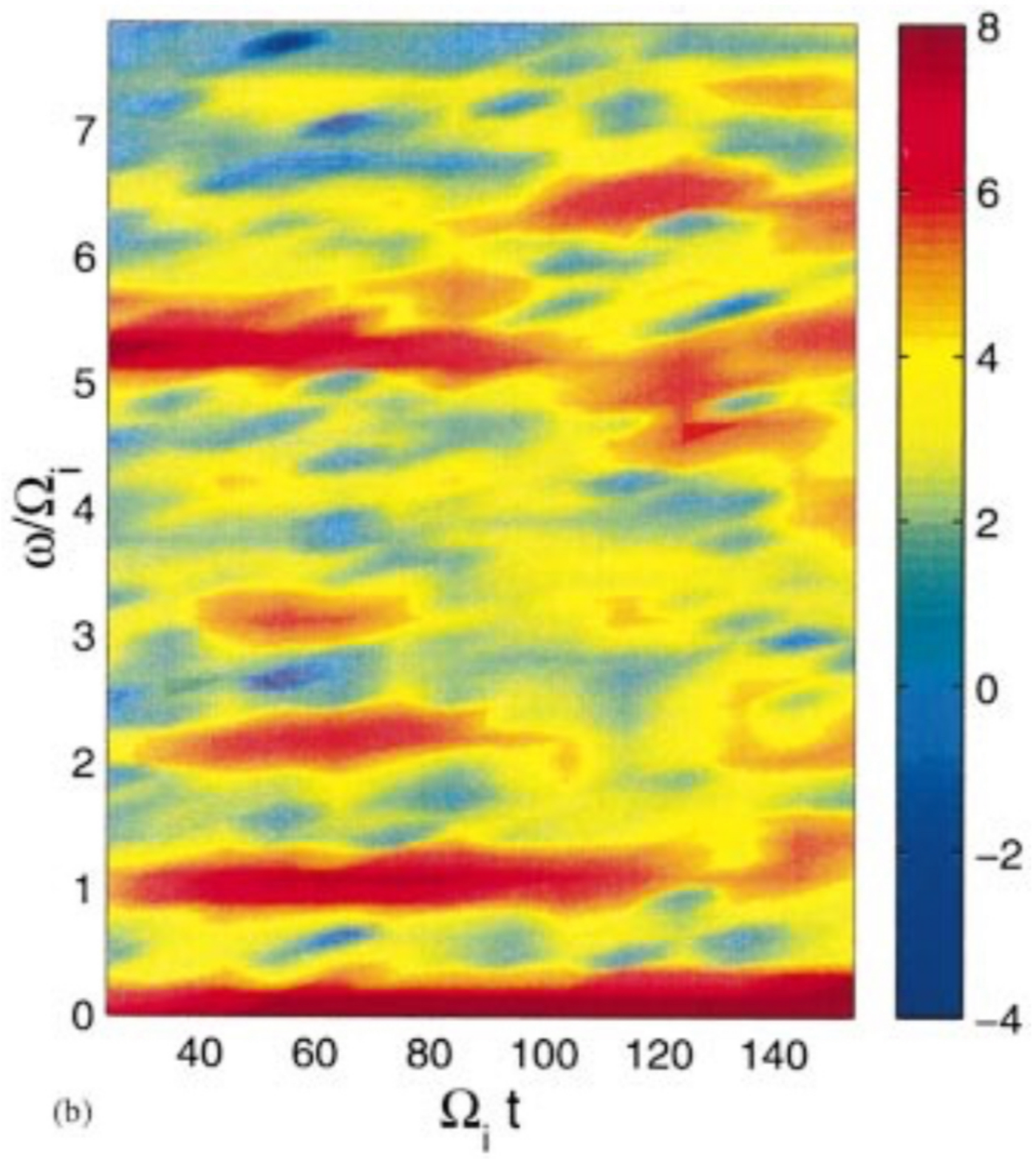}
	\includegraphics[width=0.5\textwidth]{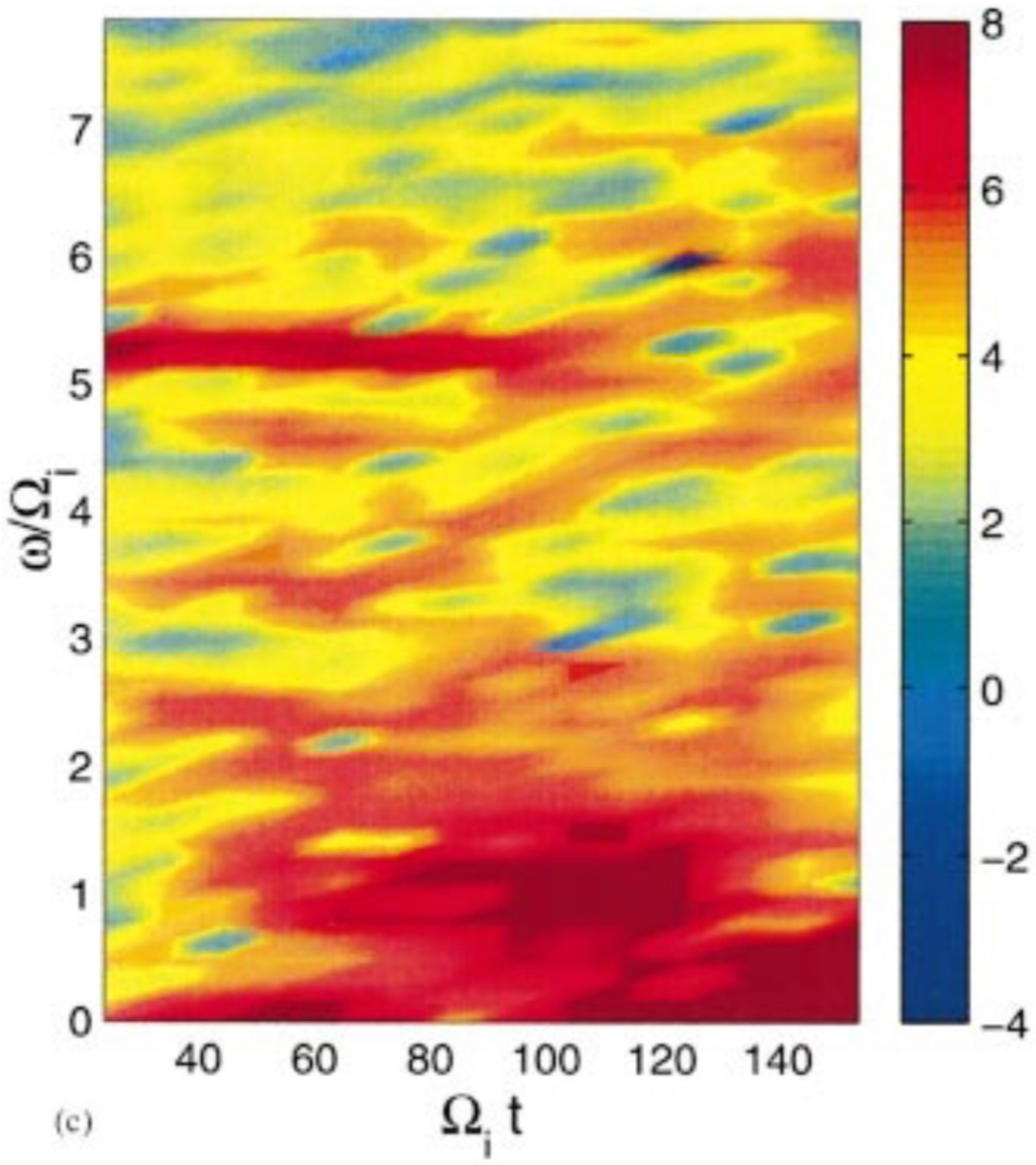}
	\caption{\label{fig:Ganguli2002_Fig3}Nonlinear spectral signature from a PIC simulation.  (left) Frequency spectrum without a transverse DC electric field.  (right) Frequency spectrum including a transverse DC electric field.  Figures reproduced from Fig. 3 of Ganguli \textit{et~al.} \cite{Ganguli:2002}.}
\end{figure}

The saturated spectral signature in the simulation without and with a uniform transverse electric field shown in Fig. \ref{fig:Ganguli2002_Fig3}.  On the left is the wave spectrum without a transverse electric field.  In this simulation several ion cyclotron harmonics are excited with discrete harmonic structure. While on the right a uniform transverse dc electric field is included with $V_{E} =0.8v_{i}$.  The washing out of the harmonic structure and broadening of the spectrum due to overlap of the discrete spectra around $\omega=0$ and multiple cyclotron harmonics becomes evident.  Larger Doppler broadening either by large $V_E$ or large bandwidth, $\Delta k_{y}$, or a combination of both, could lead to an even broader spectrum.   
	
\begin{figure}
	\includegraphics[width=0.6\textwidth]{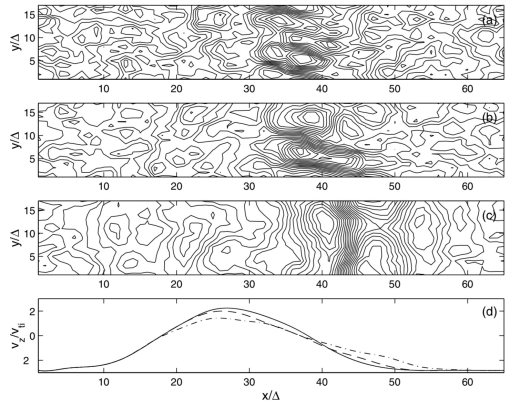}
	\caption{\label{fig:GavPRL200_Fig3} Electrostatic wave potential obtained from PIC simulations after $\Omega_i t= 40$ (a), 60 (b), 100 (c), and the corresponding ion velocity parallel to the magnetic field (d) shown by solid, dashed, and dot-dashed lines, respectively.  Reproduced from Fig. 3 of Gavrishchaka \textit{et al.} \cite{Gavrishchaka:2000}.}
\end{figure}

The meso-scale effect of the parallel flow shear driven instability (normalized by their initial values) is given in Fig. \ref{fig:GavPRL200_Fig3} (reproduced from Gavrishchaka \textit{et al.}, \cite{Gavrishchaka:2000}).  To highlight the role of shorter wavelength ion cyclotron waves the longer wavelengths are removed by using a $(64\times16)\lambda_{d}$ size simulation box in this case.   Fig. \ref{fig:GavPRL200_Fig3} illustrates that the effect of the ion cyclotron wave generation is relaxation of the flow gradient due to wave-induced viscosity.   This is similar to the effect of the transverse shear driven waves but not as strong.  This may be because the orbit modifications due to a localized transverse electric field is absent in this case.  Thus the primary conclusion is that the compression generated velocity shear either in parallel or transverse flow leads to broadband emissions accompanied by relaxation of the velocity gradient that leads to a steady state and determines the observed features that are measured by satellites.

In the above we discussed only the part of the simulation that showed the formation of the broadband spectral signature and relaxation of the velocity shear due to these waves, which is central to this article.  However, the simulation also explained a number of interesting auroral observations that are not elaborated here.  For a detailed account of these we refer to Gavrishchaka \textit{et al.} \cite{Gavrishchaka:1998,Gavrishchaka:2000} and Ganguli \textit{et al.} \cite{Ganguli:2002}.

\section{Laboratory experiments of compressed plasma behavior}
\label{sec:lab}

In Secs. \ref{sec:Equilibrium}-\ref{sec:nonlinear}, we outlined the theoretical foundation for understanding compressed plasma behavior and showed evidence of its characteristics in uncontrolled natural plasmas from \textit{in situ} data gathered from satellites.  The challenge with \textit{in situ} data is in characterization of a specific phenomenon in constantly evolving plasmas subject to uncertain external forces.  As a result, typically there are many competing theories of space plasma phenomena that are difficult to distinguish unambiguously.  Because of this difficulty, scaled laboratory experiments have become a valuable tool in understanding space plasma processes.    Not every aspect of space plasmas can be faithfully scaled in the laboratory.  Large MHD scale phenomena are especially challenging.  But others, such as cause and effects of waves and various coherent processes in the meso and micro scales, which are difficult to resolve by \textit{in situ} measurements in space, are amenable to laboratory scaling.  In the modern era, satellite clusters with multi-point measurements have been used to overcome some of the difficulties with resolving the space-time ambiguity in measurements made from a single moving platform.  While they help, they are expensive and there are still limitations of measurements made from a moving platform.  An area where laboratory experiments can contribute substantially is in the understanding of the effects of highly localized regions of strong spatial variability, such as the strong gradients over ion or electron gyroscales associated with compressed plasmas discussed in this article.  These phenomena can be scaled reasonably well in the laboratory.  The Space Chamber at the US Naval Research Laboratory (NRL) is especially designed for understanding space plasma phenomena, such as the behavior of compressed plasmas.

\begin{figure}
	\includegraphics[width=\textwidth]{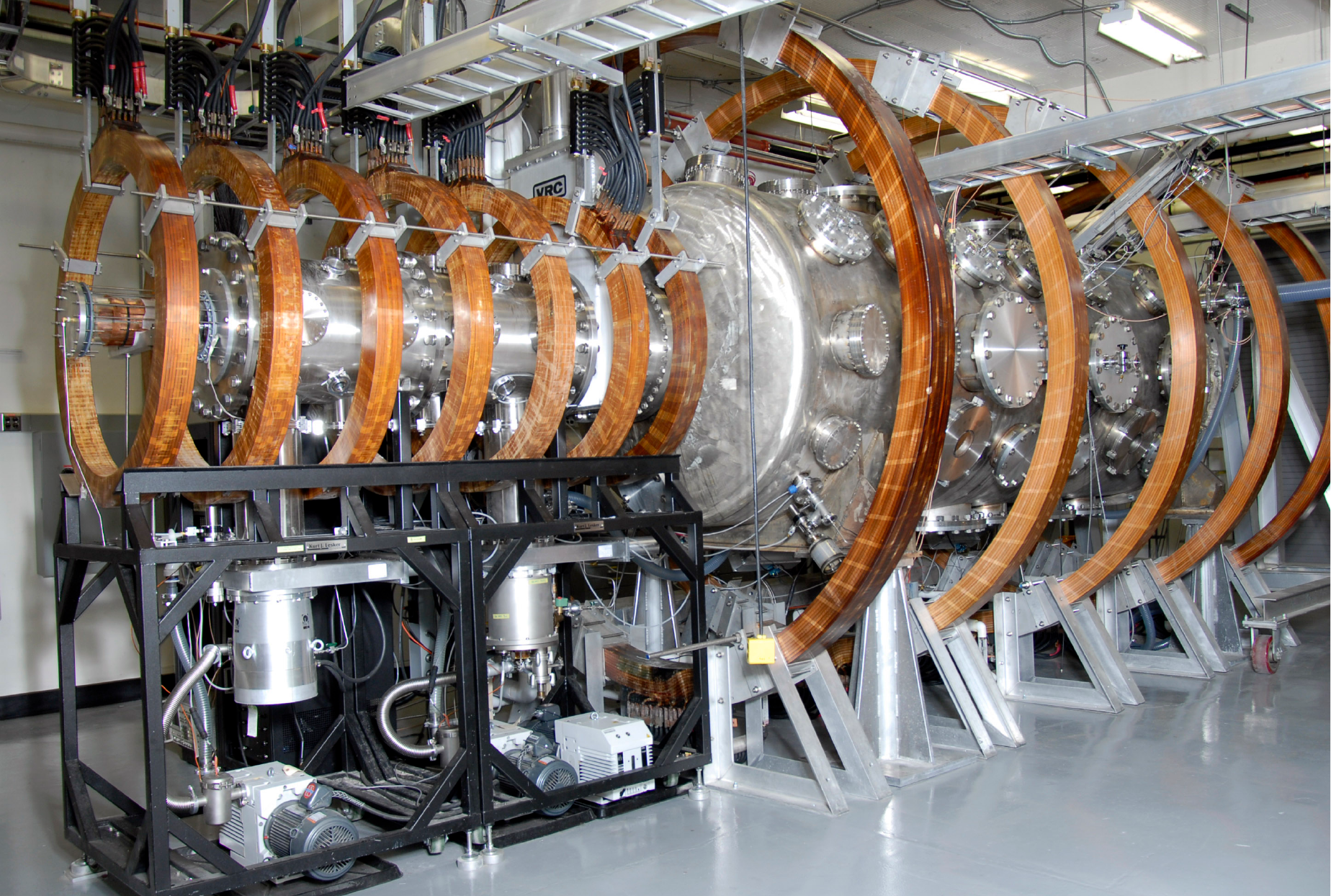}
	\caption{\label{fig:29} NRL Space Physics Simulation chamber.  Main chamber section (1.8 m by 5 m) is on the right.  Source chamber (0.55 m by 2 m) is on the left.}
\end{figure}

The NRL Space Physics Simulation Chamber (SPSC), shown in Fig. \ref{fig:29}, consists of two sections that can be operated separately or in conjunction.  The main chamber section is 1.8 m in diameter and 5 m long, while the source chamber section provides an additional 0.55 m-diameter, 2-m long experimental volume.  The steady-state magnetic field strength in the main and source chamber sections can be controlled up to 220 G and 750 G respectively, generated by 12 independently controlled water-cooled magnets capable of shaping the axial magnetic field.  Each section has a separate plasma source.  The main chamber has a 1-m x 1-m hot filament plasma source capable of generating plasmas with a range of density $n \sim  10^4 - 10^{10}$ cm$^{-3}$, electron temperature $T_{e} \sim 0.1 - 2$ eV, and ion temperature $T_i \sim 0.05$ eV.  The source chamber has a helicon source capable of generating 30-cm diameter plasmas with the following parameters: $n \sim 10^{8} - 10^{12}$ cm$^{-3}$, $T_{e} \sim  1 - 6$ eV, and $T_{i} \sim  0.1$ eV.  When the helicon plasma transitions from the source chamber to the main chamber, the plasma column diameter can be increased up to the full 1.8-m diameter of the main chamber by controlling the ratio of magnetic field strength between the two chamber sections.  The large plasma size yields up to ~150 ion gyroradii across the column.  Table \ref{table:spsc} shows the ranges of normalized plasma parameters accessible in the NRL SPSC with comparisons to those found in the ionosphere and regions of the magnetosphere.

We discuss a few experiments performed in the NRL Space Chamber and elsewhere that were designed to understand the effects of strong velocity and pressure gradients typical of compressed plasmas.  As discussed in Sec. \ref{sec:Equilibrium}, the localized electric field can be considered a surrogate for the global compression.  Thus, by studying the plasma response to localized electric fields we can glean the physical processes that characterize a compressed plasma layer.

\begin{table}
\def\arraystretch{1.5}
\begin{center}
\begin{tabular}{c c c c}
\hline 
parameter & ionosphere & RB(L=2) & NRL SPSC \\
\hline 
plasma density (cm$^{-3}$) & $10^3-10^6$ & $\sim 10^3$ &$10^4-10^{12}$ \\
electron temp. (eV) & $\sim 0.3$ & $\sim 1$ & $0.1-4$ \\
ion temp. (eV) & $\sim 0.3$ & $0.3$ & $0.05$ \\
magnetic field strength (G) & $\sim 0.3$ & $\sim 0.04$ & \begin{tabular}{@{}c@{}} up to 750 G (SC)\\ \& 250 G (MC) \end{tabular}\\
plasma freq. (Hz) & $10^5 - 10^7$ & $5\times 10^5$ & $10^6 - 10^{10}$ \\
ion gyrofrequency (Hz)  & $\sim 30$ (O$^{+}$) & $\sim 60$ (H$^{+}$) & $\sim 10^3 - 10^{5}$ (Ar$^{+}$) \\
electron gyrofrequency (Hz) &$\sim 10^{6}$ & $\sim 10^{5}$ & $10^{6}-10^{9}$ \\
\rowcolor[HTML]{ffffcc}
$\omega_{pe}/\Omega_{e}$ & 0.1-10 & $\sim 5$ & $0.01-50$ \\
\rowcolor[HTML]{ffffcc}
$\omega/\nu_{en}$ & $>1$ & $\gg 1$ & $\sim 5-600$ \\
\rowcolor[HTML]{ffffcc}
$\beta$ & $10^{-7}-10^{-4}$ & $10^{-5}$ & $10^{-7}-10^{-3}$ \\
\hline  
\end{tabular}
\caption{\label{table:spsc} Comparison of plasma parameters in the ionosphere, the Radiation Belts (RB), and the NRL SPSC.}
\end{center}	
\end{table}
\subsection{Low Frequency Limit:  Transverse Velocity Gradient}

In the 1970s, the NASA S3-3 satellite observed emissions around the ion cyclotron frequency in uniform density plasma at auroral altitudes where spatially localized DC electric fields were large \cite{Mozer1977}.  Kelly and Carlson \cite{Kelley:1977} reported intense shear in plasma flow velocity at the edge of an auroral arc associated with short wavelengths fluctuations, the origin of which was a mystery.  They noted that, “A velocity shear mechanism operating at wavelengths short in comparison with the shear scale length, such as those observed here, would be of significant geophysical importance.”  Kintner \cite{Kintner:1992} described the difficulty for exciting the current-driven ion cyclotron waves \cite{Kindel:1971} in the lower ionosphere where the magnitude of the field-aligned current is usually below the threshold and yet bulk heating of ions suspected due to ion cyclotron waves is detected.  

In addition to space observations, there were laboratory experiments, although unconnected with the space observations, reporting ion cyclotron waves correlated to localized transverse dc electric fields \cite{Sato:1986,Alport:1986}.   The generation mechanism of these ion cyclotron waves was not clear.  

These observations led to theoretical analysis at NRL, described in Sec. \ref{subsec:linear_low_freq_limit}, which suggested that the Doppler shift by a localized transverse electric field could make the energy density of the ion cyclotron waves negative in the electric field region while it is positive outside.  A flow of energy between the regions with opposite signs of wave energy density can lead to an instability \cite{Ganguli:1985a,Ganguli:1988hh}.  Because the necessary condition for instability is that the energy must flow from one region to another with opposite sign of energy density, the instability is essentially nonlocal.  It was a promising mechanism for understanding a number of mysterious observations in the auroral region including low altitude ion heating \cite{Ganguli:1985b}, which was a front burner issue of the time.  So, its validation and detailed characterization in the laboratory became an important topic.

\begin{figure}
	\includegraphics[width=\textwidth]{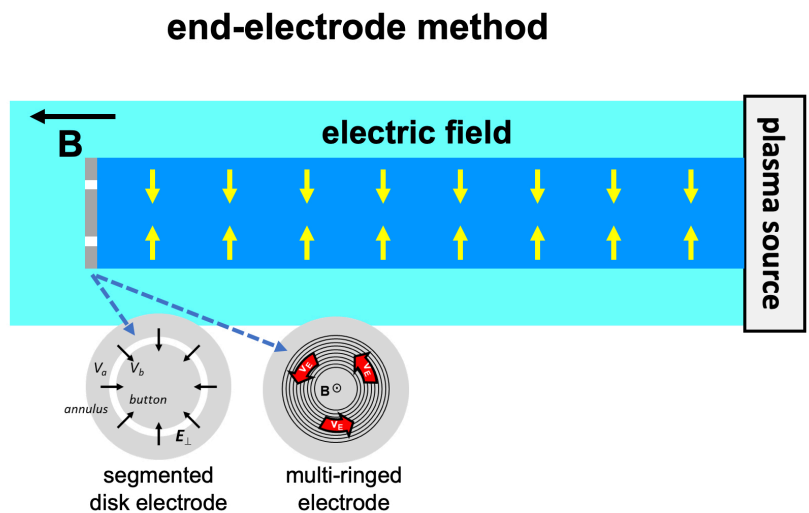}
	\caption{\label{fig:SegmentedVsRings}Segmented disk and biased multi-ringed electrode in the NRL SPSC.}
\end{figure}

\begin{figure}
	\includegraphics[width=\textwidth]{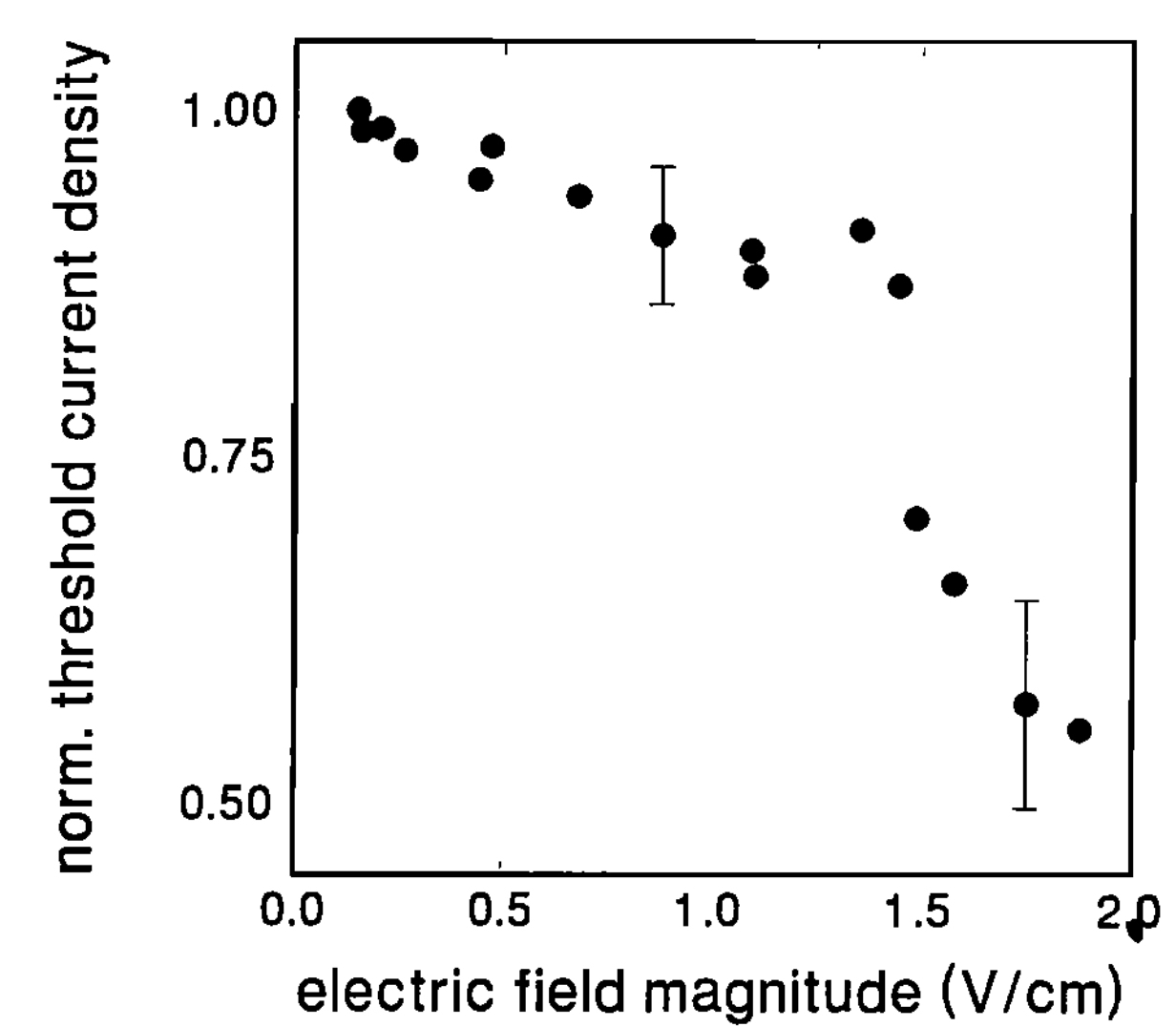}
	\caption{\label{fig:sub-threshold}Threshold value of current density as a function of transverse, localized, dc electric (TLE) field strength. Current densities are normalized to the zero-TLE-strength value. Error bars represent one standard deviation. Reproduced from Figure 2 of Amatucci \textit{et~al.} \cite{Amatucci:1994}.}
\end{figure}

\begin{figure}
	\includegraphics[width=0.6\textwidth]{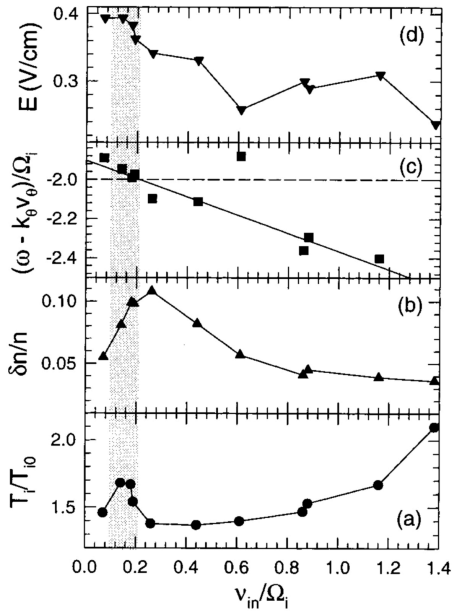}
	\caption{\label{fig:Amatucci1999-Fig3}(a) Perpendicular ion temperature $T_i /T_{i0}$ , (b) mode amplitude, (c) Doppler-shifted mode frequency, and (d) transverse electric field strength plotted as a function of the normalized ion–neutral collision frequency. A transition from a wave-heating regime ($\nu_{in}/\Omega_{i}<$0.4) to a Joule-heating regime ($\nu_{in}/\Omega_{i}>0.7$) is observed as the ion–neutral collision frequency is increased.  Reproduced from Figure 3 of Amatucci \textit{et~al.} \cite{Amatucci:1999}.}
\end{figure}

Using a segmented disc electrode, shown in Fig. \ref{fig:SegmentedVsRings}, in the West Virginia University Q-machine Amatucci \textit{et~al.} \cite{Amatucci:1994} showed that sub-threshold field-aligned current could support the ion cyclotron instability if a radially localized static electric field produced by biasing the segments is introduced (see Fig. \ref{fig:sub-threshold}).   This explained the observation of ion cyclotron waves for sub threshold currents in the auroral region noted by Kintner \cite{Kintner:1992}.  It was not possible to eliminate the axial current totally in the experiment because the inner segment of the electrode was biased and drew electrons. Subsequently, Amatucci \textit{et~al.} \cite{Amatucci:1998} demonstrated that by increasing the magnitude of the transverse electric field and virtually eliminating the axial current with biased ring electrodes (Fig. \ref{fig:SegmentedVsRings}), the electrostatic ion cyclotron waves could be sustained by a sheared transverse flow alone.  These experiments were later followed up by Tejero \textit{et~al.} \cite{Tejero:2011} to confirm the electromagnetic IEDDI \cite{Penano:1999prl}.  These waves, besides validating the theory, were shown to be efficient in ion heating \cite{Amatucci:1998} as was expected \cite{Ganguli:1985b}.   The experiment also showed that the heating profile was distinct from the typical Joule heating \cite{Amatucci:1999} as shown in Fig. \ref{fig:Amatucci1999-Fig3}.  The scale size of the electric field, $L$, was greater than the ion gyroradius, $\rho_{i}$, for these experiments.  

Characterization of the IEDDI in the laboratory was a significant contribution because it clarified the role of localized electric fields in wave generation thereby validating the theory for the origin of these waves and led to numerous applications to understand satellite observations \cite{Bonnell:1996,Liu2004,Golovchanskaya:2014a,Golovchanskaya:2014b}.  In addition, it successfully addressed a major issue in space plasmas, \textit{i.e.}, ion heating in the lower ionosphere necessary to initiate the out flow of the heavy gravitationally bound oxygen ions observed deep inside the magnetosphere \cite{Pollock:1990}.  These experiments became anchors for a comprehensive ionospheric heating model \cite{Ganguli:1994a} and inspired sounding rocket experiments to look for corroborating signatures in the ionosphere \cite{Earle:1989,Bonnell:1996,Bonnell:1997,Lundberg:2012}.  Subsequently, a comprehensive statistical survey of satellite data confirmed the importance of static transverse electric fields to wave generation in the ionosphere \cite{Hamrin:2001}.   More importantly, these early laboratory experiments started a trend in simulating space plasma phenomena in the controlled environment of the laboratory for detailed characterization that helped in the interpretation of \textit{in situ} data and develop a deeper understanding of the salient physics.

\subsection{Low Frequency Limit: Parallel Velocity Gradient}

Another intriguing issue in the ionosphere was the observations of low frequency ion acoustic-like waves \cite{Wahlund:1994} in the nearly isothermal ionosphere where the ion acoustic waves are expected to be ion Landau damped.  The origin of these low frequency waves became a much-debated issue.   As discussed in Sec. \ref{subsec:stability_due_to_par_shear} Gavrishchaka \textit{et~al.} \cite{Gavrishchaka:1998} showed that a spatial gradient in the magnetic field aligned flow could drastically lower the threshold of the ion acoustic waves by moving the phase speed of the waves away from Landau resonance.  In addition, Gavrishchaka \textit{et~al.} \cite{Gavrishchaka:2000} also showed that higher frequency waves can be triggered by spatial gradients in the parallel flow with multi-harmonic ion cyclotron emissions.  The magnitude of the gradient required for generating either of these waves was very modest.  These results could potentially explain a number of auroral observations \cite{Gavrishchaka:1999,Ganguli:2002} including the NASA FAST satellite observation of multi-ion harmonic spectrum and spiky parallel electric field structures \cite{Ergun:1998}.  Thus, validation of the Gavrishchaka \textit{et~al.} theory in the laboratory became an important issue.

In a series of Q-machine experiments with inhomogeneous magnetic field aligned flow at the University of Iowa \cite{Agrimson:2001,Agrimson:2002} and West Virginia University \cite{Teodorescu:2002a,Teodorescu:2002b} the existence of both the shear modified low frequency and the ion cyclotron frequency range fluctuations were confirmed and their signatures and properties were studied.  The experiments highlighted the critical role of the spatial gradient in the flow parallel to the magnetic field.  A similar situation can also arise in compressed plasmas in DFs as well as the plasma sheet-lobe interface, as discussed in Sec. \ref{sec:vpequi_curv}.  The laboratory validation of the theory and the characterization of the instability increased the confidence in its application to other regions of space plasmas \cite{Nykyri:2006,Slapak:2017}.

Other low frequency waves due to parallel inhomogeneous flows with a density gradient were investigated in laboratory experiments by Kaneko \textit{et~al.} \cite{Kaneko:2003,Kaneko:2005}.   They also theoretically analyzed the case and showed that drift waves can be both destabilized and stabilized by velocity shear in the parallel ion flow depending on the plasma conditions and shear strength in the parallel flow.  Similar conclusions regarding the drift wave behavior in plasma with perpendicular flow shear was discussed by Gavrishachaka \textit{et~al.}, \cite{Gavrishchaka:1996phd}.

\subsection{Intermediate Frequency Limit: Transverse Velocity Gradient}

As described in Sec. 2.1, during geomagnetically active periods, global compression of the magnetosphere by the solar wind stretches the Earth’s magnetotail and a pressure gradient builds up between the low-pressure lobe and the high-pressure plasmasheet.  The boundary between these regions exhibits a complex structure, which includes thin layers of energetic electrons confined to the outermost region of the plasmasheet \cite{Forbes:1981,Parks:1984}. Localized static electric fields in the north-south direction are observed during crossings into the plasma sheet from the lobes \cite{Cattell:1982,Orsini:1984} but their cause and effect was not known.  Also, enhanced electrostatic and electromagnetic wave activity is detected at the boundary layer \cite{Grabbe:1984,Parks:1984,Cattell:1986,Angelopoulos:1989}.

To understand the plasma sheet-lobe equilibrium properties, a kinetic description of the boundary layer was developed by Romero \textit{et~al.} \cite{Romero:1990fs}, as described in Sec. \ref{subsec:esderivoff}.  It showed that with increasing activity level, as the boundary layer scale size approaches an ion gyrodiameter, an ambipolar electric field develops across the magnetic field, which intensifies with the global compression.  As shown in Sec. \ref{sec:lineartheory}, for small enough $L$, ions effectively behave as an unmagnetized species for intermediate scales ($\Omega_{i}<\omega<\Omega_{e}$ and $k_{\perp}\rho_{i}>1>k_{\perp}\rho_{e}$) and an instability appears around the lower hybrid frequency.  The wavelength of this instability scales as $k_{\perp}L\sim 1$ where $L\gg\rho_e$ \cite{Ganguli:1988eb}, which distinguishes it from the lower-hybrid-drift instability with $k_{\perp}\rho_{e}\sim 1$ scaling.  Hence, laboratory validation and characterization of the EIH waves discussed in Section 3.3.2 became an important topic.

\begin{figure}
	\includegraphics[width=\textwidth]{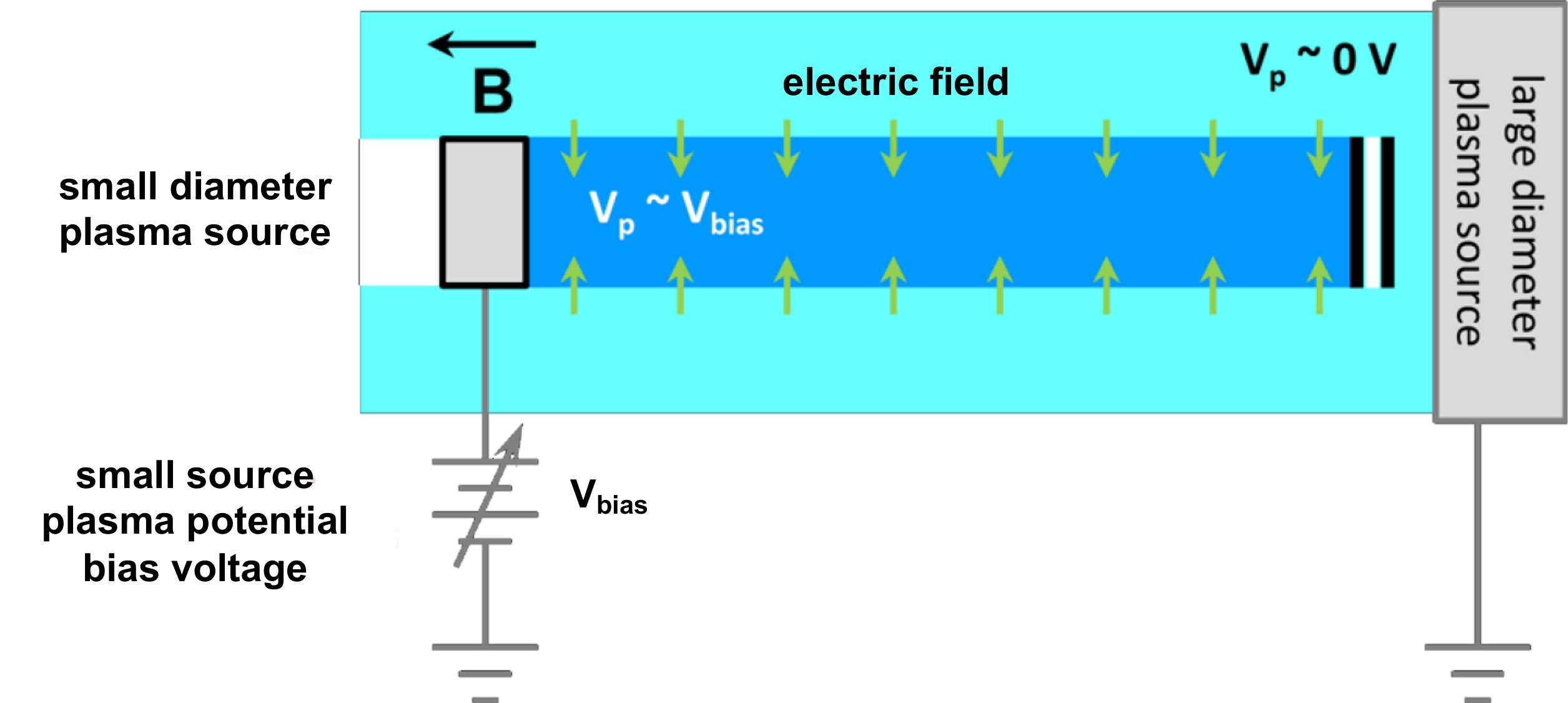}
	\includegraphics[width=0.6\textwidth]{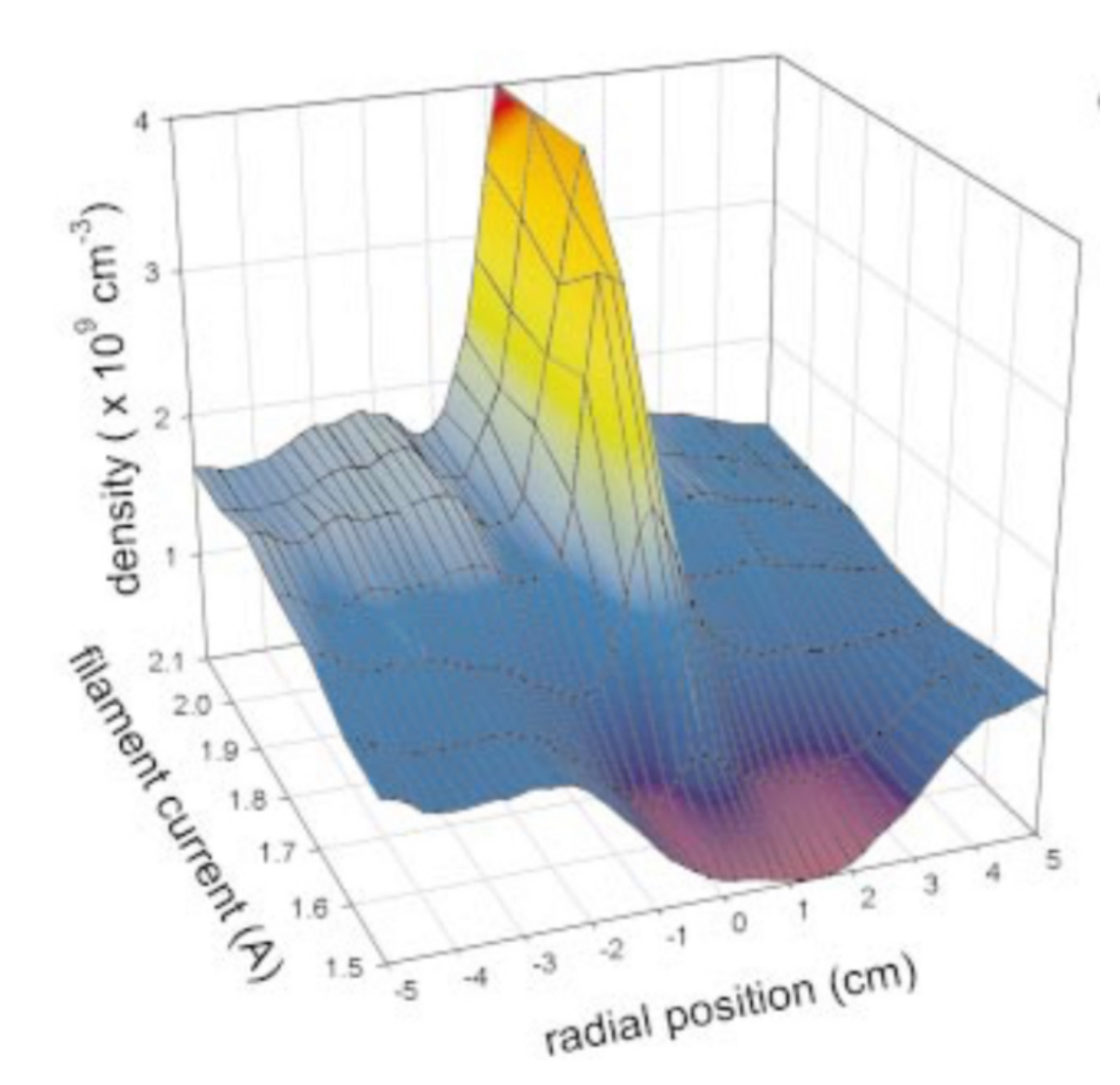}
	\caption{\label{fig:Amatucci2003_Fig3} (top)Schematic of creating localized electric fields in laboratory experiments adapted from Amatucci \textit{et~al.} \cite{Amatucci:1994}.  On the right is a large plasma source.  In front (to the left in the figure) of the large plasma source is a blocking disk that prevents plasma from the large source to stream down the center of the chamber.  On the left is a smaller source that can fill in plasma at the center.  By biasing the end plates an electric field can be created between the two plasmas.  (bottom) Measured density vs radial position in the NRL Space Physics Simulation Chamber for different filament current settings on the plasma source illustrating the experimental control over the plasma density.  (bottom) Reproduced from Fig. 3 of Amatucci \textit{et~al.}\cite{Amatucci:2003}}
\end{figure}

\begin{figure}
	\includegraphics[width=0.6\textwidth]{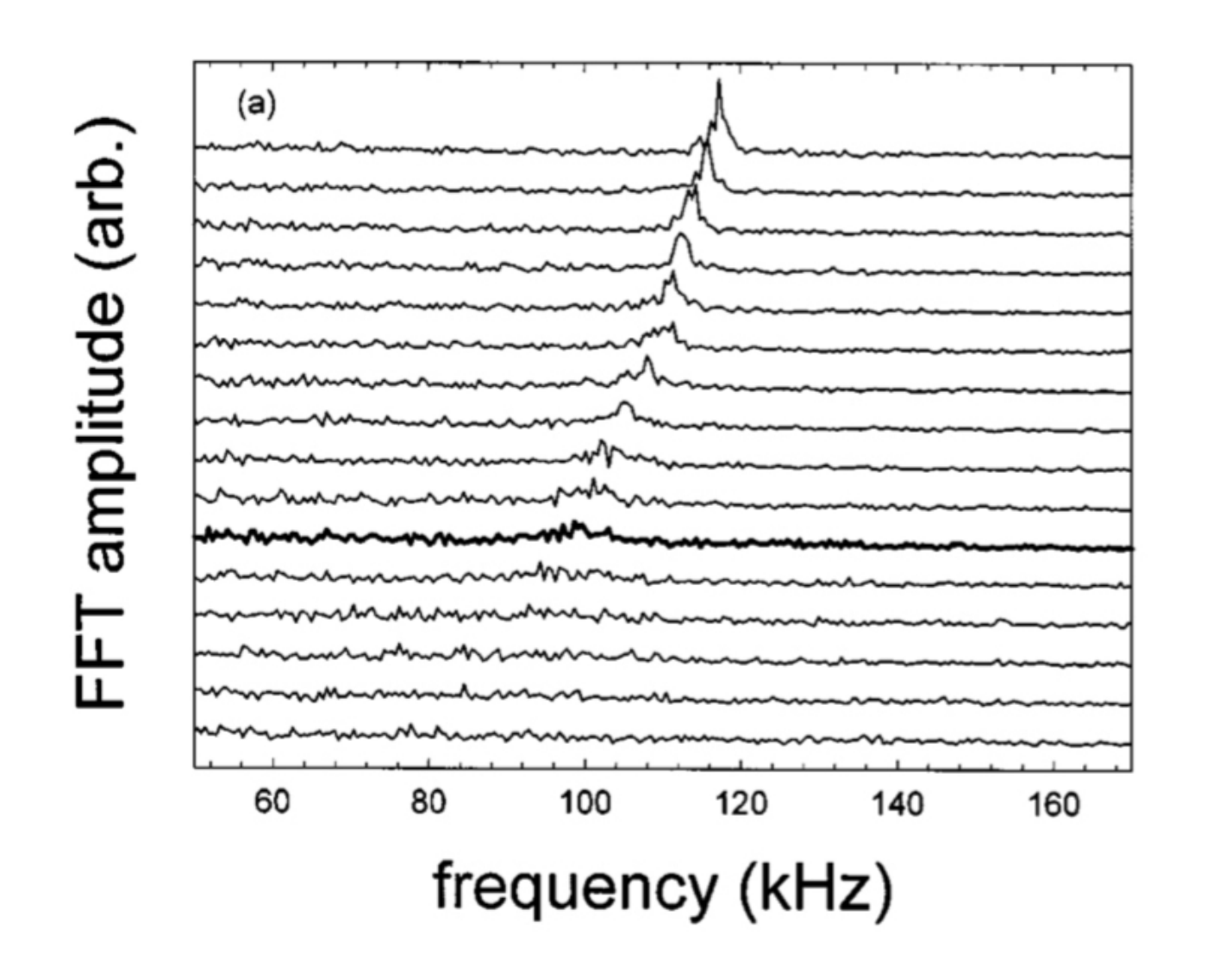}
	\caption{\label{fig:Amatucci2003_Fig8a} Stack plot of the FFT Amplitude vs Frequency as the electric bias is increased (up in the figure) showing that the EIH wave power increases as the applied electric field is increased.  Reproduced from Fig. 8a of Amatucci \textit{et~al.} \cite{Amatucci:2003}}
\end{figure}

While the basic physics of the EIH instability was verified in Japan by Matsubrara and Tanikawa \cite{Matsubara:2000} using a segmented end plate to create the localized radial electric field and then in India by Santhosh Kumar \textit{et~al.} \cite{Kumar:2002}, their experimental geometry did not correspond to the reality of the lobe-plasma sheet system.  The challenge was to produce the conditions of a stretched magnetotail in the lab where the dense plasma sheet is surrounded by tenuous lobe plasma as shown in Fig. 1a of Section \ref{subsec:vlasovpoisson}.  Amatucci \textit{et~al.} \cite{Amatucci:2003} introduced an innovative way to achieve this by using interpenetrating plasmas produced by independent sources with controllable plasma potentials and densities sketched in Fig. \ref{fig:Amatucci2003_Fig3}.  This set up was more representative of the realistic plasma sheet-lobe configuration with a boundary layer of scale size on the order or less than an ion gyroradius. The experiment demonstrated spontaneous generation of lower hybrid waves as shown in Fig. \ref{fig:Amatucci2003_Fig8a}. 

\begin{figure}
	\includegraphics[width=\textwidth]{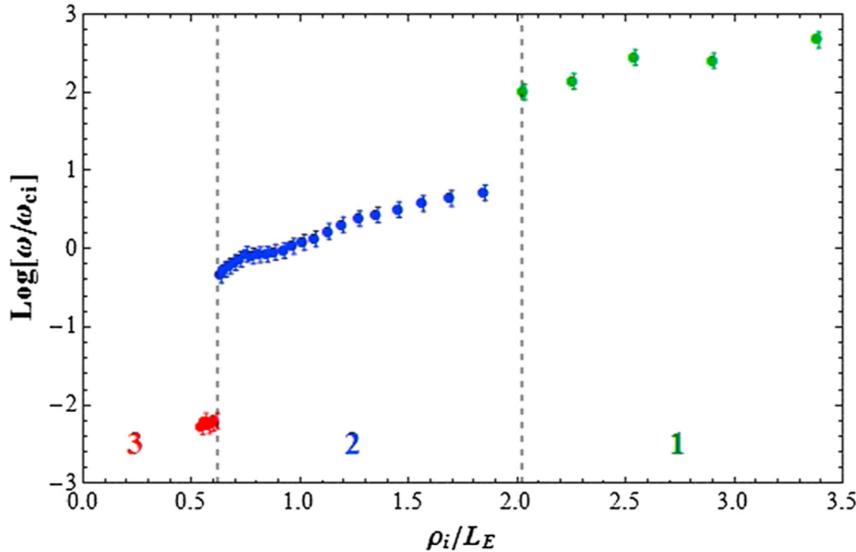}
	\caption{\label{fig:dubois2014_fig5}The log of $\omega/\Omega_{i}$ is plotted as a function of the ratio $\rho_{i}/L$ which was varied experimentally by controlling the magnitude of the magnetic field in ALEXIS.  Reproduced from Fig. 5 of DuBois \textit{et~al.} \cite{DuBois:2014}}
\end{figure}

Subsequently, DuBois \textit{et~al.} \cite{DuBois:2013,DuBois:2014} used the Amatucci method in the Auburn University Auburn Linear Experiment for Instability Studies (ALEXIS) device and varied the magnetic field to scale the ion gyroradius from larger to smaller than the electric field scale size thereby effectively simulating the variation of stress that characterizes the relaxation phase of a stressed magnetotail.  This showed the generation of a broadband emission starting from the lower hybrid frequency to less than ion cyclotron frequency differing by 5 orders of magnitudes in a single experiment as shown in Fig. \ref{fig:dubois2014_fig5}.  

The DuBois \textit{et~al.} experiment was a proof of principle of the theory \cite{Ganguli:94b} which had posited that a compressed boundary layer can relax through the emission of a hierarchy of electric field-driven waves starting from above the electron gyrofrequency to much below the ion gyrofrequency and could be the primary source for the observed broadband electrostatic noise.  Tejero \textit{et~al.} \cite{Tejero:2011} and Enloe \textit{et~al.} \cite{Enloe:2017} have subsequently shown that the plasma compression can also produce electromagnetic emissions but the wave power is primarily concentrated in the electrostatic regime \cite{Ganguli:2014ks}, consistent with the \textit{in~situ} observations \cite{Angelopoulos:1989}.   These laboratory experiments have elucidated the subtler aspects of the magnetotail dynamics, which would be difficult to discern from \textit{in~situ} measurements alone.  They also inspired new experimental research in the laboratory to understand the physics of the dipolarization fronts.

\section{Comprehensive Modeling of Space Plasma Environment}
\label{sec:ml}
Besides academic interests, the practical goal of developing a deeper understanding of space plasma processes is to improve the accuracy of space weather forecasting.  The challenge in a physics-based forecasting model is in accounting for the physics at multiple scales in a global model.  As discussed in this article, spatiotemporal processes in the space plasma environment are multi-scale.  It is not feasible to model the wide range of scales from first principles, because of computational limitations and lack of detailed initial and or boundary conditions.  Hence, success of simulations, forecasting, and interpretation of multi-scale spatiotemporal dynamics critically depends on a realistic formulation including the coupling of physical models describing processes on micro- and macro scales.  Small-scale kinetic processes could significantly influence larger-scale dynamics. However, introduction of small-scale kinetic effects as anomalous coefficients into larger-scale fluid simulations without running small-scale simulations involves empirical adjustments of coupling parameters taking into account simulation stability and other considerations. Some attempts in magnetosphere-ionosphere coupling has been made based on this concept \cite{GanguliPalm:1987,Gangulietal:1988}.  Similarly, one can use coarse-grain analogue models with just a few main elements \cite{Sharma:1995,Klimas:1996} whose characteristics are also inferred from deeper multi-scale physical models. Still such physics-based models may not be accurate enough for certain practical applications.  Recent developments in artificial intelligence (AI) and machine learning (ML) offers a new vista for deeper understanding and forecasting in the space plasma environment.

Alternatively, applied modeling of a wide range of complex systems including space weather forecasting are based on data-driven statistical and ML approaches \cite{Gleisner:1996,Gavrishchaka:2001a,Gavrishchaka:2001b,Camporeale:2019,Gopinath:2019}. Such empirical approaches could offer practical solutions with good accuracy given enough training data covering key regimes of the considered systems are available. However, performance of standard ML approaches could quickly deteriorate with severe data limitations, high dimensionality and non-stationarity \cite{Gavrishchaka:2018,Gavrishchaka:2019}. Domain-expert knowledge including physical models based on deeper understanding of the considered complex system, such as the kinetic processes discussed in this article, could play a key role in applications with severe incompleteness of training data because of natural dimensionality reduction and usage of domain-specific constraints \cite{Gavrishchaka:2018,Gavrishchaka:2019,BANERJEE20072071}. Typical practical example of the domain knowledge incorporation into ML solution is selection of model inputs and drivers using physics-based considerations \cite{Gleisner:1996,Gavrishchaka:2001a,Gavrishchaka:2001b,Gavrishchaka:2019}.  This procedure of augmenting purely data driven models with physics based models is a step towards gaining physical insight into the system.  

The most successful modern ML frameworks such as deep learning (DL) based on deep neural networks (DNNs) and boosting-based ensemble learning offer even more opportunities for efficient synergetic combination with domain-expert knowledge \cite{LeCun:2015,DengDong:2014,Hinton504,Schapire:1992,friedman2000,ChenGuestrin:2016,Gavrishchaka:2018,Gavrishchaka:2019}. First, similar to natural sciences, both techniques actively use advantages of hierarchical data and knowledge representations that are capable of crucial reduction of dependency on the training data size. This is achieved by layer-by-layer learning with automated hierarchical feature discovery and dimensionality reduction in DNNs and the intrinsically hierarchical nature of boosting algorithms where it builds a global-scale model at the first iteration and focuses on more detailed modeling of sub-populations, sub-scales and sub-regimes in subsequent iterations \cite{LeCun:2015,DengDong:2014,Hinton504,Schapire:1992,friedman2000,Gavrishchaka:2006,Gavrishchaka:2018,Gavrishchaka:2019}.  For example, in Section 2 we showed that global compression leads to ambipolar effects on ion and electron gyroscales that generate spatially localized transverse electric fields.  In Section 3 we showed the linear plasma response to such electric fields, which are much smaller scale features.  In Section 4 we showed the nonlinear evolution of these electric fields and ultimately their saturation to generate macroscopic measurable features of the larger scale dynamics that satellites measure.  These micro-macro coupling processes could be iteratively incorporated into global models to produce a much more comprehensive model of the space plasma dynamics than currently possible.  Such hierarchical physics-based knowledge could significantly improve accuracy in space weather forecasting capability.  The described nature of these algorithms creates different channels for efficient integration of many pieces of domain-expert knowledge including physics-based models, scaling and constraints. For example, collection of simplified physical models with a few adjustable empirical parameters, e.g. anomalous coefficients capturing small-scale effects, could be used as base models in boosting algorithms to create ensemble of interpretable models with boosted accuracy and stability compared to a single model \cite{Gavrishchaka:2018,Gavrishchaka:2019,BANERJEE20072071}. Alternatively, simplified physical models capturing multi-scale effects in an approximate manner can be used to generate large amounts of synthetic data for all possible regimes. Later actual data can be augmented by this synthetic data to allow a DL framework to discover robust representations that can be used to train or fine-tune DNNs or other ML models \cite{Gavrishchaka:2019}. Synergetic combination of ML algorithms and physics-based models, such as those discussed in this article and global MHD models, could be especially useful for representation and detection of rare events and regimes \cite{Senyukova:2011,Miao2020}.  Further advancements in discovery of stable and accurate hybrid solutions in complex systems modeling can be achieved by leveraging methods from computational topology which showed promising results in a wide range of applications \cite{Carlsson:2009,Edelsbrunner:2014,GARLAND201649,Miao2020}.  Until such a time when global models can capture detailed physics at all scale sizes, such hybrid modeling may be necessary for accurate space weather forecasting.

\section{Discussion and Conclusions}
\label{sec:discussion}

In this review article we have analyzed the behavior of compressed plasmas in a magnetic field, which is a configuration often encountered both in natural and laboratory plasmas.  Compression creates stress, or gradients, in the background plasma parameters.  When the scale size of the gradient across the magnetic field becomes comparable to an ion gyrodiameter a self-consistent static electric field is generated due to ambipolar kinetic effects.  This electric field is highly inhomogeneous.  Hence, the localized Doppler shift due to the $\mathbf{E}\times\mathbf{B}$ flow cannot be transformed away, which affects the dieletric properties of the plasma including the normal modes.  In addition, it affects the individual particle orbits as well as shears the mean flow velocities both transverse and along the magnetic field.  Velocity shear is a source of free energy for plasma fluctuations.  Consequently, a compressed plasma system achieves a higher energy state compared to its relaxed counterpart.  The electric field gradient, and by causality the velocity shear, that develops scales with the magnitude of compression.  Thermodynamic properties compel the plasma to seek a lower energy state.  In response, in a collisionless medium spontaneous generation of emissions follow that dissipate the velocity shear and returns the plasma to a relaxed lower energy state.  This makes compressed plasmas to be active regions with characteristic emissions.  In the space environment, these regions are relatively easy to detect and measure due to large plasma fluctuations.  The spectral signature of the emissions is typically found to be broadband in frequency with power mostly concentrated in the electrostatic regime.  Hence, they have often been referred to as the broadband electrostatic noise (BEN) in the literature.  But they are also accompanied by some electromagnetic component \cite{Angelopoulos:1989}.  As we discussed in Secs. 3 and 4, the velocity shear has the unique ability to produce such broadband signatures in which the power is mostly in the electrostatic regime but with some electromagnetic power as well.   The intensity and bandwidth of the emissions, which scale as the velocity shear, is a diagnostic of the level of compression imposed on the plasma.  This is evident from \textit{in~situ} measurements in space plasmas where broadband emissions are a hallmark of compressed plasmas found in boundary layers.


Although we used the framework provided here to analyze natural plasma processes, it is general and applicable to laboratory experiments as well as to active experiments in space.  For example, compressed plasma layers can be generated locally in the ionosphere by the ionization of exhausts or effluents discharged from rockets \cite{Bernhardt:1995} or by active chemical release experiments \cite{Ganguli:1992,Scales:1992}.  In the NASA sponsored Nickel Carbonyl Release Experiment (NICARE) \cite{Argo:1992} the introduction of electron capturing agents, such as CF3Br, SF6, Ni(CO)4, \textit{etc.}, in the ionosphere created an electron depleted region in the ionosphere surrounded by natural oxygen-electron plasma.  This generated a boundary layer of positive ions, negative ions, and electron plasma with strong spatial gradients in their densities. Experimental data indicated a large enhancement of noise level concurrent with the formation of the negative ion plasma.  This resulted in a situation similar to the natural boundary layers, discussed in Sec. 2, in which the negative ion population inside the electron-depleted region diminished to zero outside, while the electron population did the opposite in a narrow boundary layer \cite{Ganguli:1992}.  Quasi-neutrality between the electron, negative ions, and positive oxygen ions led to a strong self-consistent electrostatic potential in the boundary layer that separated the negative ion plasma from the ambient oxygen-electron plasma.  Hybrid simulations showed the formation of the boundary layer with a localized radial electric field in the intermediate ($\rho_{i}>L>\rho_{e}$) scale size and spontaneous generation of shear driven EIH waves that relaxed the boundary layer \cite{Scales:1994,Scales:1995}.

Laboratory experiments of plasma expansion due to laser ablation, in which the laser front acts as a piston to compresses the plasma, shows interesting similarity with the physics of the dipolarization fronts we discussed in Sec. 2.  Dipolarization fronts, characterized by a pressure gradient over a narrow plasma layer comparable to an ion gyroradius, are created in the aftermath of magnetic reconnection when a stretched magnetic field snaps back towards a dipolar configuration.  In a laser ablated plasma expansion across an external magnetic field similar density gradient structures with scale size comparable to an ion gyroradius accompanied with a cross-magnetic field flow are observed \cite{Mostovych:1989}.  Due to the piston-like action of the laser front both ions and electrons move with nearly the same speed across the magnetic field and hence the cross field current is negligible but there is a gradient in the intermediate scale size in the plasma flows that are generated.  Furthermore, as in the dipolarization front, waves around the lower hybrid frequency are seen, which were thought to be the lower hybrid drift waves \cite{Krall:1971} because of their association with the density gradient just as in the dipolarization front case.  However, the wavelength of the lower hybrid waves was found to be much longer than the electron gyroradius and comparable to the scale size of the cross-field flow.  As we discussed in section 3, the long wavelength signature is not consistent with the lower hybrid drift waves but similar to that expected from the EIH waves, which depend on the gradient in the flow and not on a cross-field current.  Long wavelengths are generated by nonlinear vortex merging (see Sec. 4). Peyser \textit{et~al.} \cite{Peyser:1992} analyzed a number of experimental cases and compared the data with theoretical models.  They concluded that the waves were likely to be the EIH waves; for similar reasons argued for the origin of the emissions in a dipolarization front in Secs 3 and 4.  However, due to the inability in the experiment to measure the details of the parameters, unambiguous characterization of the origin of the waves in laser ablated plasma jets was not possible.  More recent laser ablation experiments have shown the generation of waves around the lower hybrid frequency \cite{Niemann:2013} and their origin is still an open issue.

While the plasma response to velocity shears in both perpendicular and parallel flows has been studied separately their combined effect has not been analyzed.  In nature it is likely that that the velocity shear is in an arbitrary direction due to magnetic field geometry.  In Sec. 2.2.3 we showed in a simple case how this may be possible.  But in that case the scale size of the magnetic field variation was orders of magnitude larger than the electric field variation, which allowed us to cleanly separate the two scale sizes and study them individually.  Effectively, this reduced the problem to one dimension.  This may not always be possible in other instances in nature or in laboratory.  In general, the linear response will involve two or three dimensional eigenvalue conditions, which are more difficult to solve. There have been some attempts to address the combined effect of parallel and transverse velocity shear [\textit{e.g.}, Kaneko \textit{et~al.} \cite{Kaneko:2007}] but this topic remains an interesting area of research and deserves further attention.  In addition, manifestation of the velocity shear effect in a multi-species plasma, which is likely to prevail in some regions in space, is another interesting future research topic since shear effect is mass dependent and hence affects different species differently, which introduces relative differences in properties between species \cite{Gavrishchaka:1997}.

A common feature in the nonlinear evolution of a compressed plasma system is that spontaneous generation of shear driven waves relaxes the velocity gradient generated by the compression so that a balance can be achieved.  This balance, or the steady state, defines the electromagnetic plasma environment.  In addition, the shear driven waves contribute to viscosity and resistivity as feedback to the global physics and modify the meso scale plasma features.  Thus, the union of the small and large scale physics is the reality that a satellite measures, which underscores the importance of understanding both the small and large scale processes and the coupling between them as we have attempted to show through natural examples in the earth’s neighborhood plasma environment.

\begin{acknowledgements}
This work was partially supported by the Naval Research Laboratory base program and NASA grant NNH17AE70I.  Special thanks to Valeriy Gavrishchaka for reading the manuscript and for valuable discussions.
\end{acknowledgements}

%
%

\bibliographystyle{spphys}       
\bibliography{RMPP_References.bib}   

\begin{thebibliography}{100}
\providecommand{\url}[1]{{#1}}
\providecommand{\urlprefix}{URL }
\expandafter\ifx\csname urlstyle\endcsname\relax
  \providecommand{\doi}[1]{DOI \discretionary{}{}{}#1}\else
  \providecommand{\doi}{DOI \discretionary{}{}{}\begingroup
  \urlstyle{rm}\Url}\fi

\bibitem{Burch:2016fu}
J.L. Burch, T.E. Moore, R.B. Torbert, B.L. Giles, Space Science Reviews
  \textbf{199}(1-4), 5  (2016).
\newblock \doi{10.1007/s11214-015-0164-9}

\bibitem{Angelopoulos2008}
V.~Angelopoulos, Space Science Reviews \textbf{141}(1), 5 (2008).
\newblock \doi{10.1007/s11214-008-9336-1}

\bibitem{Escoubet:1997}
C.~Escoubet, R.~Schmidt, M.~Goldstein, Space Science Reviews \textbf{79}(1-2),
  11 (1997).
\newblock \doi{10.1023/a:1004923124586}

\bibitem{GORDEEV1994215}
A.~Gordeev, A.~Kingsep, L.~Rudakov, Physics Reports \textbf{243}(5), 215
  (1994).
\newblock \doi{https://doi.org/10.1016/0370-1573(94)90097-3}

\bibitem{Fu:2012fv}
H.S. Fu, Y.V. Khotyaintsev, A.~Vaivads, M.~André, S.Y. Huang,  \textbf{39}(6),
  L06105 (2012).
\newblock \doi{10.1029/2012gl051274}

\bibitem{Bernstein:1957hx}
I.B. Bernstein, J.M. Greene, M.D. Kruskal, Physical Review \textbf{108}(3), 546
   (1957).
\newblock \doi{10.1103/physrev.108.546}

\bibitem{Grad:1958}
H.~Grad, H.~Rubin, {Hydromagnetic equilibria and force-free fields}.
\newblock Tech. rep., Proceedings of the 2nd UN Conf. on the Peaceful Uses of
  Atomic Energy (1958)

\bibitem{Shafranov1966}
V.D. Shafranov, in \emph{Reviews of Plasma Physics}, vol.~2 (1966), p. 103

\bibitem{Sestero:1964}
A.~Sestero, The Physics of Fluids \textbf{7}(1), 44 (1964).
\newblock \doi{10.1063/1.1711053}.
\newblock \urlprefix\url{https://aip.scitation.org/doi/abs/10.1063/1.1711053}

\bibitem{Romero:1990fs}
H.~Romero, G.~Ganguli, P.~Palmadesso, P.B. Dusenbery, Geophysical Research
  Letters \textbf{17}(13), 2313  (1990).
\newblock \doi{10.1029/gl017i013p02313}

\bibitem{Eastman:1984}
T.E. Eastman, L.A. Frank, W.K. Peterson, W.~Lennartsson, Journal of Geophysical
  Research \textbf{89}(A3), 1553 (1984).
\newblock \doi{10.1029/ja089ia03p01553}

\bibitem{Stern:1991}
D.P. Stern, \emph{The Beginning of Substorm Research} (American Geophysical
  Union (AGU), 2013), pp. 11--14.
\newblock \doi{10.1029/GM064p0011}.
\newblock
  \urlprefix\url{https://agupubs.onlinelibrary.wiley.com/doi/abs/10.1029/GM064p0011}

\bibitem{Lui:1991}
A.T.Y. Lui, \emph{Extended Consideration of a Synthesis Model for
  Magnetospheric Substorms} (American Geophysical Union (AGU), 2013), pp.
  43--60.
\newblock \doi{10.1029/GM064p0043}.
\newblock
  \urlprefix\url{https://agupubs.onlinelibrary.wiley.com/doi/abs/10.1029/GM064p0043}

\bibitem{Grabbe:1984}
C.L. Grabbe, T.E. Eastman, Journal of Geophysical Research: Space Physics
  \textbf{89}(A6), 3865 (1984).
\newblock \doi{10.1029/JA089iA06p03865}.
\newblock
  \urlprefix\url{https://agupubs.onlinelibrary.wiley.com/doi/abs/10.1029/JA089iA06p03865}

\bibitem{Ganguli:94b}
G.~Ganguli, H.~Romero, J.~Fedder, \emph{{Interaction Between Global MHD and
  Kinetic Processes in the Magneotail}} (American Geophysical Union (AGU),
  1994), pp. 135--148.
\newblock \doi{10.1029/GM084p0135}.
\newblock
  \urlprefix\url{https://agupubs.onlinelibrary.wiley.com/doi/abs/10.1029/GM084p0135}

\bibitem{Takahashi:1988}
K.~Takahashi, E.W. Hones~Jr., Journal of Geophysical Research: Space Physics
  \textbf{93}(A8), 8558 (1988).
\newblock \doi{10.1029/JA093iA08p08558}.
\newblock
  \urlprefix\url{https://agupubs.onlinelibrary.wiley.com/doi/abs/10.1029/JA093iA08p08558}

\bibitem{Rudakov:1961}
L.I. Rudakov, R.Z. Sagdeev, Dokl. Akad. Nauk. SSR \textbf{138}, 581 (1961)

\bibitem{Pogutse:1967}
O.P. Pogutse, Zh. Eksp. Teor. Fiz. \textbf{52}, 759 (1967)

\bibitem{Coppi:1967}
B.~Coppi, M.N. Rosenbluth, R.Z. Sagdeev, The Physics of Fluids \textbf{10}(3),
  582 (1967).
\newblock \doi{10.1063/1.1762151}.
\newblock \urlprefix\url{https://aip.scitation.org/doi/abs/10.1063/1.1762151}

\bibitem{Nakamura:2002}
R.~Nakamura, W.~Baumjohann, B.~Klecker, Y.~Bogdanova, A.~Balogh, H.~Rème, J.M.
  Bosqued, I.~Dandouras, J.A. Sauvaud, K.H. Glassmeier, L.~Kistler, C.~Mouikis,
  T.L. Zhang, H.~Eichelberger, A.~Runov, Geophysical Research Letters
  \textbf{29}(20), 3 (2002).
\newblock \doi{10.1029/2002GL015763}.
\newblock
  \urlprefix\url{https://agupubs.onlinelibrary.wiley.com/doi/abs/10.1029/2002GL015763}

\bibitem{Nakamura:2009}
R.~Nakamura, A.~Retin\`o, W.~Baumjohann, M.~Volwerk, N.~Erkaev, B.~Klecker,
  E.A. Lucek, I.~Dandouras, M.~Andr\'e, Y.~Khotyaintsev, Annales Geophysicae
  \textbf{27}(4), 1743 (2009).
\newblock \doi{10.5194/angeo-27-1743-2009}.
\newblock \urlprefix\url{https://www.ann-geophys.net/27/1743/2009/}

\bibitem{Runov:2009hl}
A.~Runov, V.~Angelopoulos, M.I. Sitnov, V.A. Sergeev, J.~Bonnell, J.P.
  McFadden, D.~Larson, K.H. Glassmeier, U.~Auster, Geophysical Research Letters
  \textbf{36}(14), 5 (2009).
\newblock \doi{10.1029/2009gl038980}

\bibitem{Deng:2010kg}
X.~Deng, M.~Ashour-Abdalla, M.~Zhou, R.~Walker, M.~El-Alaoui, V.~Angelopoulos,
  R.E. Ergun, D.~Schriver, Journal of Geophysical Research: Space Physics
  \textbf{115}(A9), A09225 (2010).
\newblock \doi{10.1029/2009ja015107}

\bibitem{Angelopoulos:1992}
V.~Angelopoulos, W.~Baumjohann, C.F. Kennel, F.V. Coroniti, M.G. Kivelson,
  R.~Pellat, R.J. Walker, H.~Lühr, G.~Paschmann, Journal of Geophysical
  Research: Space Physics \textbf{97}(A4), 4027 (1992).
\newblock \doi{10.1029/91JA02701}.
\newblock
  \urlprefix\url{https://agupubs.onlinelibrary.wiley.com/doi/abs/10.1029/91JA02701}

\bibitem{ChenWolf:1993}
C.X. Chen, R.A. Wolf, Journal of Geophysical Research: Space Physics
  \textbf{98}(A12), 21409 (1993).
\newblock \doi{10.1029/93JA02080}.
\newblock
  \urlprefix\url{https://agupubs.onlinelibrary.wiley.com/doi/abs/10.1029/93JA02080}

\bibitem{Fletcher:2019kq}
A.C. Fletcher, C.~Crabtree, G.~Ganguli, D.~Malaspina, E.~Tejero, X.~Chu,
  Journal of Geophysical Research: Space Physics \textbf{0}(ja) (2019).
\newblock \doi{10.1029/2018ja026433}

\bibitem{Ganguli:2018vf}
G.~Ganguli, C.~Crabtree, A.C. Fletcher, E.~Tejero, D.~Malaspina, I.~Cohen,
  Scientific reports \textbf{8}(1), 17186 (2018).
\newblock \doi{10.1038/s41598-018-35349-9}

\bibitem{Schindler:1993}
K.~Schindler, J.~Birn, Journal of Geophysical Research: Space Physics
  \textbf{98}(A9), 15477 (1993).
\newblock \doi{10.1029/93JA01047}.
\newblock
  \urlprefix\url{https://agupubs.onlinelibrary.wiley.com/doi/abs/10.1029/93JA01047}

\bibitem{Sitnov:2006}
M.I. Sitnov, M.~Swisdak, P.N. Guzdar, A.~Runov, Journal of Geophysical
  Research: Space Physics \textbf{111}(A8), A08204 (2006).
\newblock \doi{10.1029/2005JA011517}.
\newblock
  \urlprefix\url{https://agupubs.onlinelibrary.wiley.com/doi/abs/10.1029/2005JA011517}

\bibitem{Artemyev:2019}
A.V. Artemyev, V.~Angelopoulos, A.~Runov, A.A. Petrukovich, Journal of
  Geophysical Research: Space Physics \textbf{124}(1), 264 (2019).
\newblock \doi{10.1029/2018JA026113}.
\newblock
  \urlprefix\url{https://agupubs.onlinelibrary.wiley.com/doi/abs/10.1029/2018JA026113}

\bibitem{Crabtree:2020}
C.~Crabtree, G.~Ganguli, A.~Fletcher, A.~Sen, Physics of Plasmas p. to appear
  (2020)

\bibitem{Harris:1962cw}
E.G. Harris, Il Nuovo Cimento (1955-1965) \textbf{23}(1), 115  (1962).
\newblock \doi{10.1007/bf02733547}

\bibitem{McComas:1986}
D.J. McComas, C.T. Russell, R.C. Elphic, S.J. Bame, Journal of Geophysical
  Research: Space Physics \textbf{91}(A4), 4287 (1986).
\newblock \doi{10.1029/JA091iA04p04287}.
\newblock
  \urlprefix\url{https://agupubs.onlinelibrary.wiley.com/doi/abs/10.1029/JA091iA04p04287}

\bibitem{Sergeev:1993}
V.A. Sergeev, D.G. Mitchell, C.T. Russell, D.J. Williams, Journal of
  Geophysical Research: Space Physics \textbf{98}(A10), 17345 (1993).
\newblock \doi{10.1029/93JA01151}.
\newblock
  \urlprefix\url{https://agupubs.onlinelibrary.wiley.com/doi/abs/10.1029/93JA01151}

\bibitem{Sanny:1994}
J.~Sanny, R.L. McPherron, C.T. Russell, D.N. Baker, T.I. Pulkkinen, A.~Nishida,
  Journal of Geophysical Research: Space Physics \textbf{99}(A4), 5805 (1994).
\newblock \doi{10.1029/93JA03235}.
\newblock
  \urlprefix\url{https://agupubs.onlinelibrary.wiley.com/doi/abs/10.1029/93JA03235}

\bibitem{Hoshino:1996}
M.~Hoshino, A.~Nishida, T.~Mukai, Y.~Saito, T.~Yamamoto, S.~Kokubun, Journal of
  Geophysical Research: Space Physics \textbf{101}(A11), 24775 (1996).
\newblock \doi{10.1029/96JA02313}.
\newblock
  \urlprefix\url{https://agupubs.onlinelibrary.wiley.com/doi/abs/10.1029/96JA02313}

\bibitem{Asano:2004}
Y.~Asano, T.~Mukai, M.~Hoshino, Y.~Saito, H.~Hayakawa, T.~Nagai, Journal of
  Geophysical Research: Space Physics \textbf{109}(A2) (2004).
\newblock \doi{10.1029/2003JA010114}.
\newblock
  \urlprefix\url{https://agupubs.onlinelibrary.wiley.com/doi/abs/10.1029/2003JA010114}

\bibitem{Runov:2004}
A.~Runov, V.~Sergeev, R.~Nakamura, W.~Baumjohann, Z.~V\"or\"os, M.~Volwerk,
  Y.~Asano, B.~Klecker, H.~R\`eme, A.~Balogh, Annales Geophysicae
  \textbf{22}(7), 2535 (2004).
\newblock \doi{10.5194/angeo-22-2535-2004}.
\newblock \urlprefix\url{https://www.ann-geophys.net/22/2535/2004/}

\bibitem{Schindler:2008}
K.~Schindler, M.~Hesse, Physics of Plasmas \textbf{15}(4), 042902 (2008).
\newblock \doi{10.1063/1.2907359}

\bibitem{Romero:1993ip}
H.~Romero, G.~Ganguli, Physics of Fluids B: Plasma Physics \textbf{5}(9), 3163
  (1993).
\newblock \doi{10.1063/1.860653}

\bibitem{Speiser:1965}
T.W. Speiser, Journal of Geophysical Research (1896-1977) \textbf{70}(17), 4219
  (1965).
\newblock \doi{10.1029/JZ070i017p04219}.
\newblock
  \urlprefix\url{https://agupubs.onlinelibrary.wiley.com/doi/abs/10.1029/JZ070i017p04219}

\bibitem{Chen:1986}
J.~Chen, P.J. Palmadesso, Journal of Geophysical Research: Space Physics
  \textbf{91}(A2), 1499 (1986).
\newblock \doi{10.1029/JA091iA02p01499}.
\newblock
  \urlprefix\url{https://agupubs.onlinelibrary.wiley.com/doi/abs/10.1029/JA091iA02p01499}

\bibitem{Huba:1980}
J.D. Huba, J.F. Drake, N.T. Gladd, The Physics of Fluids \textbf{23}(3), 552
  (1980).
\newblock \doi{10.1063/1.863003}.
\newblock \urlprefix\url{https://aip.scitation.org/doi/abs/10.1063/1.863003}

\bibitem{Huba:1983}
J.D. Huba, G.~Ganguli, The Physics of Fluids \textbf{26}(1), 124 (1983).
\newblock \doi{10.1063/1.864001}.
\newblock \urlprefix\url{https://aip.scitation.org/doi/abs/10.1063/1.864001}

\bibitem{Daughton:1999ex}
W.~Daughton, Physics of Plasmas \textbf{6}(4), 1329  (1999).
\newblock \doi{10.1063/1.873374}

\bibitem{Tummel:2014}
K.~Tummel, L.~Chen, Z.~Wang, X.Y. Wang, Y.~Lin, Physics of Plasmas
  \textbf{21}(5), 052104 (2014).
\newblock \doi{10.1063/1.4875720}.
\newblock \urlprefix\url{https://doi.org/10.1063/1.4875720}

\bibitem{Ganguli:1988hh}
G.~Ganguli, Y.C. Lee, P.J. Palmadesso, Physics of Fluids \textbf{31}(4), 823
  (1988).
\newblock \doi{10.1063/1.866818}

\bibitem{Chen:1992}
J.~Chen, Journal of Geophysical Research \textbf{97}(A10), 15011 (1992).
\newblock \doi{10.1029/92ja00955}

\bibitem{Gavrishchaka:1996phd}
V.~Gavrishchaka, Collective phenomenon in a magnetized plasma with a
  field-aligned drift and inhomogeneous transverse flow.
\newblock Ph.D. thesis, West Virginia University (1996)

\bibitem{Mozer1977}
F.S. Mozer, C.W. Carlson, M.K. Hudson, R.B. Torbert, B.~Parady, J.~Yatteau,
  M.C. Kelley, Phys. Rev. Lett. \textbf{38}, 292 (1977).
\newblock \doi{10.1103/PhysRevLett.38.292}.
\newblock \urlprefix\url{https://link.aps.org/doi/10.1103/PhysRevLett.38.292}

\bibitem{Gavrishchaka:1996}
V.~Gavrishchaka, M.E. Koepke, G.~Ganguli, Physics of Plasmas \textbf{3}(8),
  3091 (1996).
\newblock \doi{10.1063/1.871656}

\bibitem{Ganguli:1985a}
G.~Ganguli, Y.C. Lee, P.~Palmadesso, Physics of Fluids \textbf{28}(3), 761
  (1985).
\newblock \doi{10.1063/1.865096}

\bibitem{Penano:1999prl}
J.R. Pe\~{n}ano, G.~Ganguli, Physical Review Letters \textbf{83}(7), 1343
  (1999).
\newblock \doi{10.1103/physrevlett.83.1343}

\bibitem{Penano:2000}
J.R. Pe\~{n}ano, G.~Ganguli, Journal of Geophysical Research: Space Physics
  \textbf{105}(A4), 7441 (2000).
\newblock \doi{10.1029/1999ja000303}

\bibitem{Penano:2002}
J.R. Pe\~{n}ano, G.~Ganguli, Journal of Geophysical Research: Space Physics
  \textbf{107}(A8), SIA 14 (2002).
\newblock \doi{10.1029/2001ja000279}

\bibitem{Raleigh:1896}
L.~Raleigh, \emph{Theory of Sound}, vol.~II (MacMillan, London, 1896)

\bibitem{Drazin:1966}
P.~Drazin, L.~Howard, \emph{Advances in Applied Mechanics}, vol.~7 (Acadmemic,
  New York, 1966)

\bibitem{Keskinen:1988}
M.J. Keskinen, H.G. Mitchell, J.A. Fedder, P.~Satyanarayana, S.T. Zalesak, J.D.
  Huba, Journal of Geophysical Research: Space Physics \textbf{93}(A1), 137
  (1988).
\newblock \doi{10.1029/JA093iA01p00137}.
\newblock
  \urlprefix\url{https://agupubs.onlinelibrary.wiley.com/doi/abs/10.1029/JA093iA01p00137}

\bibitem{Satyanarayana1987}
P.~Satyanarayana, Y.C. Lee, J.D. Huba, The Physics of Fluids \textbf{30}(1), 81
  (1987).
\newblock \doi{10.1063/1.866063}.
\newblock \urlprefix\url{https://aip.scitation.org/doi/abs/10.1063/1.866063}

\bibitem{Ganguli:1997}
G.~Ganguli, Physics of Plasmas \textbf{4}(5), 1544 (1997).
\newblock \doi{10.1063/1.872285}

\bibitem{Ganguli:1988eb}
G.~Ganguli, Y.C. Lee, P.J. Palmadesso, Physics of Fluids \textbf{31}(10), 2753
  (1988).
\newblock \doi{10.1063/1.866982}

\bibitem{Liu:2018}
Y.~Liu, J.~Lei, M.~Li, Y.~Ling, J.~Yuan, Physics of Plasmas \textbf{25}(10),
  102901 (2018).
\newblock \doi{10.1063/1.5051393}.
\newblock \urlprefix\url{https://doi.org/10.1063/1.5051393}

\bibitem{Ilyasov:2015}
A.A. Ilyasov, A.A. Chernyshov, M.M. Mogilevsky, I.V. Golovchanskaya, B.V.
  Kozelov, Physics of Plasmas \textbf{22}(3), 032906 (2015).
\newblock \doi{10.1063/1.4916125}.
\newblock \urlprefix\url{https://doi.org/10.1063/1.4916125}

\bibitem{Finn:2020}
J.M. Finn, A.J. Cole, C.~Cihan, D.~Brennan, Integration of tearing layer
  equations by means of matrix riccati methods (2020)

\bibitem{Fornberg:2011}
B.~Fornberg, J.A.C. Weideman, Journal of Computational Physics
  \textbf{230}(15), 5957 (2011).
\newblock \doi{10.1016/j.jcp.2011.04.007}

\bibitem{Drummond:1962}
W.E. Drummond, M.N. Rosenbluth, Physics of Fluids \textbf{5}(12), 1507 (1962).
\newblock \doi{10.1063/1.1706559}

\bibitem{Palmadesso:1986}
P.~Palmadesso, G.~Ganguli, Y.C. Lee, \emph{A New Mechanism for Excitation of
  Waves in a Magnetoplasma II. Wave-Particle and Nonlinear Aspects} (American
  Geophysical Union (AGU), 1986), pp. 301--306.
\newblock \doi{10.1029/GM038p0301}.
\newblock
  \urlprefix\url{https://agupubs.onlinelibrary.wiley.com/doi/abs/10.1029/GM038p0301}

\bibitem{Nishikawa:1988}
K.I. Nishikawa, G.~Ganguli, Y.C. Lee, P.J. Palmadesso, Physics of Fluids
  \textbf{31}(6), 1568 (1988).
\newblock \doi{10.1063/1.866696}

\bibitem{McBride:1972}
J.B. McBride, E.~Ott, J.P. Boris, J.H. Orens, Physics of Fluids
  \textbf{15}(12), 2367 (1972).
\newblock \doi{10.1063/1.1693881}

\bibitem{Krall:1971}
N.A. Krall, P.C. Liewer, Physical Review A \textbf{4}(5), 2094 (1971).
\newblock \doi{10.1103/physreva.4.2094}

\bibitem{Romero:1994}
H.~Romero, G.~Ganguli, Geophysical Research Letters \textbf{21}(8), 645 (1994).
\newblock \doi{10.1029/93gl03385}

\bibitem{Davidson:1977}
R.C. Davidson, N.T. Gladd, C.S. Wu, J.D. Huba, Physics of Fluids
  \textbf{20}(2), 301 (1977).
\newblock \doi{10.1063/1.861867}

\bibitem{DAngelo1965}
N.~D'Angelo, Physics of Fluids \textbf{8}(9), 1748 (1965).
\newblock \doi{10.1063/1.1761496}

\bibitem{Lakhina:1987}
G.S. Lakhina, Journal of Geophysical Research \textbf{92}(A11), 12161 (1987).
\newblock \doi{10.1029/ja092ia11p12161}

\bibitem{Gavrishchaka:1998}
V.V. Gavrishchaka, S.B. Ganguli, G.I. Ganguli, Physical Review Letters
  \textbf{80}(4), 728 (1998).
\newblock \doi{10.1103/physrevlett.80.728}

\bibitem{Gavrishchaka:2000}
V.V. Gavrishchaka, G.I. Ganguli, W.A. Scales, S.P. Slinker, C.C. Chaston, J.P.
  McFadden, R.E. Ergun, C.W. Carlson, Physical Review Letters \textbf{85}(20),
  4285 (2000).
\newblock \doi{10.1103/physrevlett.85.4285}

\bibitem{Ganguli:2002}
G.~Ganguli, S.~Slinker, V.~Gavrishchaka, W.~Scales, Physics of Plasmas
  \textbf{9}(5), 2321 (2002).
\newblock \doi{10.1063/1.1445181}

\bibitem{Agrimson:2001}
E.~Agrimson, N.~D'Angelo, R.L. Merlino, Physical Review Letters
  \textbf{86}(23), 5282 (2001).
\newblock \doi{10.1103/physrevlett.86.5282}

\bibitem{Agrimson:2002}
E.P. Agrimson, N.~D'Angelo, R.L. Merlino, Physics Letters A \textbf{293}(5-6),
  260 (2002).
\newblock \doi{10.1016/s0375-9601(02)00026-9}

\bibitem{Teodorescu:2002a}
C.~Teodorescu, E.W. Reynolds, M.E. Koepke, Physical Review Letters
  \textbf{88}(18), 185003 (2002).
\newblock \doi{10.1103/physrevlett.88.185003}

\bibitem{Teodorescu:2002b}
C.~Teodorescu, E.W. Reynolds, M.E. Koepke, Physical Review Letters
  \textbf{89}(10), 105001 (2002).
\newblock \doi{10.1103/physrevlett.89.105001}

\bibitem{Kindel:1971}
J.M. Kindel, C.F. Kennel, Journal of Geophysical Research \textbf{76}(13), 3055
  (1971).
\newblock \doi{10.1029/ja076i013p03055}

\bibitem{Catto:1973}
P.J. Catto, M.N. Rosenbluth, C.S. Liu, Physics of Fluids \textbf{16}(10), 1719
  (1973).
\newblock \doi{10.1063/1.1694200}

\bibitem{Huba:1981}
J.D. Huba, Journal of Geophysical Research \textbf{86}(A11), 8991 (1981).
\newblock \doi{10.1029/ja086ia11p08991}

\bibitem{Gary:1981}
S.P. Gary, S.J. Schwartz, Journal of Geophysical Research \textbf{86}(A13),
  11139 (1981).
\newblock \doi{10.1029/ja086ia13p11139}

\bibitem{Gavrishchaka:1999}
V.V. Gavrishchaka, S.B. Ganguli, G.I. Ganguli, Journal of Geophysical Research:
  Space Physics \textbf{104}(A6), 12683 (1999).
\newblock \doi{10.1029/1999ja900094}

\bibitem{Tsang:1987}
K.T. Tsang, B.~Hafizi, Physics of Fluids \textbf{30}(3), 804 (1987).
\newblock \doi{10.1063/1.866331}

\bibitem{Romero:1992ks}
H.~Romero, G.~Ganguli, Y.C. Lee, P.J. Palmadesso, Physics of Fluids B: Plasma
  Physics \textbf{4}(7), 1708  (1992).
\newblock \doi{10.1063/1.860028}

\bibitem{Mikhailovskii:1974}
A.B. Mikhailovskii, \emph{Theory of Plasma Instabilities}, vol.~2 (Consultants
  Bureau, New York, 1974)

\bibitem{Ganguli:1994a}
G.~Ganguli, M.J. Keskinen, H.~Romero, R.~Heelis, T.~Moore, C.~Pollock, Journal
  of Geophysical Research \textbf{99}(A5), 8873 (1994).
\newblock \doi{10.1029/93ja03181}

\bibitem{DuBois:2014}
A.M. DuBois, E.~Thomas, W.E. Amatucci, G.~Ganguli, Journal of Geophysical
  Research: Space Physics \textbf{119}(7), 5624 (2014).
\newblock \doi{10.1002/2014ja020198}

\bibitem{Pritchett:1987}
P.L. Pritchett, Physics of Fluids \textbf{30}(1), 272 (1987).
\newblock \doi{10.1063/1.866187}

\bibitem{Nishikawa:1990}
K.I. Nishikawa, G.~Ganguli, Y.C. Lee, P.J. Palmadesso, Journal of Geophysical
  Research \textbf{95}(A2), 1029 (1990).
\newblock \doi{10.1029/ja095ia02p01029}

\bibitem{Amatucci:1996}
W.E. Amatucci, D.N. Walker, G.~Ganguli, J.A. Antoniades, D.~Duncan, J.H.
  Bowles, V.~Gavrishchaka, M.E. Koepke, Physical Review Letters
  \textbf{77}(10), 1978 (1996).
\newblock \doi{10.1103/physrevlett.77.1978}

\bibitem{Koepke:1994}
M.E. Koepke, W.E. Amatucci, J.J. Carroll, T.E. Sheridan, Physical Review
  Letters \textbf{72}(21), 3355 (1994).
\newblock \doi{10.1103/physrevlett.72.3355}

\bibitem{Hojo:1995}
H.~Hojo, Y.~Kishimoto, J.~Van Dam, Journal of the Physical Society of Japan
  \textbf{64}(11), 4073 (1995).
\newblock \doi{10.1143/jpsj.64.4073}

\bibitem{Kent:1969}
G.I. Kent, N.C. Jen, F.F. Chen, Physics of Fluids \textbf{12}(10), 2140 (1969).
\newblock \doi{10.1063/1.1692323}

\bibitem{Jassby:1970}
D.L. Jassby, Physical Review Letters \textbf{25}(22), 1567 (1970).
\newblock \doi{10.1103/physrevlett.25.1567}

\bibitem{Jassby:1972}
D.L. Jassby, Physics of Fluids \textbf{15}(9), 1590 (1972).
\newblock \doi{10.1063/1.1694135}

\bibitem{Penano:1998}
J.R. Pe\~{n}ano, G.~Ganguli, W.E. Amatucci, D.N. Walker, V.~Gavrishchaka,
  Physics of Plasmas \textbf{5}(12), 4377 (1998).
\newblock \doi{10.1063/1.873175}

\bibitem{Pritchett:1993}
P.L. Pritchett, Physics of Fluids B: Plasma Physics \textbf{5}(10), 3770
  (1993).
\newblock \doi{10.1063/1.860847}

\bibitem{Romero:1992a}
H.~Romero, G.~Ganguli, Y.C. Lee, Physical Review Letters \textbf{69}(24), 3503
  (1992).
\newblock \doi{10.1103/physrevlett.69.3503}

\bibitem{Lin:2019ho}
D.~Lin, W.A. Scales, G.~Ganguli, X.~Fu, C.~Crabtree, E.~Tejero, Y.~Chen, A.C.
  Fletcher, Journal of Geophysical Research: Space Physics \textbf{77}(10),
  1978 (2019).
\newblock \doi{10.1029/2019ja026815}

\bibitem{Kelley:1977}
M.C. Kelley, C.W. Carlson, Journal of Geophysical Research \textbf{82}(16),
  2343 (1977).
\newblock \doi{10.1029/ja082i016p02343}

\bibitem{Kintner:1992}
P.M. Kintner, Physics of Fluids B: Plasma Physics \textbf{4}(7), 2264 (1992).
\newblock \doi{10.1063/1.4729441}

\bibitem{Sato:1986}
N.~Sato, M.~Nakamura, R.~Hatakeyama, Physical Review Letters \textbf{57}(10),
  1227 (1986).
\newblock \doi{10.1103/physrevlett.57.1227}

\bibitem{Alport:1986}
M.J. Alport, S.L. Cartier, R.L. Merlino, Journal of Geophysical Research
  \textbf{91}(A2), 1599 (1986).
\newblock \doi{10.1029/ja091ia02p01599}

\bibitem{Ganguli:1985b}
G.~Ganguli, P.~Palmadesso, Y.C. Lee, Geophysical Research Letters
  \textbf{12}(10), 643 (1985).
\newblock \doi{10.1029/gl012i010p00643}

\bibitem{Amatucci:1994}
W.E. Amatucci, M.E. Koepke, J.J. Carroll, T.E. Sheridan, Geophysical Research
  Letters \textbf{21}(15), 1595 (1994).
\newblock \doi{10.1029/94gl00881}

\bibitem{Amatucci:1998}
W.E. Amatucci, D.N. Walker, G.~Ganguli, D.~Duncan, J.A. Antoniades, J.H.
  Bowles, V.~Gavrishchaka, M.E. Koepke, Journal of Geophysical Research: Space
  Physics \textbf{103}(A6), 11711 (1998).
\newblock \doi{10.1029/98ja00659}

\bibitem{Tejero:2011}
E.M. Tejero, W.E. Amatucci, G.~Ganguli, C.D. Cothran, C.~Crabtree, E.~Thomas,
  Physical Review Letters \textbf{106}(18), 185001 (2011).
\newblock \doi{10.1103/physrevlett.106.185001}

\bibitem{Amatucci:1999}
W.E. Amatucci, G.~Ganguli, D.N. Walker, D.~Duncan, Physics of Plasmas
  \textbf{6}(2), 619 (1999).
\newblock \doi{10.1063/1.873207}

\bibitem{Bonnell:1996}
J.~Bonnell, P.~Kintner, J.E. Wahlund, K.~Lynch, R.~Arnoldy, Geophysical
  Research Letters \textbf{23}(23), 3297 (1996).
\newblock \doi{10.1029/96gl03238}

\bibitem{Liu2004}
H.~Liu, G.~Lu, Annales Geophysicae \textbf{22}(4), 1149 (2004).
\newblock \doi{10.5194/angeo-22-1149-2004}.
\newblock \urlprefix\url{https://www.ann-geophys.net/22/1149/2004/}

\bibitem{Golovchanskaya:2014a}
I.V. Golovchanskaya, B.V. Kozelov, A.A. Chernyshov, M.M. Mogilevsky, A.A.
  Ilyasov, Physics of Plasmas \textbf{21}(8), 082903 (2014).
\newblock \doi{10.1063/1.4891668}

\bibitem{Golovchanskaya:2014b}
I.V. Golovchanskaya, B.V. Kozelov, I.V. Mingalev, M.N. Melnik, A.A. Lubchich,
  Annales Geophysicae \textbf{32}(1), 1 (2014).
\newblock \doi{10.5194/angeo-32-1-2014}

\bibitem{Pollock:1990}
C.J. Pollock, M.O. Chandler, T.E. Moore, J.H. Waite, C.R. Chappell, D.A.
  Gurnett, Journal of Geophysical Research \textbf{95}(A11), 18969 (1990).
\newblock \doi{10.1029/ja095ia11p18969}

\bibitem{Earle:1989}
G.D. Earle, M.C. Kelley, G.~Ganguli, Journal of Geophysical Research
  \textbf{94}(A11), 15321 (1989).
\newblock \doi{10.1029/ja094ia11p15321}

\bibitem{Bonnell:1997}
J.~Bonnell, Identification of broadband elf waves observed during transverse
  ion acceleration in the auroral ionosphere,.
\newblock Ph.D. thesis, Cornell University (1997)

\bibitem{Lundberg:2012}
E.T. Lundberg, P.M. Kintner, K.A. Lynch, M.R. Mella, Geophysical Research
  Letters \textbf{39}(1), L01107 (2012).
\newblock \doi{10.1029/2011gl050018}

\bibitem{Hamrin:2001}
M.~Hamrin, M.~André, G.~Ganguli, V.V. Gavrishchaka, M.E. Koepke, M.W. Zintl,
  N.~Ivchenko, T.~Karlsson, J.H. Clemmons, Journal of Geophysical Research:
  Space Physics \textbf{106}(A6), 10803 (2001).
\newblock \doi{10.1029/2001ja900003}

\bibitem{Wahlund:1994}
J.E. Wahlund, P.~Louarn, T.~Chust, H.d. Feraudy, A.~Roux, B.~Holback, P.O.
  Dovner, G.~Holmgren, Geophysical Research Letters \textbf{21}(17), 1831
  (1994).
\newblock \doi{10.1029/94gl01289}

\bibitem{Ergun:1998}
R.E. Ergun, C.W. Carlson, J.P. McFadden, F.S. Mozer, G.T. Delory, W.~Peria,
  C.C. Chaston, M.~Temerin, R.~Elphic, R.~Strangeway, R.~Pfaff, C.A. Cattell,
  D.~Klumpar, E.~Shelley, W.~Peterson, E.~Moebius, L.~Kistler, Geophysical
  Research Letters \textbf{25}(12), 2025 (1998).
\newblock \doi{10.1029/98gl00635}

\bibitem{Nykyri:2006}
K.~Nykyri, B.~Grison, P.J. Cargill, B.~Lavraud, E.~Lucek, I.~Dandouras,
  A.~Balogh, N.~Cornilleau-Wehrlin, H.~R\`eme, Annales Geophysicae
  \textbf{24}(3), 1057 (2006).
\newblock \doi{10.5194/angeo-24-1057-2006}.
\newblock \urlprefix\url{https://www.ann-geophys.net/24/1057/2006/}

\bibitem{Slapak:2017}
R.~Slapak, H.~Gunell, M.~Hamrin, Geophysical Research Letters \textbf{44}(1),
  22 (2017).
\newblock \doi{10.1002/2016gl071680}

\bibitem{Kaneko:2003}
T.~Kaneko, H.~Tsunoyama, R.~Hatakeyama, Physical Review Letters
  \textbf{90}(12), 125001 (2003).
\newblock \doi{10.1103/physrevlett.90.125001}

\bibitem{Kaneko:2005}
T.~Kaneko, E.W. Reynolds, R.~Hatakeyama, M.E. Koepke, Physics of Plasmas
  \textbf{12}(10), 102106 (2005).
\newblock \doi{10.1063/1.2102747}

\bibitem{Forbes:1981}
J.M. Forbes, Reviews of Geophysics \textbf{19}(3), 469 (1981).
\newblock \doi{10.1029/rg019i003p00469}

\bibitem{Parks:1984}
G.K. Parks, M.~McCarthy, R.J. Fitzenreiter, J.~Etcheto, K.A. Anderson, R.R.
  Anderson, T.E. Eastman, L.A. Frank, D.A. Gurnett, C.~Huang, R.P. Lin, A.T.Y.
  Lui, K.W. Ogilvie, A.~Pedersen, H.~Reme, D.J. Williams, Journal of
  Geophysical Research \textbf{89}(A10), 8885 (1984).
\newblock \doi{10.1029/ja089ia10p08885}

\bibitem{Cattell:1982}
C.A. Cattell, M.~Kim, R.P. Lin, F.S. Mozer, Geophysical Research Letters
  \textbf{9}(5), 539 (1982).
\newblock \doi{10.1029/gl009i005p00539}

\bibitem{Orsini:1984}
S.~Orsini, M.~Candidi, V.~Formisano, H.~Balsiger, A.~Ghielmetti, K.W. Ogilvie,
  Journal of Geophysical Research \textbf{89}(A3), 1573 (1984).
\newblock \doi{10.1029/ja089ia03p01573}

\bibitem{Cattell:1986}
C.A. Cattell, F.S. Mozer, E.W. Hones, R.R. Anderson, R.D. Sharp, Journal of
  Geophysical Research \textbf{91}(A5), 5663 (1986).
\newblock \doi{10.1029/ja091ia05p05663}

\bibitem{Angelopoulos:1989}
V.~Angelopoulos, R.C. Elphic, S.P. Gary, C.Y. Huang, Journal of Geophysical
  Research \textbf{94}(A11), 15373 (1989).
\newblock \doi{10.1029/ja094ia11p15373}

\bibitem{Amatucci:2003}
W.E. Amatucci, G.~Ganguli, D.N. Walker, G.~Gatling, M.~Balkey, T.~McCulloch,
  Physics of Plasmas \textbf{10}(5), 1963 (2003).
\newblock \doi{10.1063/1.1562631}

\bibitem{Matsubara:2000}
A.~Matsubara, T.~Tanikawa, Japanese Journal of Applied Physics \textbf{39}(Part
  1, No. 8), 4920 (2000).
\newblock \doi{10.1143/jjap.39.4920}

\bibitem{Kumar:2002}
T.A.S. Kumar, S.K. Mattoo, R.~Jha, Physics of Plasmas \textbf{9}(7), 2946
  (2002).
\newblock \doi{10.1063/1.1483074}

\bibitem{DuBois:2013}
A.M. DuBois, E.~Thomas, W.E. Amatucci, G.~Ganguli, Physical Review Letters
  \textbf{111}(14), 145002 (2013).
\newblock \doi{10.1103/physrevlett.111.145002}

\bibitem{Enloe:2017}
C.L. Enloe, E.M. Tejero, C.~Crabtree, G.~Ganguli, W.E. Amatucci, Physics of
  Plasmas \textbf{24}(5), 052107 (2017).
\newblock \doi{10.1063/1.4981923}

\bibitem{Ganguli:2014ks}
G.~Ganguli, E.~Tejero, C.~Crabtree, W.~Amatucci, L.~Rudakov, Physics of Plasmas
  \textbf{21}(1), 012107 (2014).
\newblock \doi{10.1063/1.4862032}

\bibitem{GanguliPalm:1987}
S.B. Ganguli, P.J. Palmadesso, Journal of Geophysical Research: Space Physics
  \textbf{92}(A8), 8673 (1987).
\newblock \doi{10.1029/JA092iA08p08673}.
\newblock
  \urlprefix\url{https://agupubs.onlinelibrary.wiley.com/doi/abs/10.1029/JA092iA08p08673}

\bibitem{Gangulietal:1988}
S.B. Ganguli, P.J. Palmadesso, H.G. Mitchell, Geophysical Research Letters
  \textbf{15}(11), 1291 (1988).
\newblock \doi{10.1029/GL015i011p01291}.
\newblock
  \urlprefix\url{https://agupubs.onlinelibrary.wiley.com/doi/abs/10.1029/GL015i011p01291}

\bibitem{Sharma:1995}
A.~Surjalal~Sharma, Reviews of Geophysics \textbf{33}(S1), 645 (1995).
\newblock \doi{10.1029/95RG00495}.
\newblock
  \urlprefix\url{https://agupubs.onlinelibrary.wiley.com/doi/abs/10.1029/95RG00495}

\bibitem{Klimas:1996}
A.J. Klimas, D.~Vassiliadis, D.N. Baker, D.A. Roberts, Journal of Geophysical
  Research: Space Physics \textbf{101}(A6), 13089 (1996).
\newblock \doi{10.1029/96JA00563}.
\newblock
  \urlprefix\url{https://agupubs.onlinelibrary.wiley.com/doi/abs/10.1029/96JA00563}

\bibitem{Gleisner:1996}
H.~Gleisner, H.~Lundstedt, P.~Wintoft, Annales Geophysicae \textbf{14}(7), 679
  (1996).
\newblock \doi{10.1007/s00585-996-0679-1}.
\newblock \urlprefix\url{https://www.ann-geophys.net/14/679/1996/}

\bibitem{Gavrishchaka:2001a}
V.V. Gavrishchaka, S.B. Ganguli, Journal of Geophysical Research: Space Physics
  \textbf{106}(A12), 29911 (2001).
\newblock \doi{10.1029/2001JA900118}.
\newblock
  \urlprefix\url{https://agupubs.onlinelibrary.wiley.com/doi/abs/10.1029/2001JA900118}

\bibitem{Gavrishchaka:2001b}
V.V. Gavrishchaka, S.B. Ganguli, Journal of Geophysical Research: Space Physics
  \textbf{106}(A4), 6247 (2001).
\newblock \doi{10.1029/2000JA900137}.
\newblock
  \urlprefix\url{https://agupubs.onlinelibrary.wiley.com/doi/abs/10.1029/2000JA900137}

\bibitem{Camporeale:2019}
E.~Camporeale, Space Weather \textbf{17}(8), 1166 (2019).
\newblock \doi{10.1029/2018SW002061}.
\newblock
  \urlprefix\url{https://agupubs.onlinelibrary.wiley.com/doi/abs/10.1029/2018SW002061}

\bibitem{Gopinath:2019}
S.~Gopinath, P.R. Prince, Journal of Earth System Science \textbf{128}(7), 172
  (2019).
\newblock \doi{10.1007/s12040-019-1194-6}.
\newblock \urlprefix\url{https://doi.org/10.1007/s12040-019-1194-6}

\bibitem{Gavrishchaka:2018}
V.~Gavrishchaka, Z.~Yang, R.~Miao, O.~Senyukova, Int. J. Mach. Learn. Comput.
  \textbf{8}(6), 549 (2018).
\newblock \doi{10.18178/ijmlc.2018.8.6.744}

\bibitem{Gavrishchaka:2019}
V.~Gavrishchaka, O.~Senyukova, M.~Koepke, Advances in Physics: X \textbf{4}(1),
  1582361 (2019).
\newblock \doi{10.1080/23746149.2019.1582361}.
\newblock \urlprefix\url{https://doi.org/10.1080/23746149.2019.1582361}

\bibitem{BANERJEE20072071}
S.~Banerjee, V.V. Gavrishchaka, Journal of Atmospheric and Solar-Terrestrial
  Physics \textbf{69}(16), 2071  (2007).
\newblock \doi{https://doi.org/10.1016/j.jastp.2007.08.004}.
\newblock
  \urlprefix\url{http://www.sciencedirect.com/science/article/pii/S1364682607002465}.
\newblock Recent Advances in the Polar Wind Theories and Observations

\bibitem{LeCun:2015}
Y.~LeCun, Y.~Bengio, G.~Hinton, Nature \textbf{521}(7553), 436 (2015).
\newblock \doi{10.1038/nature14539}.
\newblock \urlprefix\url{https://doi.org/10.1038/nature14539}

\bibitem{DengDong:2014}
L.~Deng, D.~Yu, Found. Trends Signal Process. \textbf{7}(3–4), 197–387
  (2014).
\newblock \doi{10.1561/2000000039}.
\newblock \urlprefix\url{https://doi.org/10.1561/2000000039}

\bibitem{Hinton504}
G.E. Hinton, R.R. Salakhutdinov, Science \textbf{313}(5786), 504 (2006).
\newblock \doi{10.1126/science.1127647}.
\newblock \urlprefix\url{https://science.sciencemag.org/content/313/5786/504}

\bibitem{Schapire:1992}
R.E. Schapire, The design and analysis of efficient learning algorithms.
\newblock Ph.D. thesis, Massachusetts Institute of Technology (1992)

\bibitem{friedman2000}
J.~Friedman, T.~Hastie, R.~Tibshirani, Ann. Statist. \textbf{28}(2), 337
  (2000).
\newblock \doi{10.1214/aos/1016218223}.
\newblock \urlprefix\url{https://doi.org/10.1214/aos/1016218223}

\bibitem{ChenGuestrin:2016}
T.~Chen, C.~Guestrin, in \emph{Proceedings of the 22nd ACM SIGKDD International
  Conference on Knowledge Discovery and Data Mining} (Association for Computing
  Machinery, New York, NY, USA, 2016), KDD ’16, p. 785–794.
\newblock \doi{10.1145/2939672.2939785}.
\newblock \urlprefix\url{https://doi.org/10.1145/2939672.2939785}

\bibitem{Gavrishchaka:2006}
G.V. V., \emph{Boosting-Based Frameworks in Financial Modeling: Application to
  Symbolic Volatility Forecasting} (Emerald Group Publishing Limited, 2006),
  vol. 20 Part 2, pp. 123--151.
\newblock \doi{10.1016/S0731-9053(05)20024-5}.
\newblock \urlprefix\url{https://doi.org/10.1016/S0731-9053(05)20024-5}

\bibitem{Senyukova:2011}
O.V. Senyukova, V.V. Gavrishchaka, in \emph{Computational Intelligence and
  Bioinformatics / 755: Modelling, Simulation, and Identification} (2011).
\newblock \doi{10.2316/P.2011.753-025}

\bibitem{Miao2020}
R.~{Miao}, Z.~{Yang}, V.~{Gavrishchaka}, in \emph{2020 3rd International
  Conference on Information and Computer Technologies (ICICT)} (2020), pp.
  107--113.
\newblock \doi{10.1109/ICICT50521.2020.00025}

\bibitem{Carlsson:2009}
G.~Carlsson, Bull. Amer. Math. Soc. \textbf{46} (2009).
\newblock \doi{10.1090/S0273-0979-09-01249-X}

\bibitem{Edelsbrunner:2014}
H.~Edelsbrunner, \emph{A Short Course in Computational Geometry and Topology}
  (Springer, 2014)

\bibitem{GARLAND201649}
J.~Garland, E.~Bradley, J.D. Meiss, Physica D: Nonlinear Phenomena
  \textbf{334}, 49  (2016).
\newblock \doi{https://doi.org/10.1016/j.physd.2016.03.006}.
\newblock
  \urlprefix\url{http://www.sciencedirect.com/science/article/pii/S0167278916000464}.
\newblock Topology in Dynamics, Differential Equations, and Data

\bibitem{Bernhardt:1995}
P.A. Bernhardt, G.~Ganguli, M.C. Kelley, W.E. Swartz, Journal of Geophysical
  Research \textbf{100}(A12), 23811 (1995).
\newblock \doi{10.1029/95ja02836}

\bibitem{Ganguli:1992}
G.~Ganguli, P.A. Bernhardt, W.~Scales, P.~Rodriguez, C.~Siefring, H.A. Romero,
  in \emph{Physics of Space Plasmas (1992), SPI Conference Proceedings and
  Reprint Series}, ed. by T.~Chang (Scientific Publishers, Inc.,, Cambridge,
  MA., 1993), p. 161

\bibitem{Scales:1992}
W.~Scales, P.A. Bernhardt, G.~Ganguli, in \emph{Physics of Space Plasmas
  (1992), SPI Conference Proceedings and Reprint Series}, ed. by T.~Chang
  (Scientific Publishers, Inc.,, Cambridge, MA., 1993), p. 161

\bibitem{Argo:1992}
P.~Argo, T.J. Fitzgerald, R.~Carlos, Radio Science \textbf{27}(2), 289 (1992).
\newblock \doi{10.1029/91rs02916}

\bibitem{Scales:1994}
W.A. Scales, P.A. Bernhardt, G.~Ganguli, Journal of Geophysical Research
  \textbf{99}(A1), 373 (1994).
\newblock \doi{10.1029/93ja02752}

\bibitem{Scales:1995}
W.A. Scales, P.A. Bernhardt, G.~Ganguli, Journal of Geophysical Research
  \textbf{100}(A1), 269 (1995).
\newblock \doi{10.1029/94ja02490}

\bibitem{Mostovych:1989}
A.N. Mostovych, B.H. Ripin, J.A. Stamper, Physical Review Letters
  \textbf{62}(24), 2837 (1989).
\newblock \doi{10.1103/physrevlett.62.2837}

\bibitem{Peyser:1992}
T.A. Peyser, C.K. Manka, B.H. Ripin, G.~Ganguli, Physics of Fluids B: Plasma
  Physics \textbf{4}(8), 2448 (1992).
\newblock \doi{10.1063/1.860213}

\bibitem{Niemann:2013}
C.~Niemann, W.~Gekelman, C.G. Constantin, E.T. Everson, D.B. Schaeffer, S.E.
  Clark, D.~Winske, A.B. Zylstra, P.~Pribyl, S.K.P. Tripathi, D.~Larson, S.H.
  Glenzer, A.S. Bondarenko, Physics of Plasmas \textbf{20}(1), 012108 (2013).
\newblock \doi{10.1063/1.4773911}

\bibitem{Kaneko:2007}
T.~Kaneko, K.~Hayashi, R.~Ichiki, R.~Hatakeyama, Fusion Science and Technology
  \textbf{51}(2T), 103 (2007).
\newblock \doi{10.13182/fst07-a1326}

\bibitem{Gavrishchaka:1997}
V.V. Gavrishchaka, M.E. Koepke, G.I. Ganguli, Journal of Geophysical Research:
  Space Physics \textbf{102}(A6), 11653 (1997).
\newblock \doi{10.1029/97ja00639}

\end{thebibliography}

\end{document}